%% file: MultiJetQCD.tex
\newcommand*{\ATLASLATEXPATH}{latex/}
\pdfoutput=1
\documentclass[cernpreprint,UKenglish,texlive=2011,txfonts]{\ATLASLATEXPATH atlasdoc}

\usepackage{\ATLASLATEXPATH atlaspackage}
\usepackage{multirow}
\usepackage{tikz}
\usetikzlibrary{calc}
\usepackage{\ATLASLATEXPATH atlasbiblatex}

\usepackage{\ATLASLATEXPATH atlascontribute}

\usepackage{\ATLASLATEXPATH atlasphysics}
\graphicspath{{logos/}{figures/}}

\hypersetup{pdftitle={ATLAS draft},pdfauthor={The ATLAS Collaboration}}
\input{MultiJetQCD-metadata}
\input{tex/multijetdefs}

\begin{document}

\maketitle

\tableofcontents

\section{Introduction}\label{sec:intro}
\input{tex/intro}

\section{The ATLAS detector}\label{sec:atlas}
\input{tex/atlas}

\section{Cross-section definition} \label{sec:cross_section}
\input{tex/cross_section}

\section{Monte Carlo samples}\label{sec:samples:LO}
\input{tex/samples}
\subsection{Theoretical uncertainties}\label{sec:theoUncertainties}
\input{tex/theoretical_uncertainties}

\section{Data selection and calibration}\label{sec:eventselection}
\input{tex/eventselection}

\section{Data unfolding}\label{sec:unfolding}
\input{tex/unfolding}

\section{Experimental uncertainties}\label{sec:systematics}
\input{tex/systematics}

\section{Results}\label{sec:results}
\input{tex/results}

\FloatBarrier

\section{Conclusion}\label{sec:conclusion}
\input{tex/conclusion}

\section*{Acknowledgements}
\label{sec:acknowldge}
\input{acknowledgements/Acknowledgements}

\clearpage
\appendix
\section{Tables of the measured cross sections}\label{app}

\input{tex/xs_tables.tex}

\addcontentsline{toc}{part}{References}
\printbibliography

\clearpage
\input{atlas_authlist}
\clearpage
\appendix

\end{document}

%% file: MultiJetQCD-metadata.tex
\newcommand{\ourintlumi}{{\SI{20.3}{\per\fb}}}
\newcommand{\multijettitle}{{Measurement of four-jet differential cross sections in ${\bf \sqrt{s}=8}$\,TeV proton--proton collisions using the ATLAS detector}}
\AtlasTitle{\multijettitle}

\author{The ATLAS Collaboration}

\AtlasRefCode{STDM-2014-14}

\AtlasNote{ATL-COM-PHYS-2015-135}

\PreprintIdNumber{CERN-PH-2015-181}

 \AtlasJournal{JHEP}

\newcommand\multijetabstract{{Differential cross sections for the production of at least four jets have been measured in
proton--proton collisions at $\sqrt{s} = 8$~TeV at the Large Hadron Collider using
the ATLAS detector.
Events are selected if the four \antikt\ $R=0.4$ jets with the largest
transverse momentum (\pt) within the rapidity range $|y|<2.8$ are well separated
($\dRmin>0.65$), all have $\pT>64$~GeV, and include at least one jet
with $\pt>100$~GeV.
The dataset corresponds to an integrated luminosity of
\ourintlumi. The cross sections, corrected for detector effects, are compared to leading-order and next-to-leading-order calculations as a function of the jet momenta, invariant
masses, minimum and maximum opening angles and other kinematic variables.
}} 
\AtlasAbstract{%
\multijetabstract
}

\AtlasCoverSupportingNote{ATL-COM-PHYS-2014-1363}{https://cds.cern.ch/record/1956329}

\AtlasCoverAnalysisTeam{Alan Barr, Mireia Crispin-Ortuzar, Zach Marshall, Matthew Mondragon, Ricardo Piegaia, Sabrina Sacerdoti, Anna Sfyrla, Christopher J. Young}

\AtlasCoverEdBoardMember{Shima Shimizu~(chair), Joey Houston, Francois Corriveau, Lee Sawyer}

\AtlasCoverEgroupEditors{atlas-STDM-2014-14-editors@cern.ch}

\AtlasCoverEgroupEdBoard{atlas-STDM-2014-14-editorial-board@cern.ch}

%% file: tex/multijetdefs.tex
\newcommand\tikzmark[1]{\tikz[remember picture,overlay] \coordinate (#1);}

\newcommand{\figref}[1]{figure~\ref{#1}}
\newcommand{\figsref}[1]{figures~\ref{#1}}
\newcommand{\secref}[1]{section~\ref{#1}}
\newcommand{\secsref}[1]{sections~\ref{#1}}
\newcommand{\tabref}[1]{table~\ref{#1}}

\newcommand{\TwoToTwo}     {\ensuremath{2 \rightarrow 2}}

\newcommand{\TwoToFour}     {\ensuremath{2 \rightarrow 4}}

\newcommand{\pti}     {\ensuremath{p_{\text{T}}^{(i)}}}
\newcommand{\ptOne}     {\ensuremath{p_{\text{T}}^{(1)}}}
\newcommand{\ptTwo}     {\ensuremath{p_{\text{T}}^{(2)}}}
\newcommand{\ptThree}     {\ensuremath{p_{\text{T}}^{(3)}}}
\newcommand{\ptFour}     {\ensuremath{p_{\text{T}}^{(4)}}}
\newcommand{\mjjjj}     {\ensuremath{m_{\text{4j}}}}
\newcommand{\mjjminvar} {\ensuremath{m_{\text{2j}}^{\min}}}
\newcommand{\mjjminratio}{\ensuremath{\mjjminvar/\mjjjj}}
\newcommand{\dRmin}     {\ensuremath{\Delta R_{\text{4j}}^{\min}}}
\newcommand{\minDphiij} {\ensuremath{\Delta\phi_{\text{2j}}^{\min}}}
\newcommand{\minDphiijk}{\ensuremath{\Delta\phi_{\text{3j}}^{\min}}}
\newcommand{\minDrapij} {\ensuremath{\Delta y_{\text{2j}}^{\min}}}
\newcommand{\minDrapijk}{\ensuremath{\Delta y_{\text{3j}}^{\min}}}
\newcommand{\maxDrapij} {\ensuremath{\Delta y_{\text{2j}}^{\max}}}
\newcommand{\sumptcent} {\ensuremath{\Sigma p_{\text{T}}^{\text{central}}}}

\newcommand{\akt}       {\ensuremath{{\textrm{anti-}}k_{t}}\xspace}

\newcommand{\antikt}    {\akt}

\newcommand{\mc}[1]{{\small\tt{#1}}}
\newcommand{\mcc}[1]{{\small\textsc{#1}}}
\newcommand{\Pythia}{{\mcc{Pythia}}}
\newcommand{\Pythiav}[1]{{\small\textsc{Pythia\,{#1}}}}
\newcommand{\Herwigpp}{{\mcc{Herwig++}}}
\newcommand{\Herwigppv}[1]{\mcc{\Herwigpp\,{#1}}}

\newcommand{\Madgraph}{{\mcc{MadGraph}}}

\newcommand{\MadgraphPythia}{{\mcc{MadGraph}+\Pythia}}

\newcommand{\Comix}{{\mcc{Comix}}}
\newcommand{\Sherpa}{{\mcc{Sherpa}}}

\newcommand{\BlackHatSherpa}{{\mcc{BlackHat/Sherpa}}}
\newcommand{\BlackHat}{{\mcc{BlackHat}}}
\newcommand{\HEJ}{{\mcc{HEJ}}}
\newcommand{\NJET}{{\mcc{NJet}}}
\newcommand{\NJETSherpa}{{\mcc{NJet/Sherpa}}}

\newcommand{\pQCD}{\textit{p}\textrm{QCD}}

\def\lsim{\mathrel{\rlap{\lower4pt\hbox{\hskip1pt$\sim$}}
    \raise1pt\hbox{$<$}}}                
\def\gsim{\mathrel{\rlap{\lower4pt\hbox{\hskip1pt$\sim$}}
    \raise1pt\hbox{$>$}}}                

%% file: tex/intro.tex
The production of particle jets at hadron colliders such as the Large Hadron
Collider (LHC)~\cite{Evans:2008zzb}  provides a fertile testing ground for the theory describing strong interactions,
Quantum Chromodynamics (QCD).  In QCD, jet production is
interpreted as the fragmentation of quarks and gluons produced in the
scattering process followed by their subsequent hadronisation.  At
high transverse momenta (\pT) the scattering of partons can be
calculated using perturbative QCD (\pQCD) and experimental
jet measurements are directly related to the scattering of quarks and
gluons.  The large cross sections for such processes allow
for differential measurements in a wide kinematic range
and stringent testing of the underlying theory.

This analysis studies events where at least four jets are produced in a
hard-scatter process. These events are of particular interest as the corresponding Feynman diagrams
require several vertices even at leading-order (LO) in the strong coupling
constant \alphas. The current state-of-the-art theoretical predictions for such
processes are at next-to-leading-order in \alphas (next-to-leading-order perturbative
QCD, NLO \pQCD{})~\cite{Berger:2008sj,Bern:2011ep}, and they have recently been
combined with parton shower (PS) simulations~\cite{Hoeche:2012yf}.
An alternative approach is taken by generators which provide a matrix element
(ME) for the hardest $2\rightarrow2\,$ process while the rest of the jets are provided by
a PS model, which implements a resummation of the leading-logarithmic terms
(e.g. \Pythia\,~8~\cite{Sjostrand:2007gs} and \Herwigpp~\cite{Bahr:2008pv}).
It is also interesting to test multi-leg (i.e., $2\rightarrow n$) LO \pQCD\ generators 
(e.g. \Sherpa~\cite{Gleisberg:2008ta} or \Madgraph~\cite{Alwall:2014hca}), since
they may provide adequate descriptions of the data in specific kinematic regions
and have the advantage of being less computationally expensive than NLO
calculations. 

It is interesting to note that the previous ATLAS measurement of multi-jet production at ${\sqrt{s}=7}$\,TeV ~\cite{Collaboration:2011tq}
indicates that predictions may differ from data by $\sim 30$\% even at NLO~\cite{badger2014next}.
This work explores a variety of kinematic regimes and topological
distributions to test the validity of QCD calculations, including the PS
approximation and the necessity of higher-order ME in Monte Carlo (MC)
generators. 

Additionally, four-jet events
represent a background to many other processes at hadron colliders. Hence, the
predictive power of the QCD calculations, in particular their ability to
reproduce the shapes of the distributions studied  in this analysis, is of general interest. While 
searches for new phenomena in multi-jet events use data-driven techniques to estimate the contribution from QCD
events, as was done for example in ref.~\cite{Aad:2013wta}, these methods are
tested in MC simulations. The accuracy of the theoretical predictions remains
therefore important. 

Three-jet events have been measured differentially by many experiments. Indeed
it was observations of such events that heralded the discovery of the
gluon~\cite{Brandelik:1979bd,PhysRevLett.43.830,1980PhLB...91..142B,1979PhLB...86..418B}.
More recently, at the LHC, ATLAS has measured the three-jet cross section
differentially~\cite{Aad:2014rma} and CMS has used the ratio of three to two jet
events to measure \alphas~\cite{Chatrchyan:2013txa}. Event shape variables have also been measured, showing sensitivity to higher-order \pQCD{} effects~\cite{Aad:2012np,Khachatryan:2014ika}. Multi-jet cross sections have been measured previously at CMS~\cite{Khachatryan:2015xwa}, ATLAS~\cite{Collaboration:2011tq}, CDF~\cite{CDFmultijet,CDFmultijet2} and D0~\cite{D0multijet,D0multijet2}, although with smaller datasets and/or lower energy, and generally focussed on different observables. 

This paper presents the differential cross sections for events with at least
four jets, studied as a function of a variety of kinematic and topological variables
which include momenta, masses and angles. 
Events are selected if the four \antikt\ $R=0.4$ jets with the largest
transverse momentum within the rapidity range $|y|<2.8$ are well separated,
all have $\pt>64$~GeV, and include at least one jet with $\pt>100$~GeV.
The measurements are corrected for
detector effects. The variables are binned in the leading jet
\pt\ and the total invariant mass, such that different regimes and
configurations can be tested. The measurements are sensitive to 
the various mass scales in an event, the presence of forward jets, or the
azimuthal configuration of the jets -- that is, one jet recoiling against three, or two
recoiling against two. 

The structure of the paper is as follows. Section~\ref{sec:atlas} introduces the
ATLAS detector. The observables and phase space of interest are defined in
\secref{sec:cross_section}. The MC simulation samples studied in this work are
summarised in \secref{sec:samples:LO}, while the theory predictions and their
uncertainties are described in \secref{sec:samples:NLO}. The trigger, jet calibration and
data cleaning are presented in \secref{sec:eventselection}.  The unfolding
of detector effects is detailed in \secref{sec:unfolding}.
Section~\ref{sec:systematics} provides the experimental uncertainties included in
the final distributions. 
Finally, the results are shown in \secref{sec:results} and the conclusions are drawn in \secref{sec:conclusion}.

%% file: tex/atlas.tex
The ATLAS experiment~\cite{Aad:2008zzm} is a multi-purpose particle
physics detector with a forward-backward symmetric cylindrical
geometry and nearly 4$\pi$ coverage in solid angle, with instrumentation up to
$|\eta|=4.9$.\footnote{\scriptsize{ATLAS uses a right-handed Cartesian coordinate system with its origin
at the nominal interaction point~(IP) in the centre of the detector.
The $z$-axis is taken along the beam pipe, and the $x$-axis points from the
IP to the centre of the LHC ring. Cylindrical coordinates $(r,\phi)$ are used in the
transverse plane, $\phi$ being the azimuthal angle around the beam
pipe. The rapidity $y$ is defined by $\frac{1}{2}\ln\frac{E+p_z}{E-p_z}$,
the pseudorapidity in terms of the polar angle
$\theta$ as $\eta=-\ln\tan(\theta/2)$.}}

The layout of the detector is based on four superconducting magnet systems, 
which comprise a thin solenoid surrounding the inner tracking detectors~(ID) 
and a barrel and two end-cap toroids generating the magnetic field for a large muon spectrometer. 
The calorimeters are located between the ID and the muon system.  
The lead/liquid-argon~(LAr) electromagnetic~(EM) calorimeter is split into
two regions: the barrel ($|\eta|< 1.475$) and the end-cap ($1.375 < |\eta|< 3.2$). The hadronic
calorimeter is divided into four regions: the barrel ($|\eta|< 0.8$) and the
extended barrel ($0.8 < |\eta|< 1.7$) made of scintillator/steel, the end-cap
($1.5 < |\eta|< 3.2$) with LAr/copper modules, and the forward calorimeter
 ($3.1 < |\eta| < 4.9$) composed of LAr/copper and LAr/tungsten modules.

A three-level trigger system~\cite{Aad:2012xs} is used to select events for further analysis. 
The first level (L1) of the trigger reduces the event rate to less than \SI{75}{\kHz} using
hardware-based trigger algorithms acting on a subset of detector
information. The second level (L2) uses fast online algorithms, while the final trigger
stage, called the Event Filter~(EF), uses reconstruction software with algorithms similar to
the offline versions. The last two  software-based trigger levels, referred to
collectively as the High-Level Trigger~(HLT), further reduce the 
event rate to about \SI{400}{\Hz}.

%% file: tex/cross_section.tex
\begin{table*}
\centering
\renewcommand\arraystretch{1.7}
\begin{tabular}{c|c|p{6cm}}
\hline
Name & Definition & Comment \\
\hline
$\pti$    & Transverse momentum of the $i$th jet & Sorted descending in \pt \\
\HT          & $\sum\limits_{i=1}^{4} \pti$ &Scalar sum of the \pt of the four jets \\
\mjjjj\       & $\left(\left(\sum\limits_{i=1}^4 E_i\right)^2 - \left(\sum\limits_{i=1}^4 \mathbf{p}_i\right)^2 \right)^{1/2}$ & Invariant mass of the four jets \\
\mjjminratio\   & $\min_{\substack{{i,j} \in [1,4] \\ i \neq j}} \left(\left(E_i + E_j \right)^2 - \left( \mathbf{p}_i + \mathbf{p}_j \right)^2 \right)^{1/2}$ / \mjjjj\  &  Minimum invariant mass of two jets relative to invariant mass of four jets  \\
\minDphiij\   & $\min_{\substack{i,j \in [1,4]\\i \neq j}}\left(|\phi_i-\phi_j|\right)$ & Minimum azimuthal separation of two jets \\
\minDrapij\   &  $\min_{\substack{i,j \in [1,4]\\i \neq j}}\left(|y_i-y_j|\right)$ & Minimum rapidity separation of two jets \\
\minDphiijk\  & $\min_{\substack{i,j,k \in [1,4]\\i \neq j \neq k}}\left(|\phi_i-\phi_j|+|\phi_j-\phi_k|\right)$ & Minimum azimuthal separation between any three jets \\
\minDrapijk\  &  $\min_{\substack{i,j,k \in [1,4]\\i \neq j \neq k}}\left(|y_i-y_j|+|y_j-y_k|\right)$ & Minimum rapidity separation between  any three jets \\
\maxDrapij\   & $\Delta y^{\max}_{ij} = \max_{i,j \in [1,4]}\left(| y_{i}- y_{j}|\right)$ & Maximum rapidity difference between two jets \\
\sumptcent\   & $|p_{\rm T}^{c}|+|p_{\rm T}^{d}|$ & If \maxDrapij\ is defined by jets $a$ and $b$, this is the scalar sum of the \pt\ of the other two jets, $c$ and $d$ (`central' jets)  \\
\hline
\end{tabular}
\caption{\label{tab:variables}
Definitions of the various kinematic variables measured. Only the four 
jets with the largest \pt{} are considered in all cases. 
}
\end{table*}

This measurement uses jets reconstructed with the \antikt
algorithm~\cite{Cacciari:2008gp} with four-momentum recombination as implemented in the
FastJet package~\cite{fastjet}. The radius parameter is $R=0.4$. 

Cross sections are calculated for events with at least four jets within the rapidity range $|y|<2.8$. 
Out of those four jets, the leading one must have $\pt>100$~GeV, while the next three must have $\pt>64$~GeV. 
In addition, these four jets must be well separated from one another by $\dRmin>0.65$, 
where $\dRmin\ = \min_{\substack{i,j\in[1,4]\\ i \neq j}}(\Delta R_{ij})$, 
and $\Delta R_{ij} = (|y_i-y_j|^2 + |\phi_i-\phi_j|^2)^{1/2}$. This set of criteria is also referred to 
as the `inclusive analysis cuts' to differentiate them from the cases where additional
requirements are made, for example on the invariant mass of the four leading jets. 
The inclusive analysis cuts are mainly motivated by the triggers used to select events, described in~\secref{sec:trigger}.

Cross sections are measured differentially as a function of the kinematic variables defined in \tabref{tab:variables}; 
the list includes momentum variables, mass variables and angular variables. 
The only jets used in all cases are the four leading ones in \pt.
The observables were selected
for their sensitivity to differences between different Monte Carlo models of QCD processes and their ability to describe
the dynamics of the events.
For example, the \HT\ variable is often used to set the scale of multi-jet
processes. The four-jet invariant mass \mjjjj\ is representative of the largest
energy scale in the event whereas \mjjminvar{}, the minimum dijet invariant
mass, probes the smallest jet-splitting scale. The ratio $\mjjminvar/\mjjjj$ therefore provides information about the
range of energy scales relevant to the QCD calculation. The \minDphiij\ and
\minDrapij\ variables quantify the minimum angular separation between any two
jets. The azimuthal variable \minDphiijk{} distinguishes events with pairs of
nearby jets (which have large \minDphiijk) from the recoil of three jets against one
(leading to small \minDphiijk\ values). The rapidity variable \minDrapijk\ works
in a similar way. The \maxDrapij\ and \sumptcent\ variables are designed to be
sensitive to events with forward jets. In order to build \sumptcent, first the
two jets with the largest rapidity interval in the event are identified, and
then the scalar sum of the \pt of the remaining two jets is calculated. 

Different phase-space regions are probed by binning the variables in regions defined by a lower bound on \ptOne\ and \mjjjj.
This allows one to distinguish between the two types of topologies characterised by \minDphiijk,
or to track the position of the leading jet with respect to the forward--backward pair in the \sumptcent\ variables.
Table \ref{tab:bins} summarises all the phase-space regions considered in the analysis for each of the variables. 

The resulting differential cross-section distributions are corrected for
detector effects (\textit{unfolding}) and taken to the so-called particle-jet level, or simply `particle level'. In the MC simulations used in the unfolding procedure, particle jets are built from
particles with a proper lifetime $\tau$ satisfying $c\tau>10$ mm, including
muons and neutrinos from hadron decays. The event selection described above is applied to particle jets to define the phase space of the unfolded results.

\begin{table*}[h!]
\centering
\renewcommand\arraystretch{1.3}
\begin{tabular}{c | c | c | c | c | c}
\hline
Observable    & $\dRmin>...$ & $\ptFour>...$~[GeV] &  $\ptOne>...$~[GeV] & $\mjjjj>...$~[GeV] & $\maxDrapij>...$ \\
\hline
$\pti$        & \tikzmark{c} & \tikzmark{a} & 100 & - & - \\
\HT           &   &   &  100 & -  & - \\
\mjjjj\       &   &   & 100 & -  & - \\ 
\mjjminratio\ &   &   &  100 & 500, 1000, 1500, 2000 & - \\
\minDphiij\     &   &  & 100, 400, 700, 1000 & - & - \\
\minDrapij\     &   &  & 100, 400, 700, 1000 & - & - \\
\minDphiijk\    &   &  & 100, 400, 700, 1000 & - & - \\
\minDrapijk\    &   &  & 100, 400, 700, 1000 & - & -  \\
\maxDrapij\     &   &  &  100, 250, 400, 550 & - & - \\ 
\sumptcent\     & \tikzmark{d}  & \tikzmark{b} &  100, 250, 400, 550 & - & 1, 2, 3, 4\\
\hline
\end{tabular}
\caption{\label{tab:bins} Summary of the analysed phase-space regions, including
the \ptOne, \mjjjj\ and \maxDrapij\ bins into which each of the differential cross-section
measurements is split (a dash indicates when the cut is not applied on a
variable). The \dRmin\ and \ptFour\ requirements, specified in the
second and third columns respectively, apply to all variables. The observables
are defined in \tabref{tab:variables}.}

\tikz[remember picture,overlay]
{\draw ([yshift=1ex]a) -- (b) node[midway,fill=white] {64};}

\tikz[remember picture,overlay]
{\draw ([yshift=1ex]c) -- (d) node[midway,fill=white] {0.65};}

\end{table*}

Double parton interactions have not been investigated independently, so the measurement is inclusive in this respect.
They are expected to contribute 1\% or less to the results.

%% file: tex/samples.tex
Monte Carlo samples are used to estimate experimental systematic uncertainties,
deconvolve detector effects, and provide predictions to be compared with the
data.
Leading-order Monte Carlo samples are used for all three purposes. A set of
theoretical calculations at higher orders, described in~\secref{sec:samples:NLO}, are also compared to
the data.
The full list of generators is shown in~\tabref{tab:mc}. 

\begin{table}
\centering\renewcommand\arraystretch{1.3}
\footnotesize
\begin{tabular}{l|l|l|l|l|l|l}
\hline
{\bf Name} & {\bf Hard scattering} & {\bf LO/NLO} & {\bf PDF} & {\bf PS/UE} & {\bf Tune} & {\bf Factor}\\
\hline
Pythia & \Pythiav{8} & LO (\TwoToTwo) & \mc{CT10} & \Pythiav{8} & \mc{AU2-CT10} & 0.6 \\ 
Herwig++  & \Herwigpp\ & LO (\TwoToTwo) & \mc{CTEQ6L1} & \Herwigpp & \mc{UE-EE-3-CTEQ6L1} & 1.4 \\ 
MadGraph+Pythia & \Madgraph\ & LO (\TwoToFour) & \mc{CTEQ6L1} & \Pythiav{6}  & \mc{AUET2B-CTEQ6L1} & 1.1 \\ 
HEJ & \HEJ\ & All $^{\dagger}$ & \mc{CT10} & --- & ---  &  0.9 \\
BlackHat/Sherpa & \BlackHatSherpa\ &  NLO (\TwoToFour) &\mc{CT10} & --- & --- &  --- \\
NJet/Sherpa & \NJETSherpa\ & NLO (\TwoToFour) & \mc{CT10} & --- & --- &  --- \\
\hline
\end{tabular}
\caption{\label{tab:mc}
The generators used for comparison against the data
are listed, together with the parton distribution functions (PDFs), 
PS algorithms, underlying event (UE) and parameter tunes.
Each MC prediction is multiplied by a normalisation factor (last column) as described 
in~\secref{sec:samples:normalisation}, except \BlackHatSherpa\ and \NJETSherpa.
$(\dagger)$~The \HEJ\ sample is based on an approximation to all orders in \alphas.
}
\end{table}

The samples used in the experimental studies comprise two LO $2 \rightarrow 2$ generators, \Pythiav{8.160}~\cite{Sjostrand:2007gs} and \Herwigppv{2.5.2}~\cite{Bahr:2008pv},
and the LO multi-leg generator \mc{\Madgraph5\,v1.5.12}~\cite{Alwall:2014hca}.
As described in the introduction, LO generators are still widely used in searches for new physics, which motivates the comparison of their predictions to the data.

Both \Pythia{} and \Herwigpp{} employ leading-logarithmic PS models matched to LO
ME calculations. \Pythia{} uses a PS algorithm based on \pt ordering,
while the PS model implemented in \Herwigpp{}
follows an angular ordering. The ME calculation provided by \Madgraph{} contains up to four outgoing partons in the ME. It is matched
to a PS generated with \Pythiav{6.427}~\cite{Sjostrand:2006za} using
the shower $k_{\rm t}$-jet MLM matching~\cite{hoche2006matching}, where the
jet--parton matching scale is set to 20 GeV.
Hadronisation effects are included via the string model in the case of the
\Pythia{} and \Madgraph{} samples~\cite{Sjostrand:2006za}, or the cluster model~\cite{Corcella:2002jc} in events simulated with \Herwigpp{}.
The parton distribution functions (PDFs) used are the NLO \mc{CT10}~\cite{Lai:2010vv} 
or the LO distributions of~\mc{CTEQ6L1}~\cite{Pumplin:2002vw}
as shown in \tabref{tab:mc}.

Simulations of the underlying event, including multiple parton
interactions, are included in all three LO samples.
The parameter tunes employed are the ATLAS tunes
\mc{AU2}~\cite{ATL-PHYS-PUB-2012-003}
and \mc{AUET2B}~\cite{ATLAS:2011zja} for \Pythia\ and \Madgraph\ respectively, and
the \Herwigpp{} tune \mc{UE-EE-3}~\cite{Gieseke:2012ft}.

The multiple $pp$ collisions within the same and neighbouring bunch crossings
(pile-up) are simulated as additional inelastic $pp$ collisions using \Pythiav{8}. 
Finally, the interaction of particles with the ATLAS detector is simulated
using a GEANT4-based program~\cite{Agostinelli:2002hh,Aad:2010ah}.

\section{Theoretical predictions}
\label{sec:samples:NLO}

The results of the measurement are compared to NLO predictions, in addition to 
the LO samples described in \secref{sec:samples:LO}. These are calculated using
\BlackHatSherpa~\cite{Berger:2008sj,Bern:2011ep} and
\NJETSherpa~\cite{badger2013numerical,Badger2013965}, and have been provided by
their authors. They are both fixed-order calculations with no PS and no hadronisation. Therefore, the results are presented at the parton-jet
level, that is, using jets built from partons instead of hadrons.
For the high-\pT phase space covered in this analysis, non-perturbative corrections are expected to be small~\cite{Aad:2014qxa,Aad:2014vwa}.
\BlackHat\ performs one-loop virtual corrections using the unitarity method
and on-shell recursion. The remaining terms of the full NLO computation are
obtained with \mc{AMEGIC++}~\cite{Krauss:2001iv,gleisberg2008automating}, part
of \Sherpa. 
\NJET\ makes a numerical evaluation of the one-loop virtual corrections to multi-jet production in massless QCD. 
The Born matrix elements are evaluated with the \Comix\
generator~\cite{Gleisberg:2008fv,Hoeche:2012xx} within \Sherpa.
\Sherpa\ also performs the phase-space integration and infra-red subtraction via the Catani-Seymour dipole formalism.
Both the \BlackHatSherpa\ and \NJETSherpa\ predictions use the \mc{CT10} PDFs. 

The results are also compared to
predictions provided by \HEJ\ ~\cite{andersen2010constructing,andersen2010factorization,andersen2010high}. \HEJ\ is
a fully exclusive Monte Carlo event generator based on a perturbative cross-section calculation which approximates the hard-scattering
ME to all orders in the strong coupling constant \alphas for jet multiplicities of two or greater. The approximation is exact
in the limit of large separation in rapidity between partons. The calculation uses the \mc{CT10} PDFs. 
As in the case of the NLO predictions, no PS or hadronisation are included.  

The different predictions tested are expected to display various levels of
agreement in different kinematic configurations.
The generators which combine \TwoToTwo\ parton matrix elements (MEs) with parton showers (PSs)
are in principle not expected to provide a good description of the data,
particularly in regions where the additional jets are neither soft nor collinear. 
A previous measurement of multi-jet cross sections at 7~TeV by the ATLAS
Collaboration~\cite{Collaboration:2011tq} found that the cross section predicted
by MC models typically disagreed with the data by $\mathcal{O}(40\%)$. It also
found disagreements of up to 50\% in the shape of the differential cross section
measured as a function of \ptOne\ or \HT. Nevertheless, there are also examples
of exceptional cases where these calculations perform well, which adds interest
to the measurement; for example, the same 7~TeV ATLAS paper observed that the
shape of the \ptFour\ distribution was described by \Pythia\ within just 10\%.
It is also interesting to test whether PSs based on an angular
ordering perform better in angular variables such as \minDphiij\ or \minDphiijk\
than those using momentum ordering. In contrast to PS predictions,
multi-leg matrix element calculations matched to parton showers (ME+PS) were
seen at 7~TeV to significantly improve the accuracy of the cross-section
calculation and the shapes of the momentum observables. In the present analysis, such calculations
are expected to perform better in events with additional high-\pt\ jets and/or
large combined invariant masses of jets. This is also the type of scenario where
\HEJ\ is expected to perform well, since it provides an all-order description of
processes with more than two hard jets, and it is designed to capture the hard,
wide-angle emissions which a standalone PS approach would miss.
Variables such as \maxDrapij, \minDrapijk\ or \sumptcent\ were included in the
analysis with this purpose in mind. Finally, the fixed-order, four-jet NLO
predictions are expected to provide a better estimation of the cross sections
than the LO calculations. Interestingly, studies at 7~TeV found that the NLO
cross section for four-jet events was $\sim30\%$ higher than the data~\cite{badger2014next}.

\subsection{Normalisation}
\label{sec:samples:normalisation}

To facilitate comparison with the data, the cross sections predicted by the LO generators as well
as \HEJ\ are multiplied by a scale factor. The factor is such that the integrated number of events
in the region $\SI{500}{GeV} < \ptOne < \SI{1.5}{TeV}$ which satisfy the inclusive analysis cuts
in \secref{sec:cross_section} is equal to the corresponding number in data. 
The full set of normalisation factors is shown in \tabref{tab:mc}.
No scale factor is ascribed to \BlackHatSherpa\ and \NJETSherpa\ 
 such that the level of agreement with data can be assessed in light of the theoretical uncertainties, 
as discussed in section~\secref{sec:theoUncertainties}. 

%% file: tex/theoretical_uncertainties.tex
Theoretical uncertainties have been computed for \HEJ\ and the NLO predictions. The 
sensitivity of the \HEJ\ calculation to higher-order corrections was determined by the authors of the calculation by varying independently
the renormalisation and factorisation scales by factors of $\sqrt{2}$, 2, $1/\sqrt{2}$ and $1/2$ around the central value of \HT/2. 
The total uncertainty is the result of taking the envelope of all the variations. 
The typical size of the uncertainty is $^{+50\%}_{-30\%}$, and it is not
drawn on the figures for clarity. 

The central value of the renormalisation and factorisation scales used in the
\NJETSherpa\ and \BlackHatSherpa\ samples is also \HT/2. Scale uncertainties are evaluated for \NJETSherpa\ by simultaneously varying both scales by factors of 1/2 and 2. PDF uncertainties are
obtained by reweighting the distributions for all the PDF error
sets using LHAPDF~\cite{buckley2014lhapdf6}, following the recommendations from
ref.~\cite{botje2011pdf4lhc}. The additional PDF sets include variations in the value of \alphas. 
The sum in quadrature of the resulting scale and
PDF variations defines the NLO theoretical uncertainty included in the result
figures in \secref{sec:results}. The uncertainty is dominated by the scale component due to 
the rapid drop of the cross section with decreasing values of the renormalisation and factorisation scales. 
As a result, the uncertainty is significantly asymmetric.

%% file: tex/eventselection.tex
The data sample used was taken during the period from
March to December 2012 with the LHC operating at a $pp$ centre-of-mass
energy of $\sqrt{s}=8~\TeV$.
The application of data-quality requirements results in an integrated luminosity
of $20.3\,\ifb$.

\subsection{Trigger}\label{sec:trigger}

The events used in this analysis are selected by a combination of four
jet triggers, consisting of the three usual levels and defined in terms of the jets produced in the event.
The hardware-based L1 trigger provides a fast decision based on
the energy measured by the calorimeter. The L2 trigger performs a simple jet reconstruction procedure
in the geometric regions identified by the L1 trigger.
The final decision taken by the EF trigger is made using jets from the region of $|\eta|<3.2$, and
 reconstructed from topological clusters~\cite{Lampl:2008} using the \antikt\ algorithm with $R=0.4$.

\begin{figure}
\begin{center}
\includegraphics[width=0.4\textwidth]{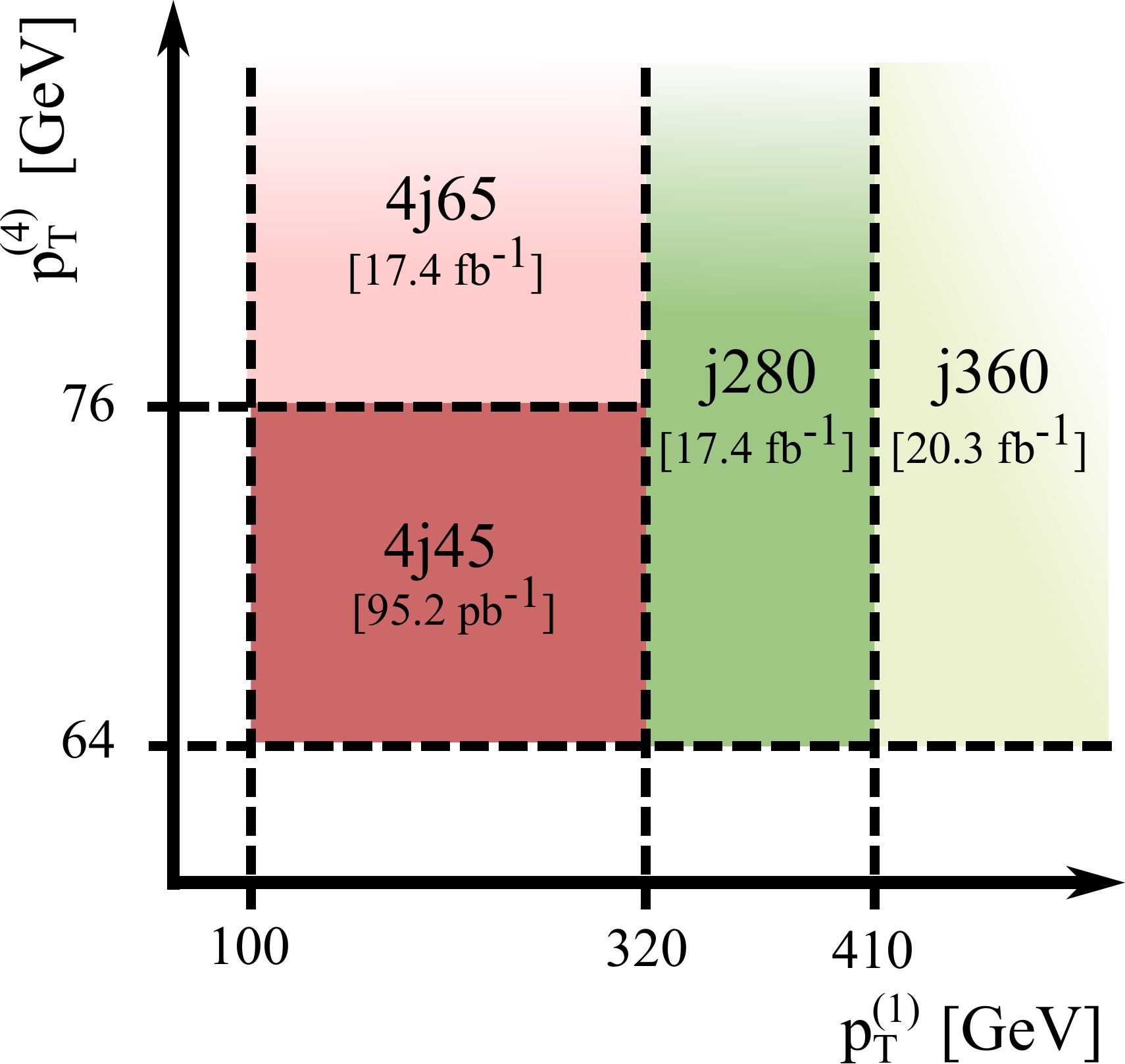}
\end{center}
\caption{\label{fig:trigger}
Schematic of the kinematic regions in which the four different jet triggers are
used, including the total luminosity that each of them recorded. The term 4j45
(4j65) refers to a trigger requiring at least four jets with $\pt>45$~GeV
(65 GeV), where the \pt is measured at the EF level of the triggering system. The term j280
(j360) refers to a trigger requiring at least one jet with $\pt> 280$~GeV
(360 GeV) at the EF level. The horizontal and vertical axes correspond to \ptOne\ and
\ptFour\ respectively, both calculated at the offline level (i.e., including the
full object calibration). 
}
\end{figure}

The four different triggers used in this paper are shown in \figref{fig:trigger}.
Two of the triggers select events with at least four jets, 
while the remaining two select events with at least one jet at a higher \pt\ threshold.
Events are split into the four non-overlapping kinematic regions shown
in \figref{fig:trigger}, requiring at least four well-separated jets with
varying \pt thresholds in order to apply the corresponding trigger. This
ensures trigger efficiencies greater than 99\% for any event passing the
inclusive analysis cuts.
The small residual loss of data due to trigger inefficiency 
is corrected as a function of jet \pt{} 
using the techniques described in \secref{sec:unfolding}.

As noted in \figref{fig:trigger}, three out of the four triggers only recorded a fraction of the total dataset. 
The contributions from the events selected by those three triggers are scaled by
the inverse of the corresponding fraction.
\subsection{Jet reconstruction and calibration}

Jets are reconstructed using the
anti-$k_t$ jet algorithm~\cite{Cacciari:2008gp} with four-momentum
recombination and radius parameter $R=0.4$. 
The inputs to the jet algorithm are locally-calibrated 
topological clusters of calorimeter cells~\cite{Lampl:2008},
which reconstruct the three-dimensional shower topology of each
particle entering the calorimeter. 

ATLAS has developed several jet calibration schemes~\cite{atlas-csc}
with different levels of complexity and different sensitivities to
systematic effects. In this analysis the local cluster weighting (LCW) calibration~\cite{Lampl:2008}
method is used, which classifies topological clusters as either being of
electromagnetic or hadronic origin.
Based on this classification, specific energy corrections are applied, improving
the jet energy resolution. The final jet energy calibration, generally referred
to as the jet energy scale, corrects the average calorimeter response to
reproduce the energy of the true particle jet.

The jet energy scale and resolution 
have been measured in $pp$ collision data 
using techniques described in 
references~\cite{Aad:2014bia, ATLAS-CONF-2015-002, ATLAS-CONF-2015-017}.
The effects of pile-up on jet energies are accounted for by a jet-area-based 
correction~\cite{Cacciari:2007fd} prior to the final calibration, where the area of
the jet is defined in $\eta$--$\phi$ space.
Jets are then calibrated to the hadronic energy scale
using $\pT$- and $\eta$-dependent calibration factors based on MC
 simulations, and their response is corrected based on several
observables that are sensitive to fragmentation effects.
A residual calibration is applied to take into account differences 
between data and MC simulation based on a combination of several 
in-situ techniques~\cite{Aad:2014bia}.

\subsection{Data quality criteria}\label{sec:cuts}

Before applying the selection that defines the kinematic region of interest,
events are required to pass the trigger, as described in~\secref{sec:trigger},
and to contain a primary vertex with at least two tracks.
Events which contain energy deposits in the calorimeter consistent with noise, 
or with incomplete event data, are rejected. In addition, events containing
jets pointing to problematic calorimeter regions, or originating from non-collision
background, cosmic rays or detector effects, are vetoed. These cleaning procedures
are emulated in the MC simulation used to correct for experimental effects, as is 
discussed in detail in \secref{sec:unfolding}.

No attempt is made to exclude jets that result from
photons or leptons impacting the calorimeter,
nor are the contributions from such signatures corrected for.
Events containing photons or $\tau$ leptons are 
expected to contribute less than 0.1\% to the cross sections under study.

\begin{figure}
\subfloat[\ptOne]{\includegraphics[width=0.48\textwidth]{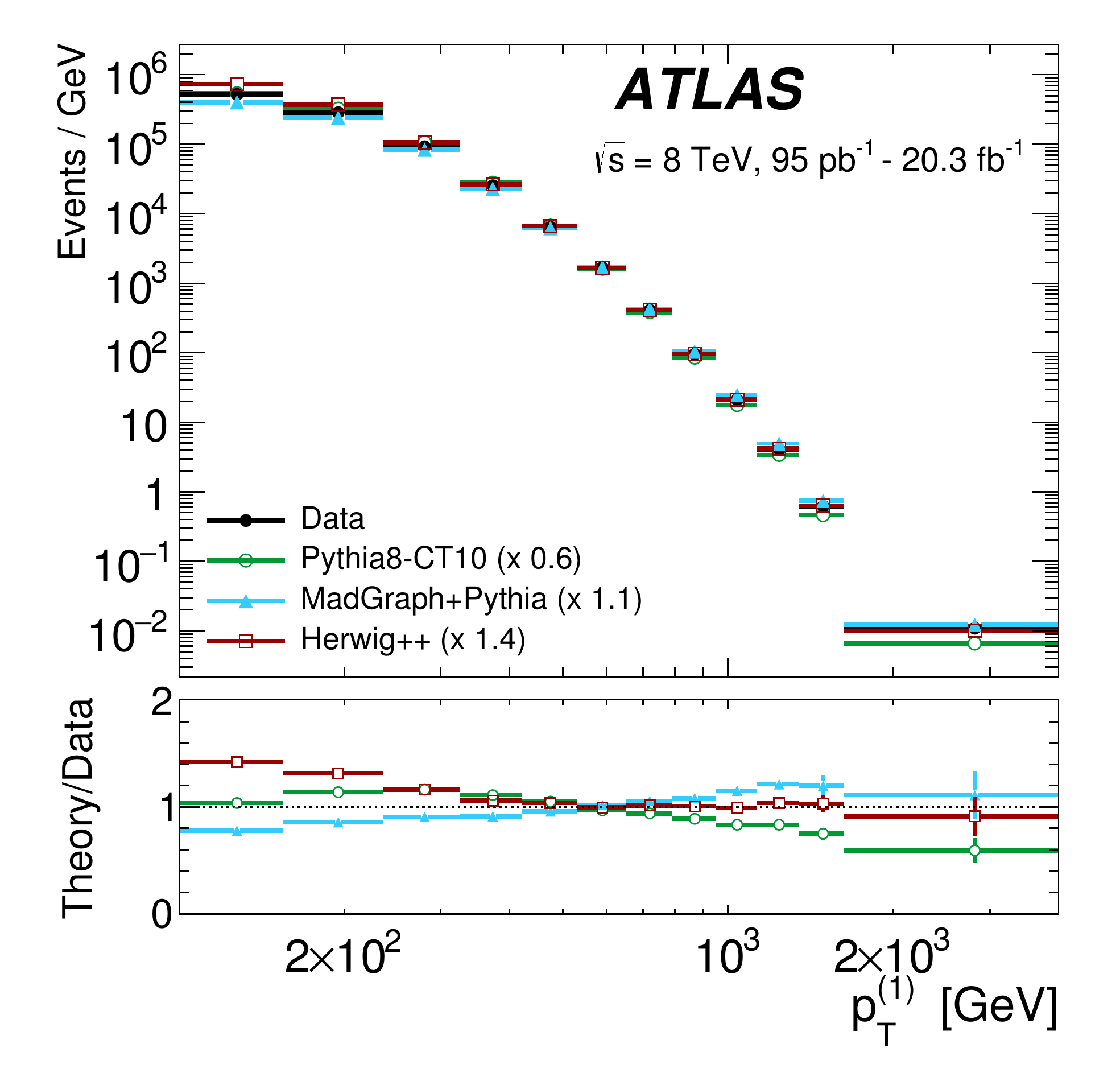}}
\subfloat[\ptFour]{\includegraphics[width=0.48\textwidth]{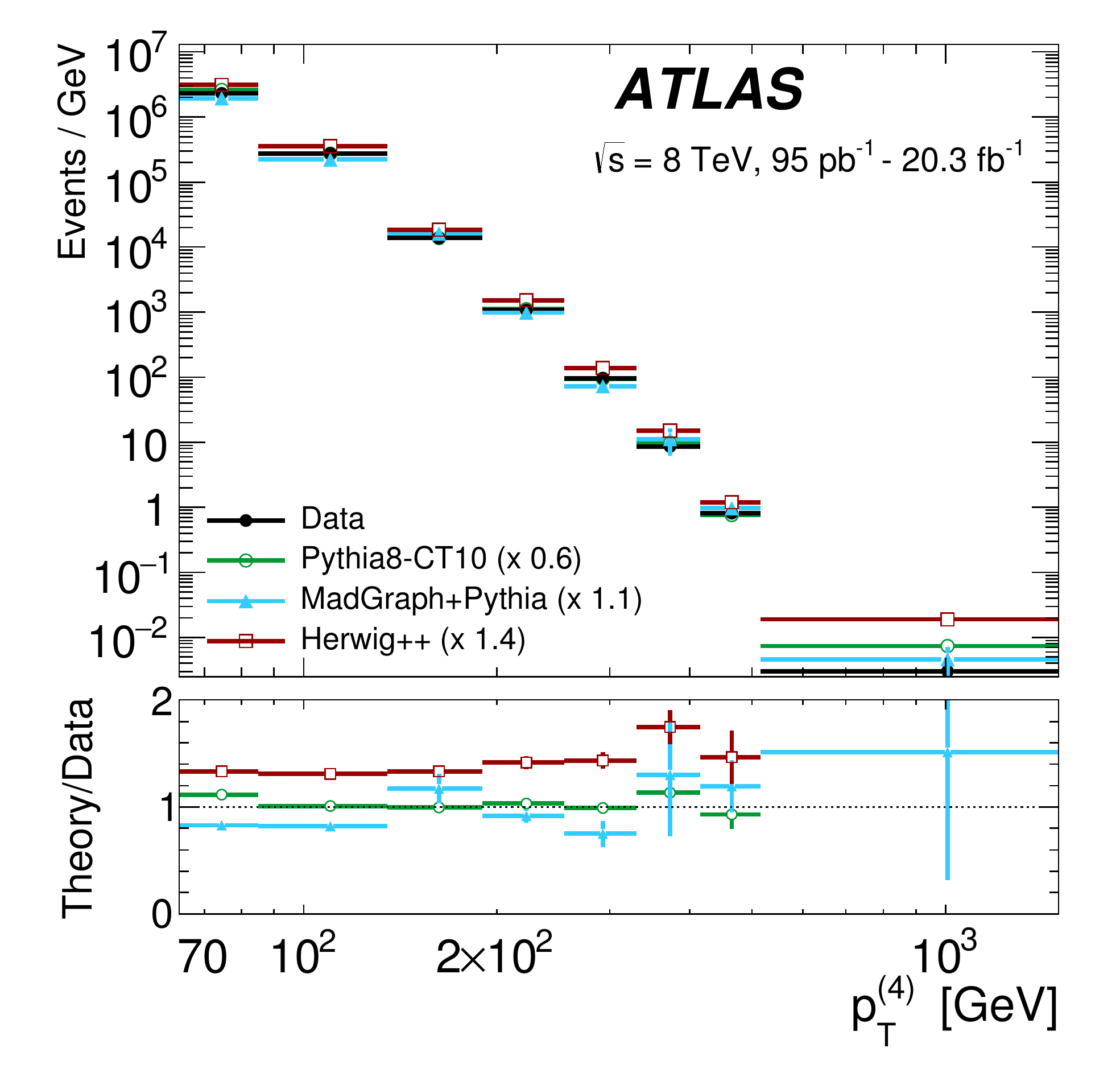}}
\caption{\label{fig:control_plots}
Detector-level distributions of (a) \ptOne\ and (b) \ptFour\ for data
and for example MC predictions.
The MC predictions have passed through detector simulation.
The lower panel in each plot shows the ratios of the MC predictions to data. For
better comparison, the predictions are multiplied by the factors indicated in
the legend.
}
\end{figure}

Distributions of two example variables (\ptOne\ and \ptFour)
can be seen at the detector level (i.e. prior to unfolding detector effects) in 
\figref{fig:control_plots}. Different sets of points correspond to the data and the different MC generators,
which are normalised to data with the scale factors indicated in
\tabref{tab:mc}. These are constant factors used to facilitate the comparison
with data, as described
in~\secref{sec:samples:normalisation}. Given that the generators have only LO or even only
leading-logarithmic accuracy, the observed agreement is reasonable.

%% file: tex/unfolding.tex
Cross sections are measured differentially in several variables, each of which
is binned in \ptOne\ or \mjjjj. Each of the corresponding distributions is
individually unfolded to deconvolve detector effects such as inefficiencies and resolutions. 
The unfolding is performed using the Bayesian Iterative
method~\cite{DAgostini:1994zf,dagostini_improved}, as 
implemented in the \texttt{RooUnfold} package~\cite{roounfold}. The algorithm
builds an unfolding matrix starting with an initial prior probability
distribution taken from MC simulation, and improves it iteratively. The method
takes into account migrations between bins. It also corrects the results for the
presence of events which pass the selection at reconstructed-level but not at the particle level; and for detector inefficiencies, which have the opposite effect. 
The number of iterations is optimised in order to minimise the size of the
statistical and systematic uncertainties. A lower number of iterations results
in a higher dependence on the MC simulation, whereas higher values give larger statistical
uncertainties. For the analysis presented in this paper, two iterations are used.

The data are unfolded to the particle-jet level using the \Pythia{} MC simulation to build the unfolding matrix. 
In order to construct the matrix, events are required to pass the inclusive analysis cuts
at both the reconstructed and particle levels. The cuts require that events
 have at least four jets within $|y|<2.8$,
with $\ptOne>$100~GeV and $\ptTwo, \ptThree, \ptFour>64$~GeV. 
The four leading jets must in addition be separated by $\dRmin>0.65$. For observables
requiring additional kinematic cuts, these are also applied both at the reconstructed
and particle levels.
No spatial matching is performed between reconstructed-level and particle-level jets.

The correlation between the observables before and after the incorporation of
experimental effects tends to be higher for \pt-based variables, such as \HT{}.
In the case of angular variables, such as \minDphiij{}, the correlation is
weakened due to cases where energy resolution effects lead to re-ordering of the jet \pt{}.
Nevertheless, even in the case of such angular variables the entries far from
the diagonal of the correlation matrix are significantly smaller
than the diagonal elements. The binning is derived from an optimisation
procedure such that the purity of the bins is between 70\% and 90\%, and 
the statistical uncertainty of the measurement is $\lesssim 10$\%. The purity is defined as the fractional number of
events per bin which do not migrate to other bins after the detector simulation, calculated with respect to
the number of events which pass the particle-level cuts.

The possible presence of biases in the unfolded spectra due to MC
mismodelling of the reconstructed-level spectrum is evaluated using a
data-driven closure test. In this study, the MC distributions are reweighted to
match the shape of those obtained from the data, and then unfolded using the
same unfolding matrix as for the data. A data-driven systematic uncertainty is
computed by comparing the result obtained from this procedure and the original
reweighted particle-level MC distributions. With two iterations of the unfolding algorithm, this systematic uncertainty is found to be negligible.

A second unfolding uncertainty is evaluated to account for the model dependence
of the efficiency with which both the reconstructed- and particle-level cuts are
satisfied in each MC event.
The systematic uncertainty is derived from the differences between 
the efficiencies calculated with \Herwigpp{} and those calculated using
\Pythia{}. The resulting uncertainty is found to be subdominant in most cases,
with typical sizes of 2--10\%. The uncertainty is rebinned and smoothed, such that its
statistical uncertainty is smaller than 40\%. 

The statistical uncertainties are calculated with pseudo-experiments. For each
pseudo-experiment, the data and MC distributions are reweighted event by event
following a Poisson distribution centred at one. Each resulting
Poisson replica of the data is unfolded using the corresponding fluctuated
unfolding matrix. The random numbers for the pseudo-experiments are generated
using unique seeds, following the same scheme used by the inclusive jet
\cite{Aad:2014vwa}, dijet \cite{atlas-dijet} and three-jet \cite{Aad:2014rma}
measurements at $\sqrt{s}=7$\,TeV, to allow for possible future combination of
results with the same dataset used for this analysis.

The integral of the unfolded distributions, corresponding to the cross section
in the fiducial range determined by the inclusive analysis cuts, was compared for 
all the variables defined in the same region of phase space and found to agree with each other within 0.5\%.

%% file: tex/systematics.tex
Several sources of experimental uncertainty are considered in this analysis. 
Those arising from the unfolding procedure are described in
\secref{sec:unfolding}. This section presents the uncertainties which arise
from the jet energy scale (JES), jet energy resolution (JER), jet angular
resolution and integrated luminosity. The dominant source of uncertainty
in this measurement is the JES.

The uncertainty in the JES calibration is determined in the central detector region by
exploiting the transverse momentum balance in $Z+$jet,
$\gamma+$jet or multi-jet events, which are measured in situ. The uncertainties in the
energy of the reference object are propagated to the jet whose energy scale is
being probed. The uncertainty in the central region is propagated to the forward
region using dijet systems balanced in transverse momentum. The procedure is
described in detail in ref.~\cite{Aad:2014bia}.

The total JES uncertainty is decomposed into eighteen components, which account
for the uncertainty in the jet energy scale calibration itself, as well as
uncertainties due to the pile-up subtraction procedure, parton flavour differences between
samples, $b$-jet energy scale and punch-through. 
Each of these uncertainties is incorporated as a coherent shift of the scale of
the jets in the MC simulation. The energies and transverse momenta of all jets with $\pt>20$~GeV
and $|y|<2.8$ are varied up and down by one standard deviation of each
uncertainty component; these components are asymmetric, i.e. the values of the
upwards and downwards variations are different. The shifts are then propagated through the unfolding. The unfolded distributions corresponding to the systematically
varied spectra are compared one by one to the nominal ones, and the difference
taken as the unfolded-level uncertainty due to that JES uncertainty component.
The total JES uncertainty is obtained by summing all such contributions
quadratically, respecting the sign of the variations in the event yields; that
is, positive and negative event yield variations are added independently.

Statistical uncertainties on each of the JES uncertainty components are obtained
by creating Poisson replicas of the systematically varied spectra, obtained as
explained in \secref{sec:unfolding}. Such statistical uncertainties are used to
evaluate the significance of the uncertainty for each component and for each bin of
all the differential distributions. As in the case of the unfolding uncertainty, the
unfolded-level uncertainty due to each JES component is then rebinned and
smoothed using a Gaussian kernel regression in order to get statistical
uncertainties smaller than $40\%$ in all bins. The typical size of the JES
uncertainty is 4--15\%.

Jets may be affected by additional energy originating from pile-up interactions.
This effect is corrected for as part of the jet energy calibration. The
distributions were binned in different ranges of the average number of
interactions per bunch crossing in order to test the possible presence of
residual effects. No significant deviations were observed, therefore no
uncertainty associated with pile-up mismodelling was considered beyond the pile-up uncertainty already included in the jet calibration procedure. 

The JER has been measured in data using dijet events \cite{Aad:2012ag}, and an
uncertainty was derived from the differences seen between data and MC
prediction. In general, the energy resolution observed in data is somewhat worse
than that in MC simulations.
The uncertainty on the observables can
therefore be evaluated by smearing the energy of the reconstructed jets in the MC simulation.
After applying this smearing to the jets, an alternative unfolding matrix is
derived and used to unfold the nominal MC prediction. Then the MC distribution
is unfolded using both the nominal and the smeared matrices, and the difference
between the two is symmetrised and taken as the JER systematic uncertainty. The
typical size of this uncertainty is 1--10\% of the cross section.

The jet angular resolution was estimated in MC simulation for the pseudorapidity and $\phi$ by
matching spatially jets at the reconstructed and particle level, and found to be $\lsim 2\%$. 
This is in agreement with in-situ measurements, so no systematic uncertainty is assigned.

Finally, the uncertainty on the integrated luminosity is $\pm 2.8$\%. It is
derived following the same methodology as that detailed in
ref.~\cite{Aad:2013ucp}.

\begin{figure}
\subfloat[][\HT]{\includegraphics[width=0.5\textwidth]{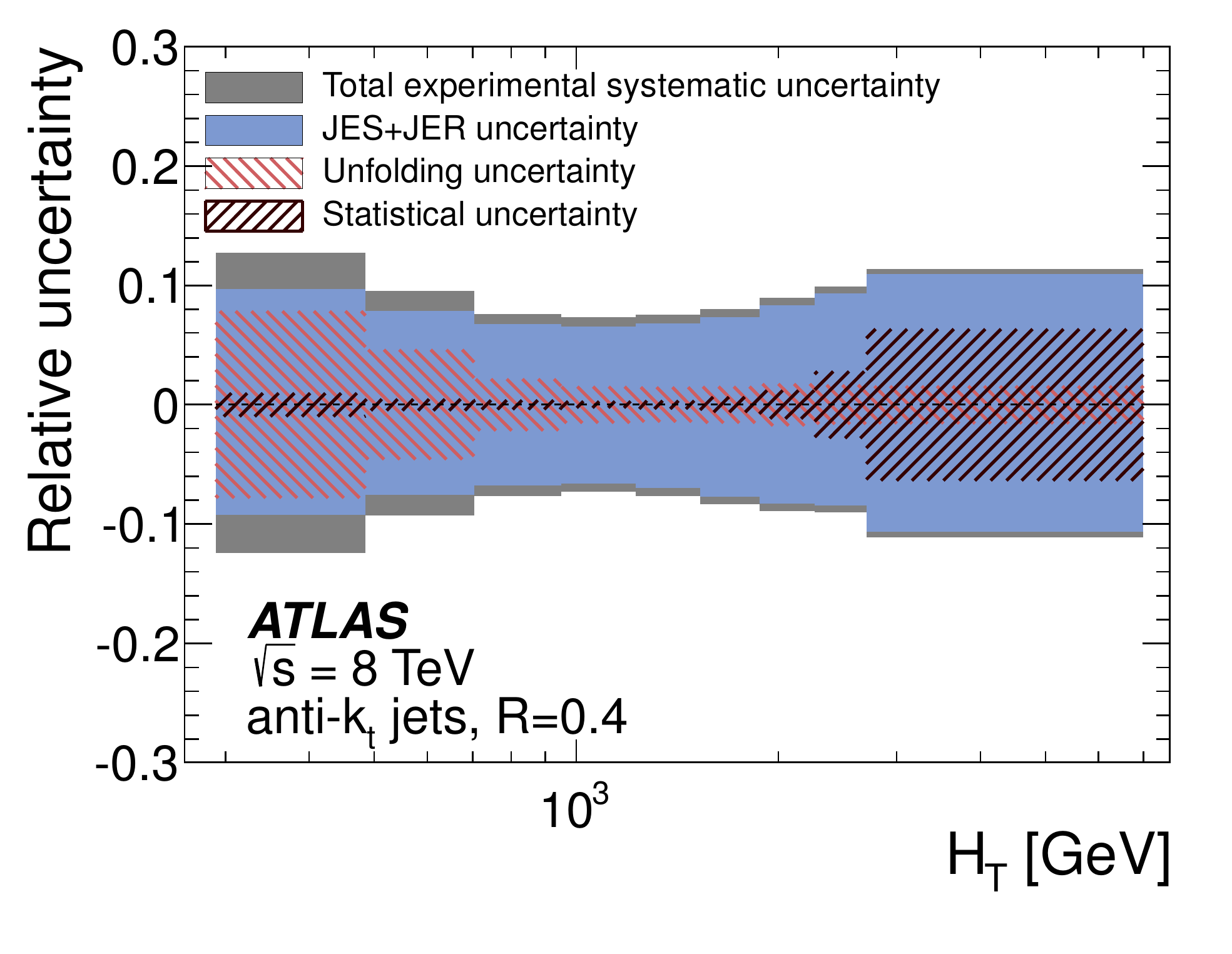}}
\subfloat[][\minDphiij]{\includegraphics[width=0.5\textwidth]{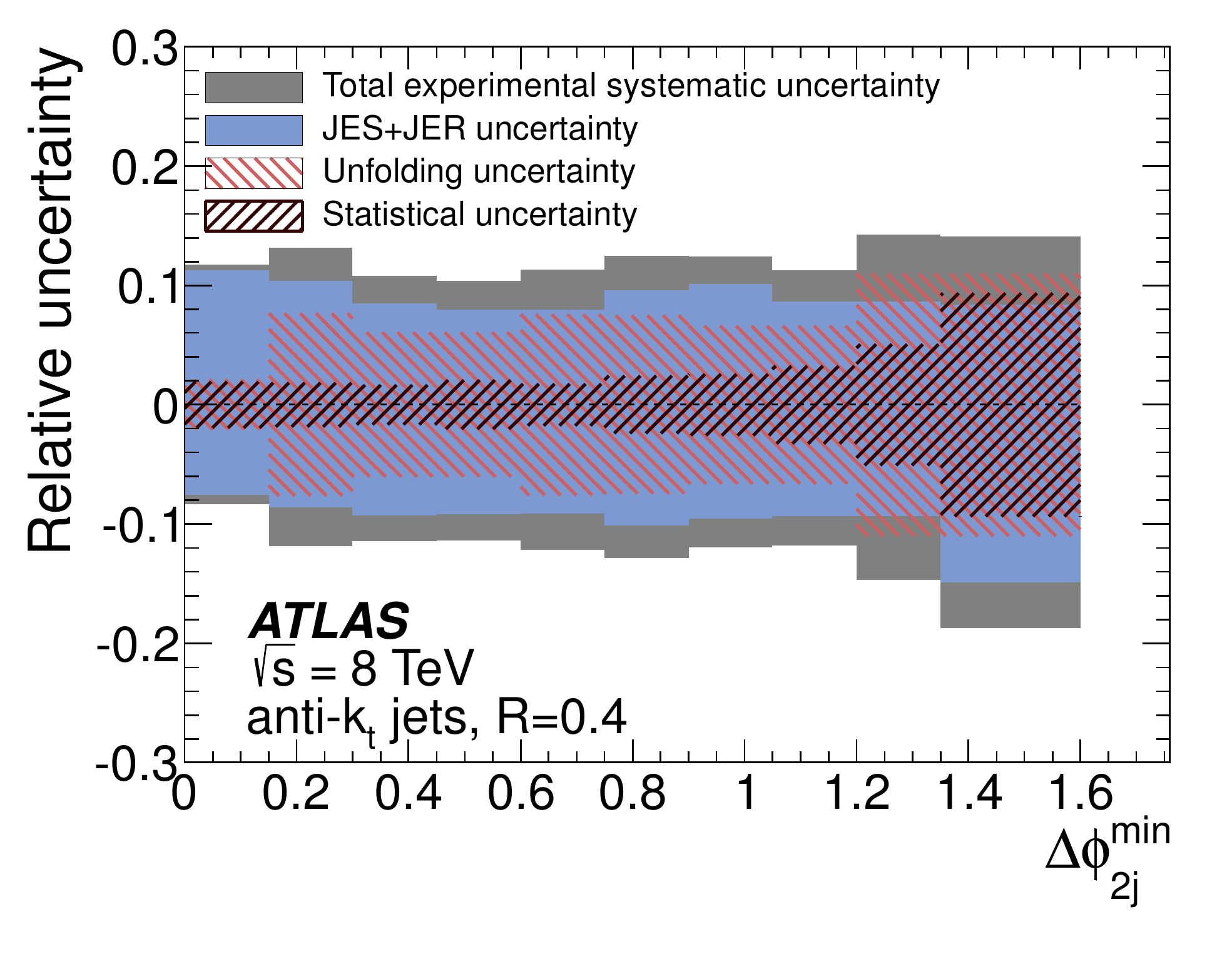}}
\caption{Total systematic uncertainty in the four-jet cross section measurement for \antikt
  $R=0.4$ jets as a function of (a) \HT and (b) \minDphiij. In both cases the
  event selection corresponds to the inclusive analysis cuts, namely
  $\ptFour>64$~GeV, $\ptOne>100$~GeV and $\dRmin>0.65$. Separate bands show the
jet energy scale (JES) and resolution (JER), and the  unfolding uncertainty, as well as the
combined total systematic uncertainty resulting from adding in quadrature all
the components. The total statistical uncertainty of the unfolded data spectrum
is also shown. The luminosity uncertainty is not shown separately but is included in the total uncertainty band.}
\label{fig:total_exp_uncert}
\end{figure}

Two examples of the values of the total experimental systematic uncertainty are
shown in \figref{fig:total_exp_uncert} for two representative variables, namely \HT and \minDphiij. The jet energy scale and resolution uncertainties dominate in the majority of bins, being larger at the high and low ends of the \HT\ spectrum. The unfolding uncertainty is nearly as large at low values of the jet momenta, and it is therefore an important contribution in most of the \minDphiij\ bins. 

%% file: tex/results.tex
\label{unfolding:results}

\begin{figure}
\begin{center}
\includegraphics[width=0.9\textwidth]{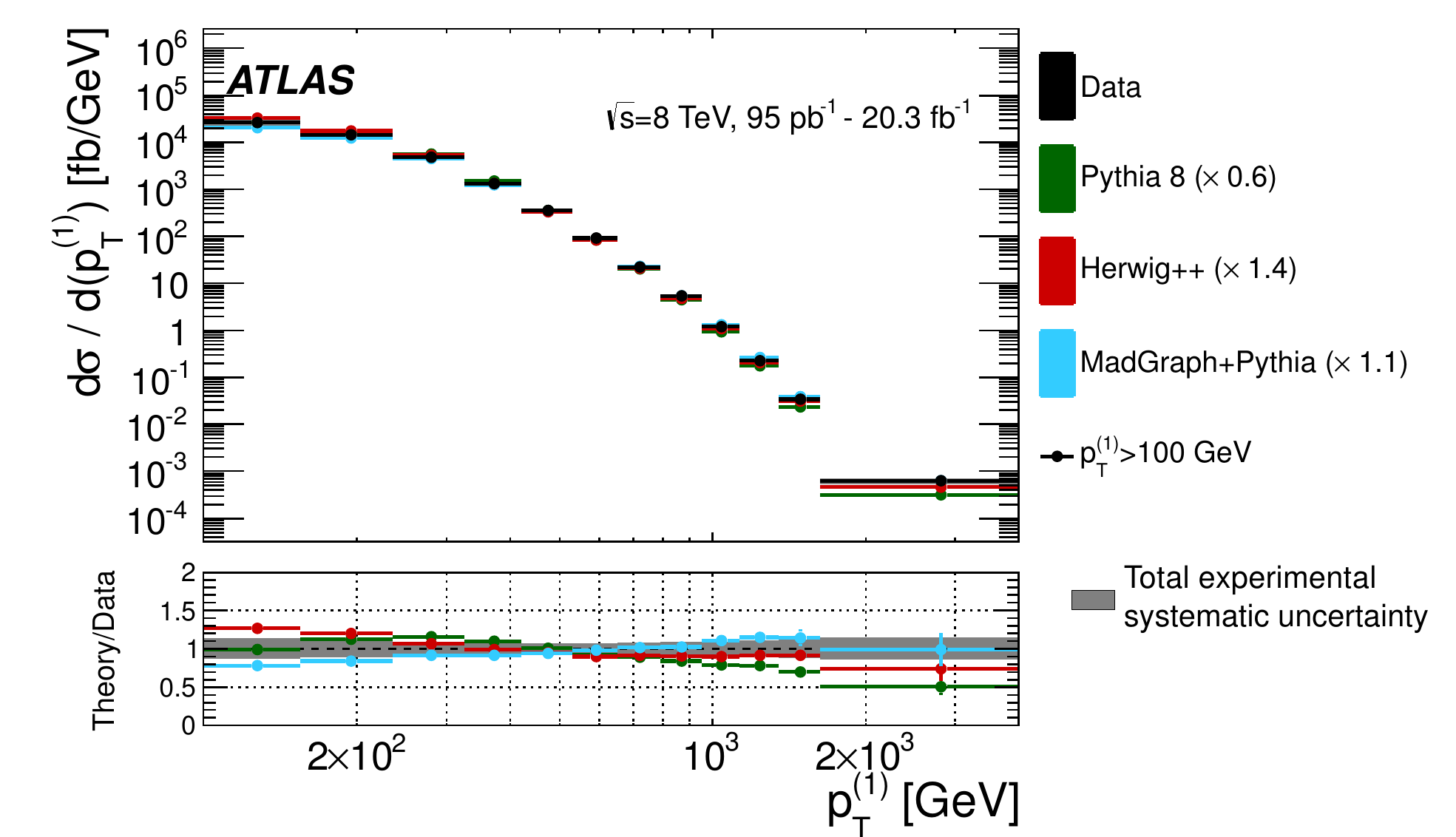}\\
\vspace{0.5cm}
\includegraphics[width=0.9\textwidth]{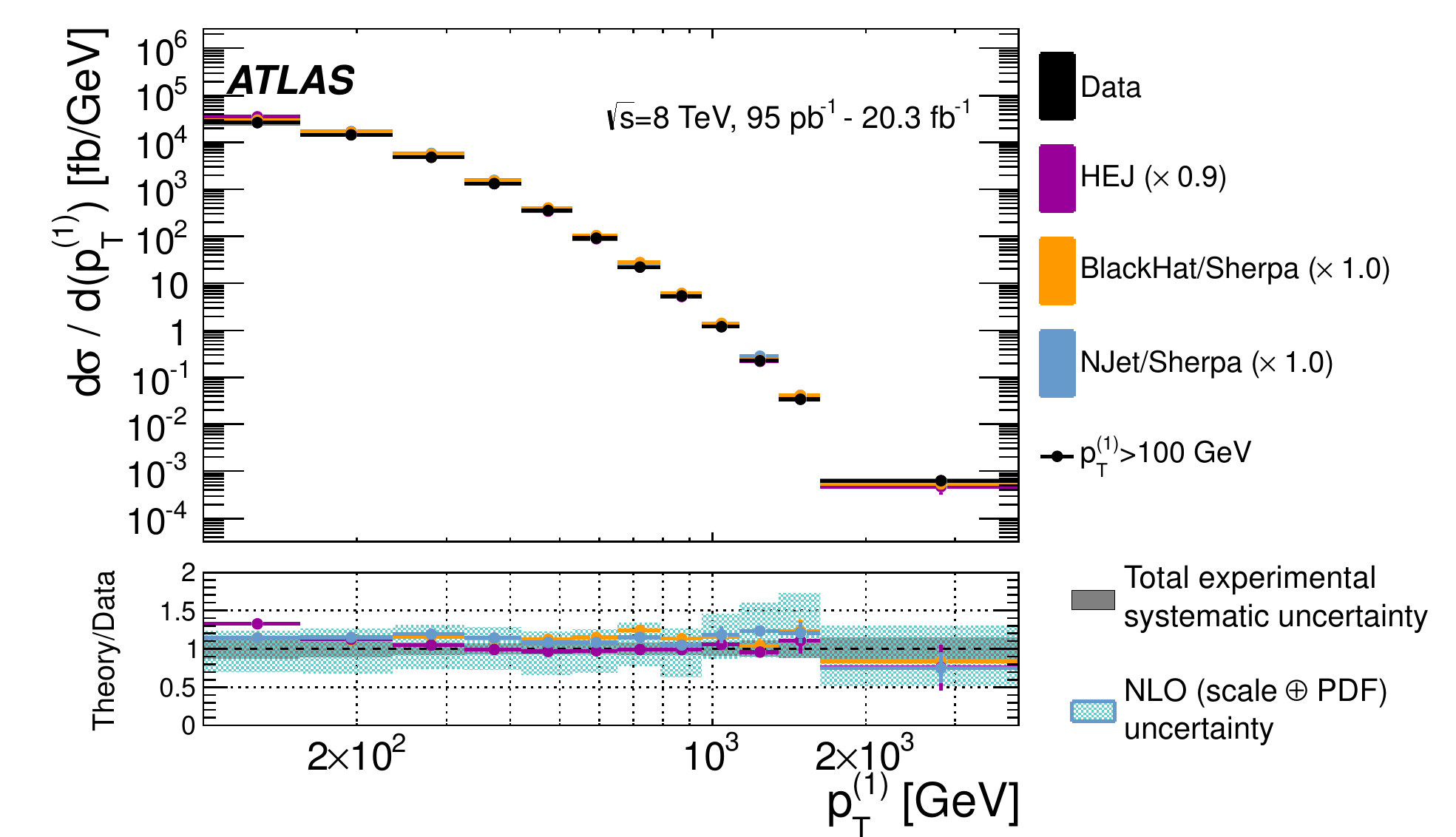}\\
\caption{The four-jet differential cross section as a function of leading jet
  \pt (\ptOne), compared to different theoretical predictions: \Pythia,
  \Herwigpp\ and \MadgraphPythia\ (top), and \HEJ, \NJETSherpa\ and
  \BlackHatSherpa\ (bottom). For better comparison, the predictions are
  multiplied by the factors indicated in the legend. In each figure, the top
  panel shows the full spectra and the bottom panel the ratios of the different
  predictions to the data. The solid band represents the total experimental
  systematic uncertainty centred at one. The patterned band represents the NLO
  scale and PDF uncertainties calculated from \NJETSherpa\ centred at the
  nominal \NJETSherpa\ values.  The scale uncertainties for \HEJ\ (not drawn)
  are typically $^{+50\%}_{-30\%}$. The ratio curves are formed by the central
  values with vertical uncertainty lines resulting from the propagation of the
  statistical uncertainties of the predictions and those of the unfolded data
  spectrum.
}
\label{fig:unfolded_jetpt0}
\end{center}
\end{figure}

\begin{figure}
\begin{center}
\includegraphics[width=1.0\textwidth]{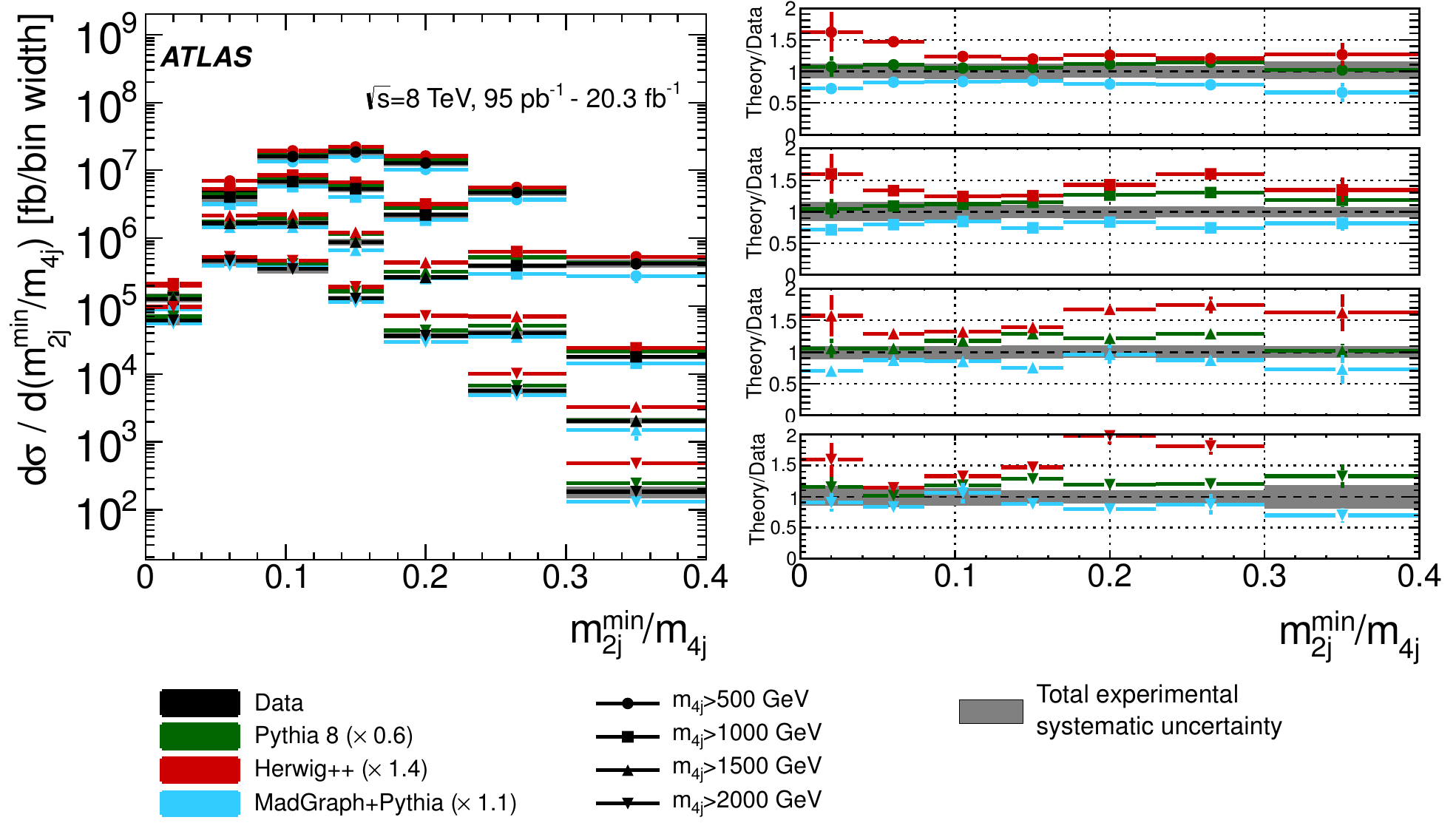}\\
\vspace{0.5cm}
\includegraphics[width=1.0\textwidth]{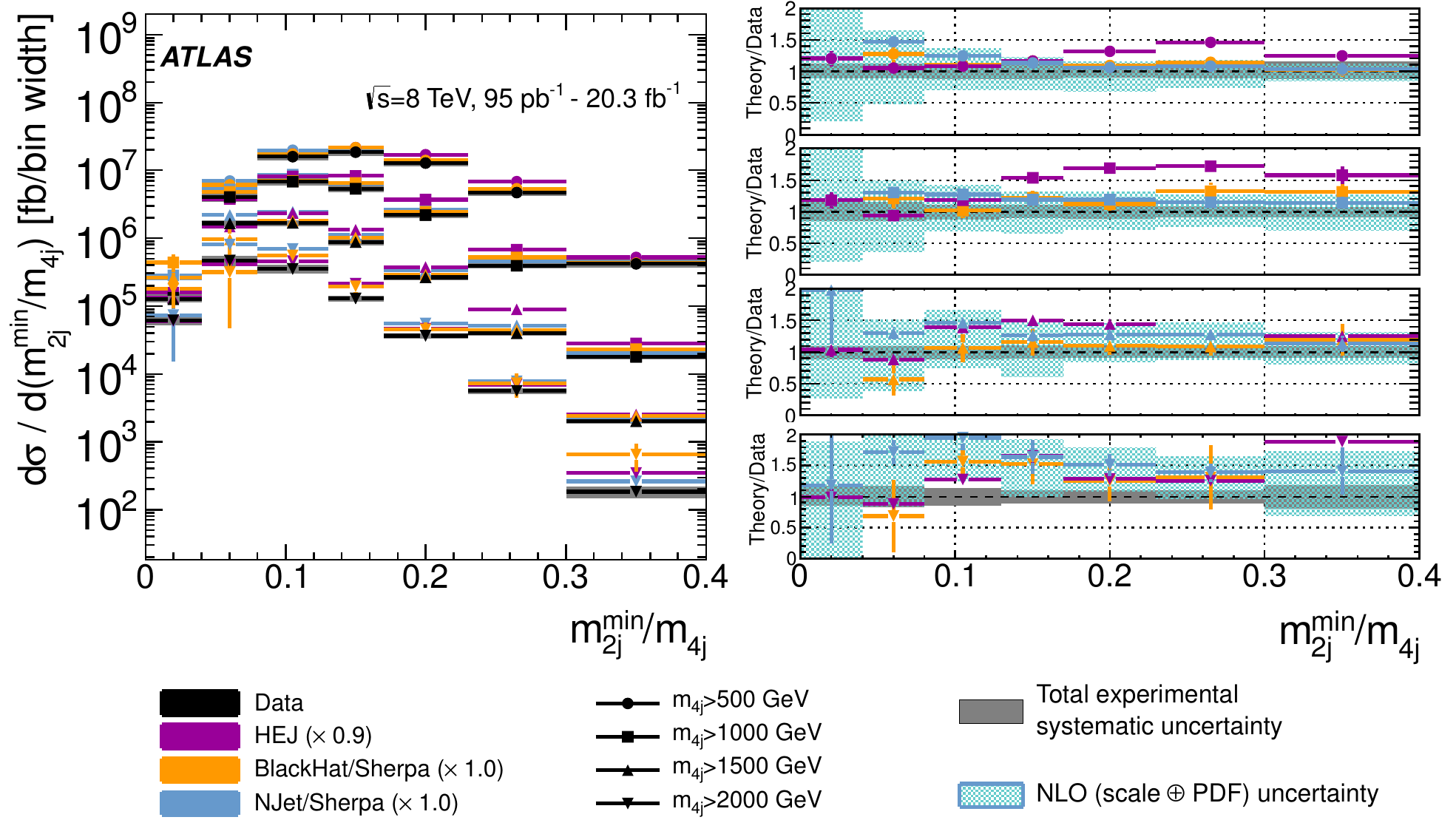}\\
\caption{Unfolded four-jet differential cross section as a function of
  \mjjminratio, compared to different theoretical predictions: \Pythia,
  \Herwigpp\ and \MadgraphPythia\ (top), and \HEJ, \NJETSherpa\ and
  \BlackHatSherpa\ (bottom). For better comparison, the predictions are
  multiplied by the factors indicated in the legend. In each figure, the left
  panel shows the full spectra and the right panel the ratios of the different
  predictions to the data, divided according to the selection criterion applied
  to \mjjjj. The solid band represents the total experimental systematic
  uncertainty centred at one. The patterned band represents the NLO scale and
  PDF uncertainties calculated from \NJETSherpa\ centred at the nominal
  \NJETSherpa\ values. The scale uncertainties for \HEJ\ (not drawn) are
  typically $^{+50\%}_{-30\%}$.  The ratio curves are formed by the central
values and vertical uncertainty lines resulting from the propagation of the
statistical uncertainties of the predictions and those of the unfolded data
spectrum.}
\label{fig:unfolded_mjjminratio}
\end{center}
\end{figure}

The various differential cross sections measured in events with at least
four jets are shown in \figsref{fig:unfolded_jetpt0} to
\ref{fig:unfolded_sumptcent4} for jets reconstructed with the \antikt algorithm
with $R=0.4$. The observables used for the measurements are defined in
\tabref{tab:variables}. The measurements are performed for a wide range of
 jet transverse momenta from
64 GeV to several TeV, spanning two orders of magnitude in \pt\ and over seven
orders of magnitude in cross section. The measured cross sections are
corrected for all detector effects using the unfolding procedure described in
\secref{sec:unfolding}. 
The theoretical predictions described in~\secsref{sec:samples:LO} 
and \ref{sec:samples:NLO} are compared to the unfolded results.

\paragraph{Summary of the results} The scale factors applied to LO generators
(see~\secref{sec:samples:normalisation}) 
are found to vary between $0.6$ and $1.4$, as shown previously in
\tabref{tab:mc}. Not all generators describe the shape
of \ptOne\ correctly, so these scale factors should not be seen as a measure of
the level of agreement between MC simulation and data, which may vary as a function of the cuts in \ptOne\ and \mjjjj. 
The cross section predicted by \BlackHatSherpa\ and \NJETSherpa\ is larger than
that measured in data, but overall the difference is covered by the scale and PDF 
uncertainties evaluated using \NJETSherpa, with only a few exceptions. 
\BlackHatSherpa\ and \NJETSherpa\ give identical results within 
statistical uncertainties; therefore only one of the two (\NJETSherpa) 
is discussed in the following, for simplicity. It is nevertheless
interesting to compare experimental results with two different implementations
of the same NLO \pQCD\ calculations as an additional cross-check. 

In general, an excellent description of both the shape and the normalisation of
the variables is given by \NJETSherpa. The small differences found are covered
by theoretical and statistical uncertainties in almost all cases; only the tails
of \ptFour\ and \maxDrapij\ hint at deviations from the measured distribution.
\MadgraphPythia\ describes the data very well in most regions of phase space,
the most significant discrepancy being in the slopes of \ptOne\ and \ptTwo and
derived variables.
\HEJ\ also provides a good description of most variables; the most significant
discrepancy occurs for the angular variables \minDrapij\ and \maxDrapij\ when
\ptOne\ is small. However when \ptOne\ is large, HEJ describes \maxDrapij\
better than \NJETSherpa, which highlights one of the strengths of this
calculation.
The \TwoToTwo\ ME calculations matched to parton showers provide
different levels of agreement depending on the variable studied; the only
variable whose shape is reasonably well described by both \Pythia\ and
\Herwigpp\ is \HT.

The following discussion is based on the results obtained after applying the particular choice of normalisation of the theoretical predictions as explained at the beginning of this section. \NJETSherpa, which generally gives very good agreement with the data, is only discussed for those cases where some deviations are present.

\paragraph{Momentum variables} 
The momentum variables comprise the \pt\ of the four leading jets and \HT.
Part of the importance of these variables lies in their wide use in analyses,
alone or as inputs to more complex observables. They are also interesting in
themselves: it has been shown that the ratio of the NLO to the LO predictions is
relatively flat across the \ptOne\ spectrum with a maximum variation of
approximately 25\%~\cite{badger2014next}. Perhaps surprisingly, the PS
 description of \ptFour\ was found to be better than that of
\ptOne\ in the 7~TeV multi-jet measurement published by ATLAS~\cite{Collaboration:2011tq}.

Figures~\ref{fig:unfolded_jetpt0} to \ref{fig:unfolded_jetpt3} show the \pt{}
distributions of the leading four jets. All the LO generators show a slope with
respect to the data in the leading jet \pt\ (\figref{fig:unfolded_jetpt0}).
The ratios of \Herwigpp\ and \HEJ\ to data are remarkably flat above
$\sim500$~GeV and $\sim300$~GeV respectively. \MadgraphPythia\ is within the experimental uncertainties
above $\sim300$~GeV, and it is the only one with a positive slope in the
ratio to data.

The subleading jet \pt\ (\figref{fig:unfolded_jetpt1}) is well described by
\HEJ, while the LO generators show similar trends to those in \ptOne. 
\MadgraphPythia\ describes both \ptThree and \ptFour well, as shown in
\figsref{fig:unfolded_jetpt2} and \ref{fig:unfolded_jetpt3}.
As the 7~TeV results suggested, \Pythia\ gives a good description of the
distribution of \ptFour.
\HEJ\ and \Herwigpp\ overestimate the number of events with high \ptFour. \NJETSherpa\ shows a similar
trend at high \ptFour,  but the discrepancy is mostly covered by the
theoretical uncertainties. \HT, shown in \figref{fig:unfolded_HT}, exhibits features similar to those in \ptOne.

In summary, \Pythia\ and \Herwigpp\ tend to describe the \pt\
spectrum of the leading jets with similar levels of agreement, whereas \Pythia\ is better at describing
\ptFour. \MadgraphPythia\ does a reasonable job for all of them,
while \HEJ\ and \NJETSherpa\ are very good for the leading jets and less so for
\ptFour. This could perhaps be improved by matching the calculations to PSs.

\paragraph{Mass variables} Mass variables are widely used in physics searches,
and they are also sensitive to events with large separations between jets, which
puts the \HEJ\ and \MadgraphPythia\ predictions to the test, as they are expected
to be especially accurate in this regime. 

The distribution of the total invariant mass \mjjjj\ is studied in \figref{fig:unfolded_mjjjj}. 
\Pythia\ and \MadgraphPythia\ describe the data very well. \Herwigpp\ describes
the shape of the data between 1~TeV and 3--6~TeV.
\HEJ\ is mostly compatible with the measurement, but the ratio to data has a bump structure in
the region of approximately 1 to 2~TeV. This feature is also shared by
\NJETSherpa, but the differences with respect to the data are covered by the NLO
uncertainties.

The description of different splitting scales is tested in
\figref{fig:unfolded_mjjminratio} through the variable \mjjminratio. This
distribution is well described by \Pythia, whereas \Herwigpp\ gets
worse with increasing \mjjjj, consistently overestimating the two ends of the 
\mjjminratio\ spectrum. \MadgraphPythia\ provides a very good description, with a flat
ratio for all the \mjjjj\ cuts. The \HEJ\ prediction shows trends similar to those of \Herwigpp\ at
higher values of \mjjjj. These differences are covered in all cases by the large
associated scale uncertainty. \NJETSherpa\ overestimates the number of events in
the very first bin, possibly due to the lack of a PS, but otherwise
agrees with the data within the theoretical uncertainties.

Overall, \MadgraphPythia\  provides the best description of mass variables. 

\paragraph{Angular variables} Similarly to mass variables, angular variables are
able to test the description of events with small- and wide-angle radiation. In
addition, they can also provide information on the global spatial distribution
of the jets. High-\pt, large-angle radiation should be well captured by the
ME+PS description of \MadgraphPythia, or the all-orders approximation of \HEJ\
-- particularly the rapidity variables \minDrapij, \maxDrapij\ and
\minDrapijk. PS generators are expected to perform poorly at large angles, given that they only contain two hard jets, and the rest is left to the soft- and collinear-enhanced PS. The fixed-order NLO prediction of \NJETSherpa\ should provide a very good description of these variables too, as long as they are far from the infrared limit. This is indeed the case, and therefore no detailed comments about its performance are given here. 

Figure \ref{fig:unfolded_minDphiij} compares the distributions of \minDphiij\ for different cuts in \ptOne. \Pythia\ has a small downwards slope with respect to the data in all the \ptOne\ ranges. \MadgraphPythia\ also shows a small slope. The other generators, both LO and NLO, reproduce the data very well. \Herwigpp, in particular, provides a very good description of the data.  

The \minDphiijk\ spectrum is shown in \figref{fig:unfolded_minDphiijk}. The
different \ptOne\ cuts change the spatial distribution of the events, such that
at low \ptOne\ most events contain two jets recoiling against two, while at high
\ptOne\ the events where one jet recoils against three dominate. In general, the description of the data improves as \ptOne\ increases. For \Pythia, the number of events where one jet recoils against three (low \minDphiijk) is significantly overestimated when \ptOne\ is low; as \ptOne\ increases, the agreement improves such that the $\ptOne>1000$~GeV region is very well described.
\MadgraphPythia, \Herwigpp\ and \HEJ\ are mostly in good agreement with data.

Figure \ref{fig:unfolded_minDrapij} compares the distributions of \minDrapij\
with data. This variable is remarkably well described by \Pythia, showing no
significant trend. \MadgraphPythia\ mostly underestimates high \minDrapij\
values, while \Herwigpp\ has a tendency to underestimate the low values. \HEJ\ overestimates
the number of events with high \minDrapij\ values at low \ptOne, but describes
the data very well at larger values of \ptOne.

For the variable \minDrapijk, presented in \figref{fig:unfolded_minDrapijk}, 
the predictions provided by \Pythia\ and \Herwigpp\ show in general a positive
slope with respect to the data.
\MadgraphPythia\ reproduces the shape of the data well, as does \HEJ\ for
$\ptOne>400$~GeV. However, for smaller values of $\ptOne$ \HEJ\ overestimates the number of
events at the end of the spectrum, as was the case for \minDrapij.

The variable \maxDrapij, shown in \figref{fig:unfolded_maxDrapij}, is very well
described by \HEJ\ in events with $\ptOne>400$~GeV. The ratios to data in both \Pythia\ and \Herwigpp\ have upwards slopes in all \ptOne\ bins. \MadgraphPythia\ provides mostly a good description of the data, with a tendency to underestimate the extremes of the distribution. Interestingly, \NJETSherpa\ seems to overestimate the number of events in the tail, although it is a statistically limited region and the comparison with \BlackHatSherpa\ is not conclusive.

In summary: \NJETSherpa\ mostly agrees with the data within the uncertainties,
but its ratio to data has an upwards trend in the tail of \maxDrapij. \HEJ\ provides a very good description of all angular variables for the region $\ptOne>400$~GeV, as expected, but shows significant discrepancies with respect to the data in all the rapidity variables for lower \ptOne\ values. It is important to keep in mind, though, that the associated scale uncertainties are large. 
\MadgraphPythia\ describes all the data well, apart from the tail of \minDrapij\ and the extreme values of \maxDrapij, which it underestimates. \Herwigpp\ gives good descriptions of the $\phi$ variables, but fails at describing the rapidity variables. \Pythia\ has some problems describing both three-jet variables, as well as \maxDrapij. These variables highlight the need to combine four-jet ME calculations with parton showers. 

\paragraph{\sumptcent\ variables} The variables setting a minimum
forward--backward rapidity interval and measuring the total \pt{} of the central
jets (\sumptcent) were defined to test the framework of \HEJ. \HEJ\ has been
designed to describe events with two jets significantly separated in rapidity
with additional, central, high-\pt radiation. These variables are also useful to
describe the spatial configuration of the events, as they represent the
forward--backward rapidity span of the jets, and whether the leading jet is
among the two central ones or not. The NLO predictions and \MadgraphPythia\ are
also expected to be successful in this regime, whereas the \TwoToTwo\ generators
with PSs are expected to be less suitable.

The variable \sumptcent\ is studied for values of \maxDrapij\ larger than 1, 2,
3 or 4, and for different cuts in \ptOne. In most cases, the description of the
observable worsens significantly with increasing \maxDrapij\ and \ptOne. Figures
\ref{fig:unfolded_sumptcent1} to \ref{fig:unfolded_sumptcent4} correspond to the
results for $\maxDrapij>1,2,3,4$.

The generators with \TwoToTwo\ MEs have problems describing the
data around the threshold values where the contribution from different jets
changes, which results in kinks in the ratio distributions. One such transition occurs
 at the \sumptcent\ value for which the leading jet is first allowed to be central. For
$\ptOne>400$~GeV, this happens at $\sumptcent>464$~GeV, at which point there is a
major jump in \Pythia\ in the second ratio plot of
\figref{fig:unfolded_sumptcent1}. \Pythia\ gives in general the most discrepant
prediction, with kinks in the ratio to data at the transition points that reach
differences of 70\% at high \ptOne, as well as global slopes. \Herwigpp\
describes the data very well at lower \maxDrapij\ values, but as \maxDrapij\ grows
its normalisation worsens, as well as the shape -- particularly at high \ptOne.

\MadgraphPythia\ provides an excellent description of the \sumptcent\
variables, especially at low \ptOne. The agreement deteriorates at high \ptOne,
but it is not very much affected by the changes between different jet
configurations, providing overall a very good description of the shapes.  Most
distributions are well described by \HEJ, especially the high \sumptcent\
region; the low \sumptcent\ region shows more shape differences, which get worse
at large \maxDrapij. This is compatible with similar observations made in
previous ATLAS measurements performed with 7~TeV data~\cite{Aad:2014pua}, where it was also shown that the agreement was significantly improved after interfacing \HEJ\ with a PS generator. \NJETSherpa\ has a tendency to overestimate the number of events with very low \sumptcent, which may be correlated with the \ptFour\ discrepancy discussed earlier. 
It provides a very good description of the data otherwise. 

Tables~\ref{tab:xs_0JetPt} to~\ref{tab:xs_RapGap4SumPtCentral_pt550} in
appendix~\ref{app} contain the numerical values of the measured
differential cross sections and their corresponding uncertainties. The quoted
values correspond to the average differential cross sections over the bin ranges
given.

\newcommand\otherdetails{The other details are as for \figref{fig:unfolded_jetpt0}. }
\newcommand\missNLO{Some points in the ratio curves for \NJETSherpa\ fall
outside the $y$-axis range, and thus the NLO uncertainty is shown partially, or
not shown, in these particular bins.}
\begin{figure}
\begin{center}
\includegraphics[width=0.9\textwidth]{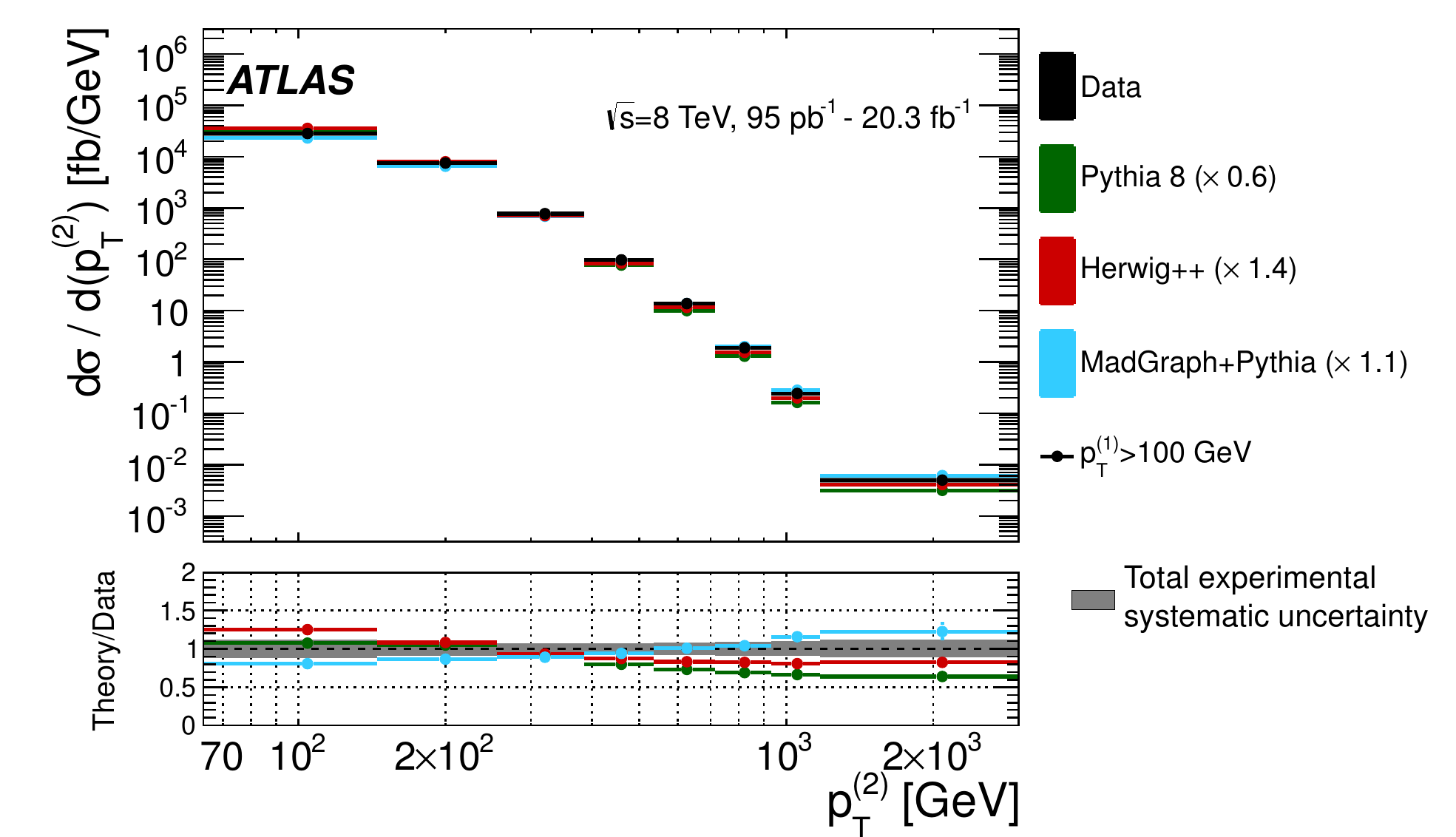}\\
\vspace{0.5cm}
\includegraphics[width=0.9\textwidth]{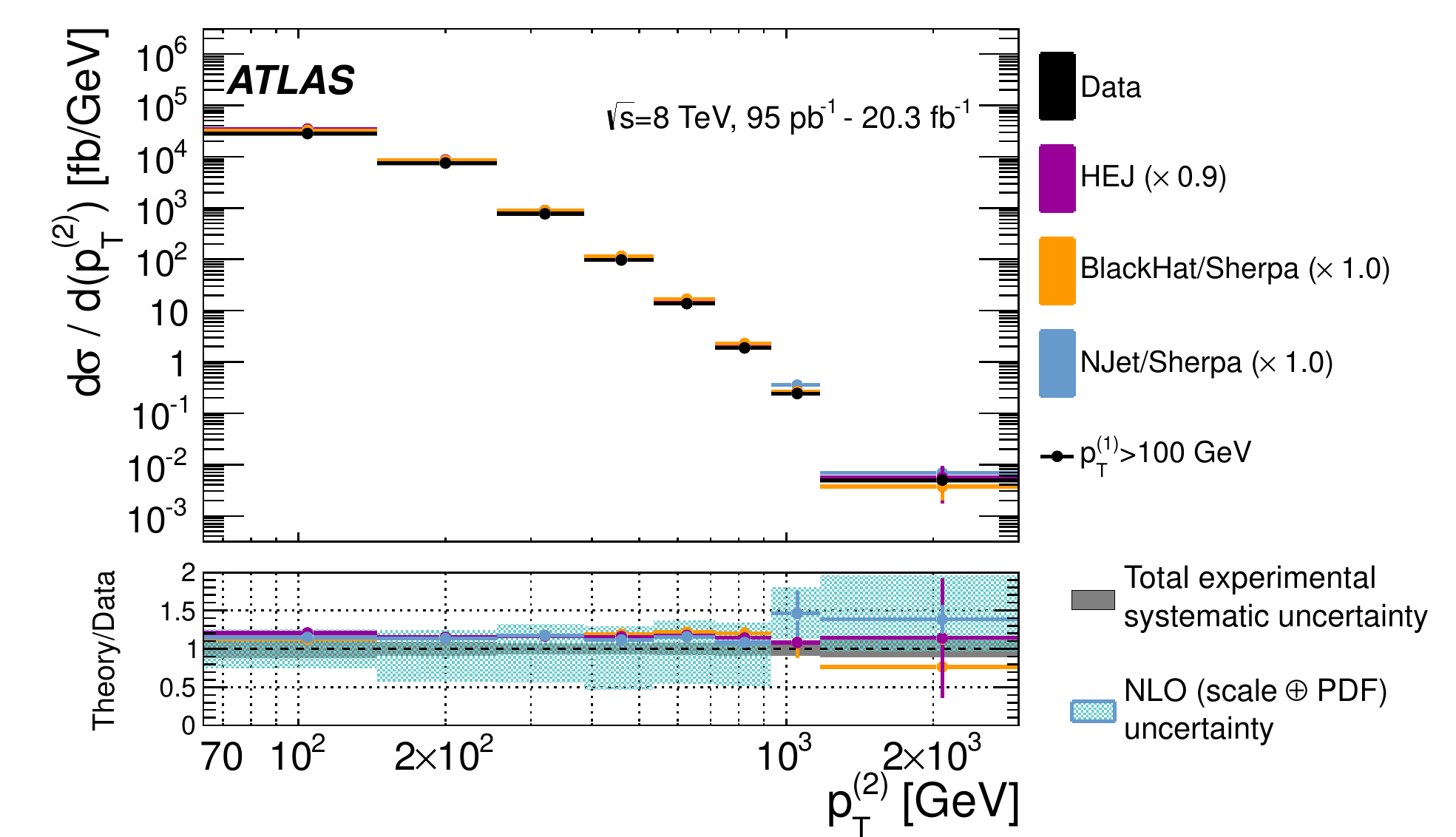}\\
\caption{
Unfolded four-jet differential cross section as a function of \ptTwo, compared to different theoretical predictions. \otherdetails}
\label{fig:unfolded_jetpt1}
\end{center}
\end{figure}

\begin{figure}
\begin{center}
\includegraphics[width=0.9\textwidth]{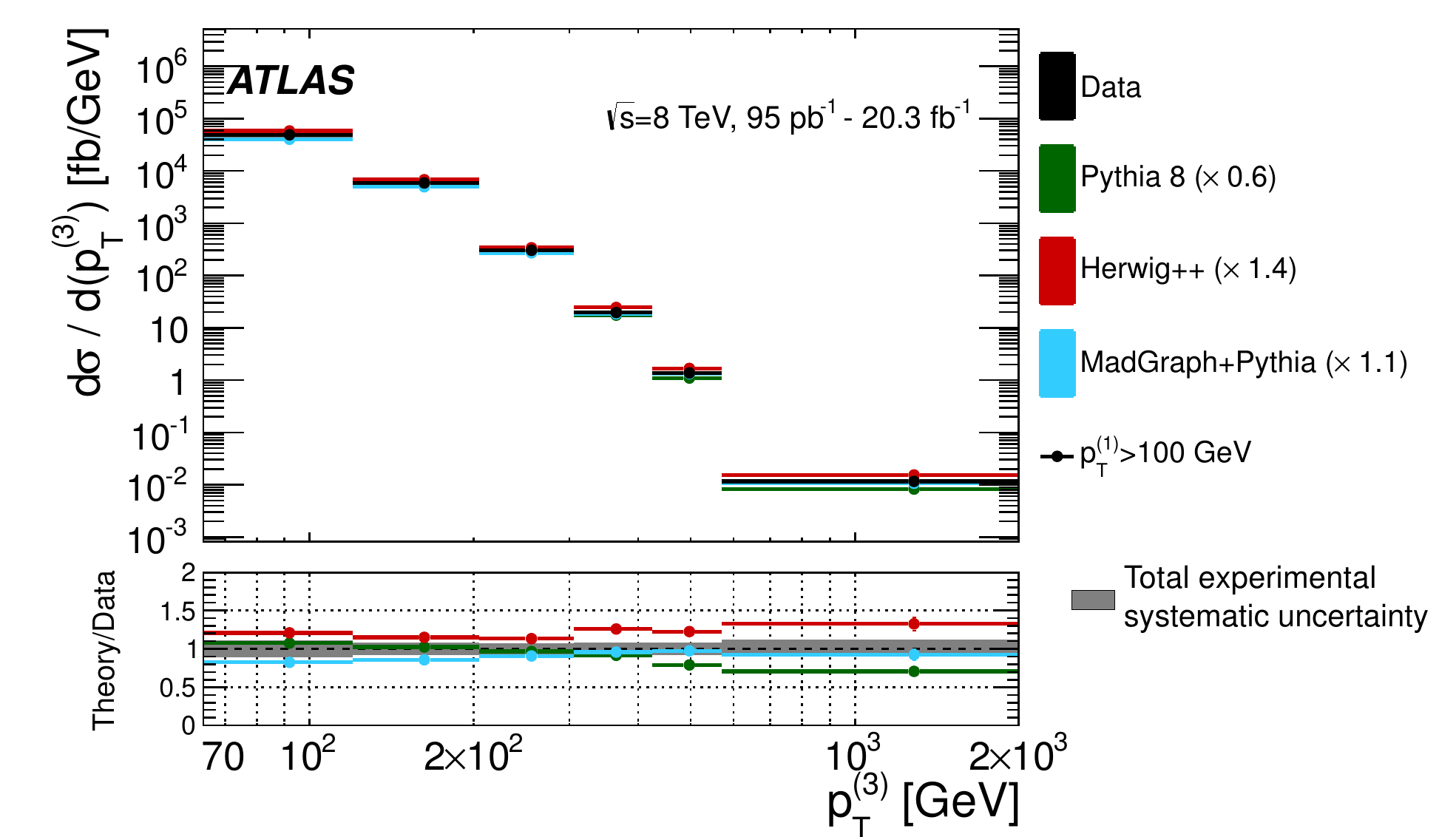}\\
\vspace{0.5cm}
\includegraphics[width=0.9\textwidth]{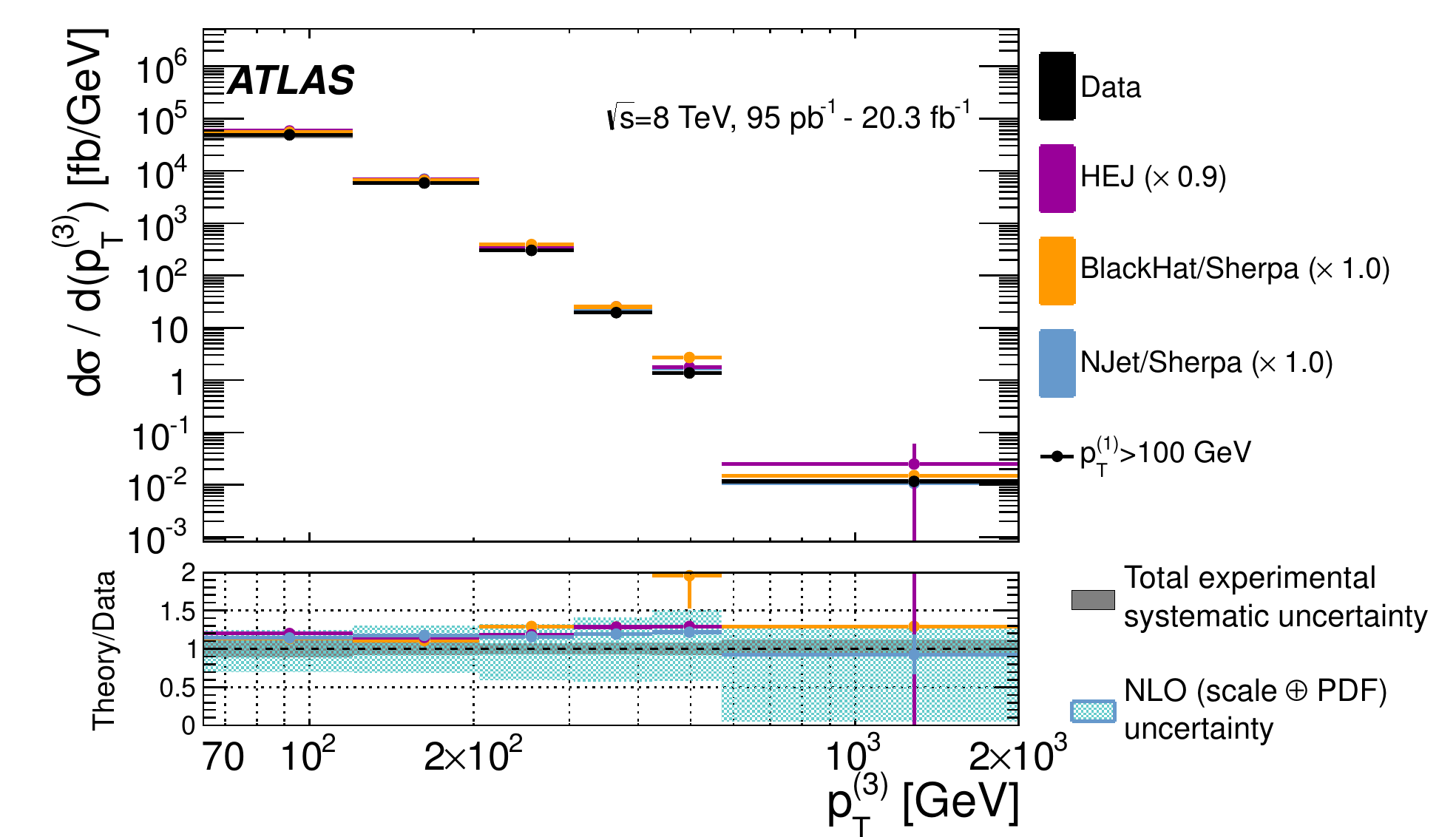}\\
\caption{Unfolded four-jet differential cross section as a function of \ptThree, compared to different theoretical predictions. \otherdetails}
\label{fig:unfolded_jetpt2}
\end{center}
\end{figure}

\begin{figure}
\begin{center}
\includegraphics[width=0.9\textwidth]{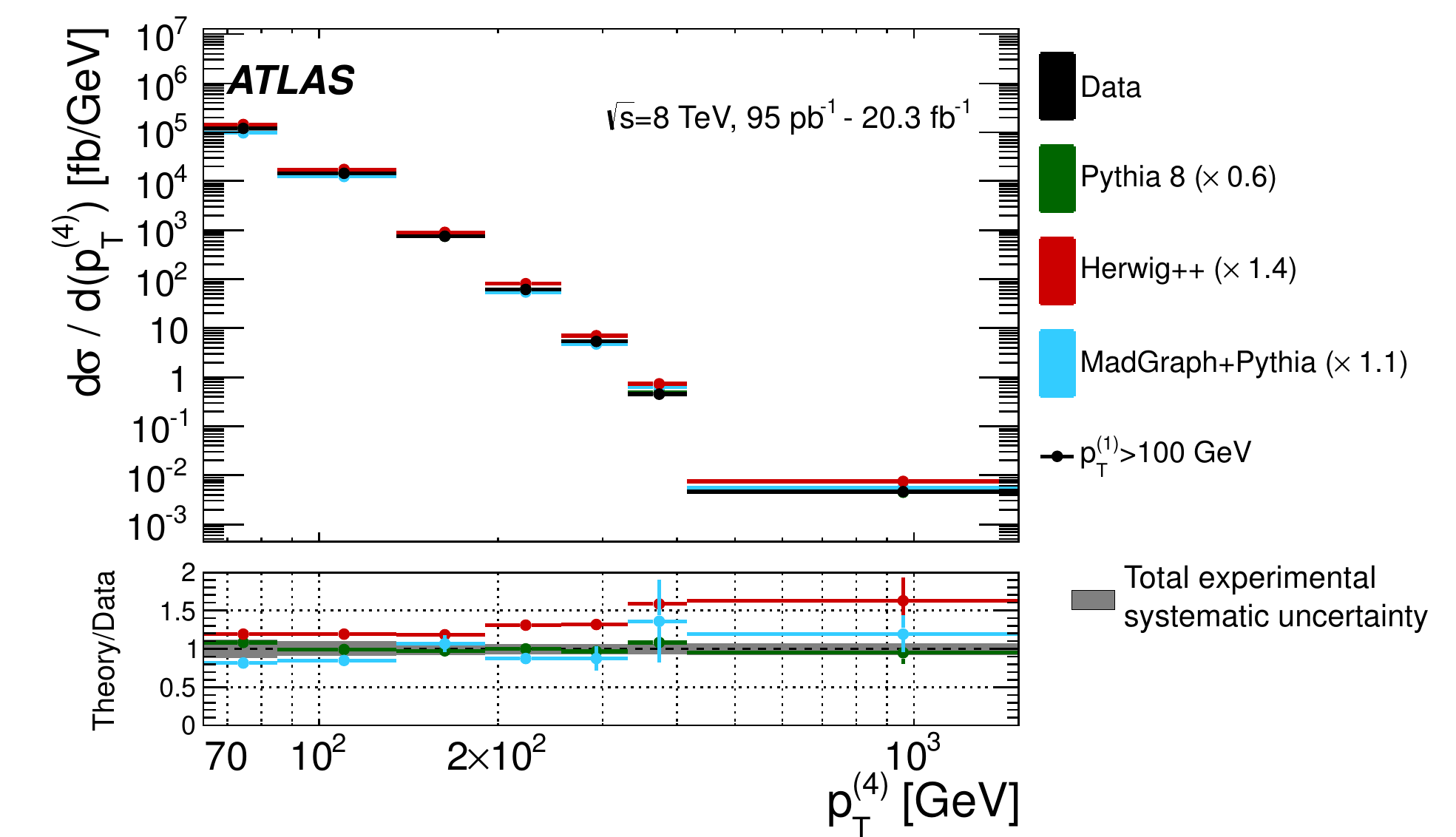}\\
\vspace{0.5cm}
\includegraphics[width=0.9\textwidth]{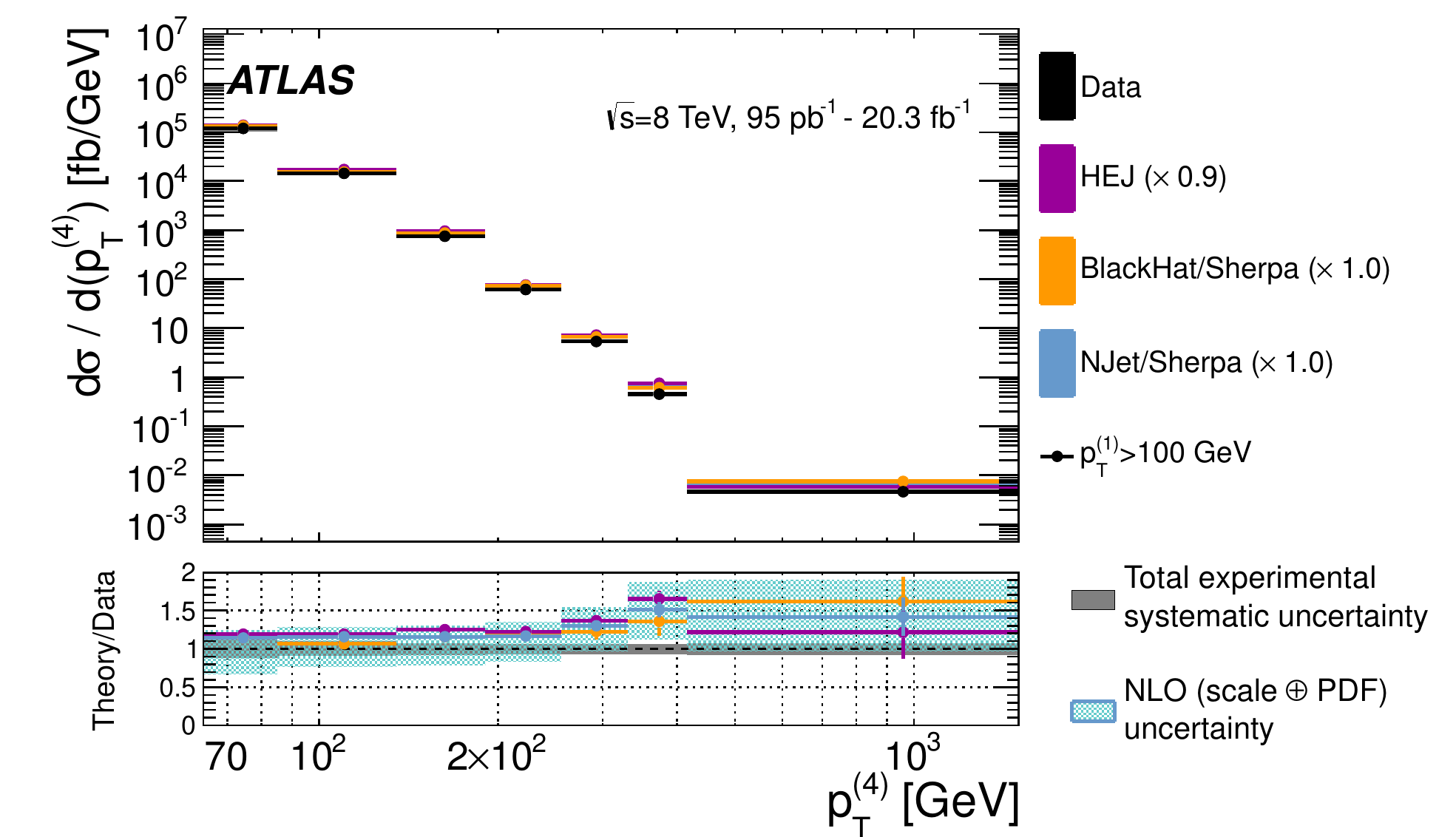}\\
\caption{Unfolded four-jet differential cross section as a function of \ptFour, compared to different theoretical predictions. \otherdetails}
\label{fig:unfolded_jetpt3}
\end{center}
\end{figure}

\clearpage

\begin{figure}
\begin{center}
\includegraphics[width=0.9\textwidth]{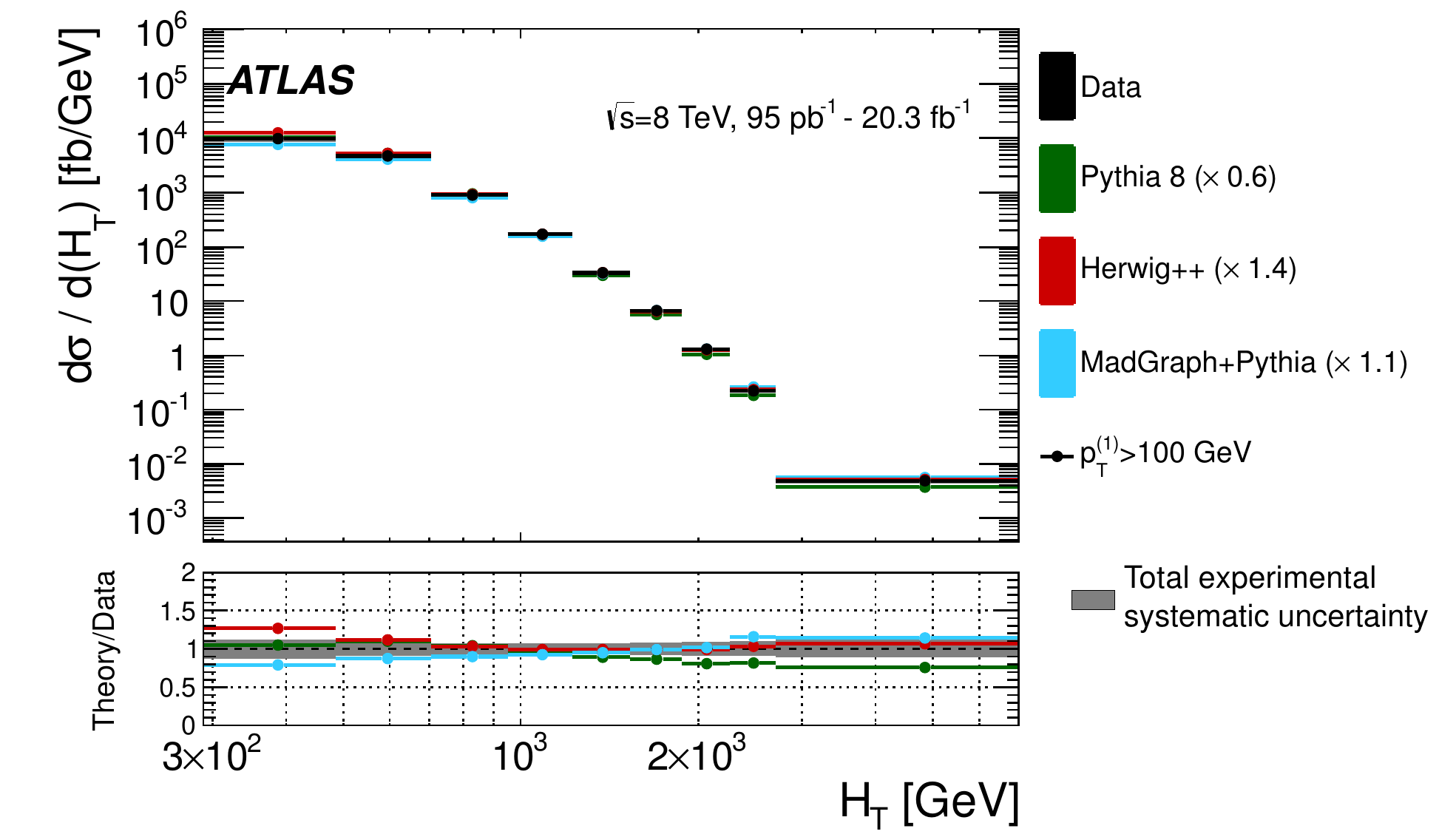}\\
\vspace{0.5cm}
\includegraphics[width=0.9\textwidth]{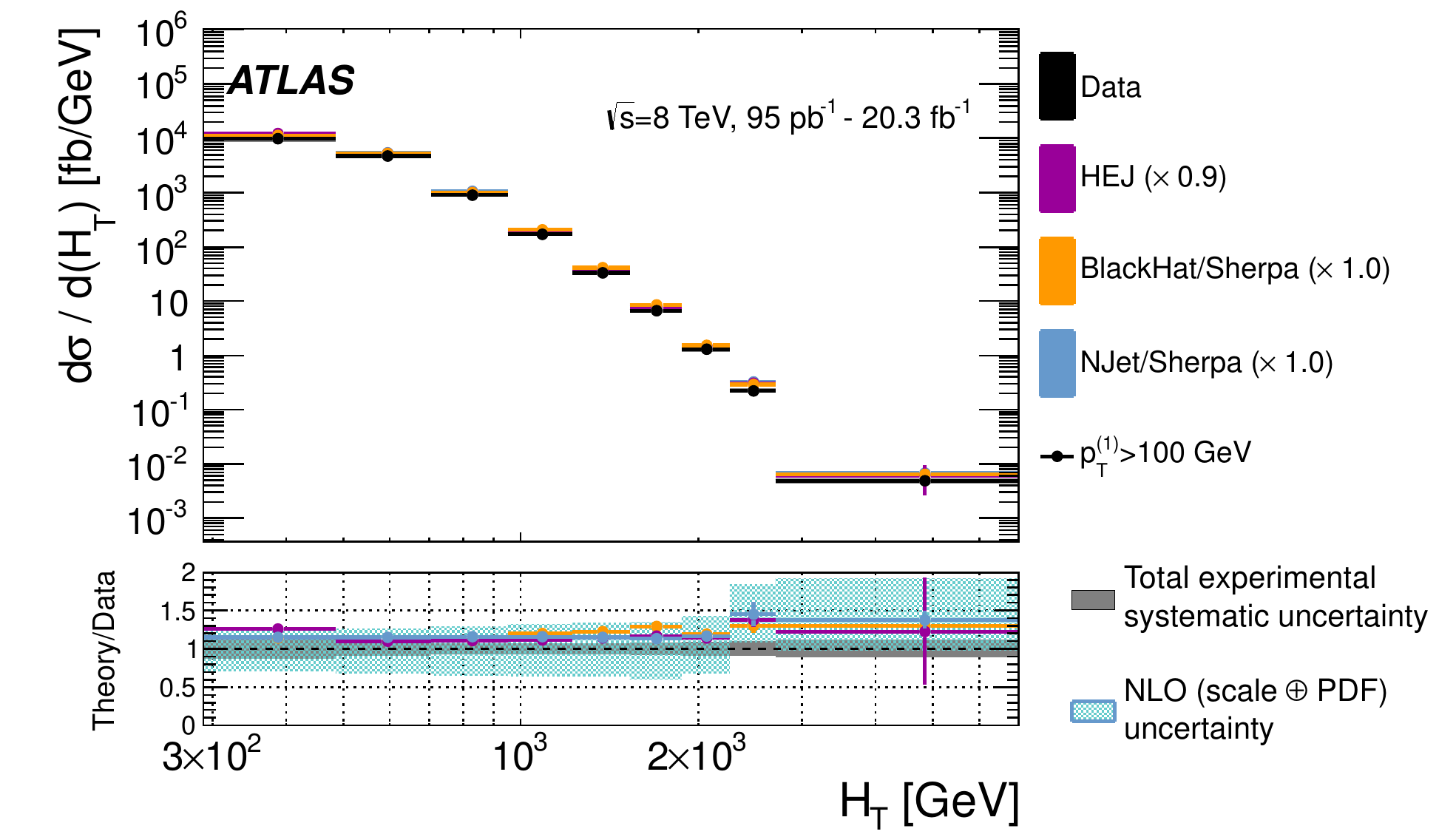}\\
\caption{Unfolded four-jet differential cross section as a function of \HT, compared to different theoretical predictions. \otherdetails}
\label{fig:unfolded_HT}
\end{center}
\end{figure}

\begin{figure}
\begin{center}
\includegraphics[width=0.9\textwidth]{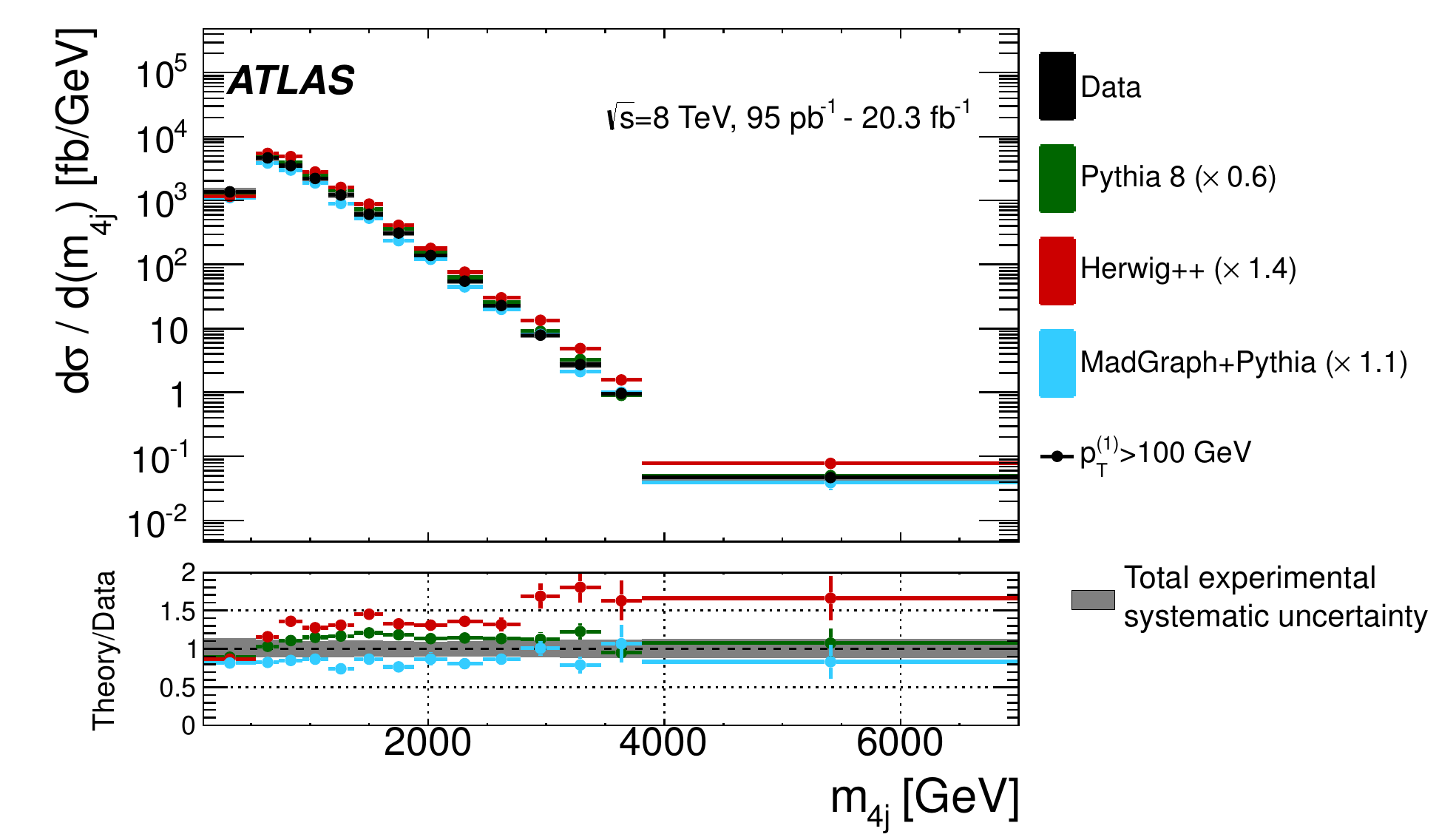}\\
\vspace{0.5cm}
\includegraphics[width=0.9\textwidth]{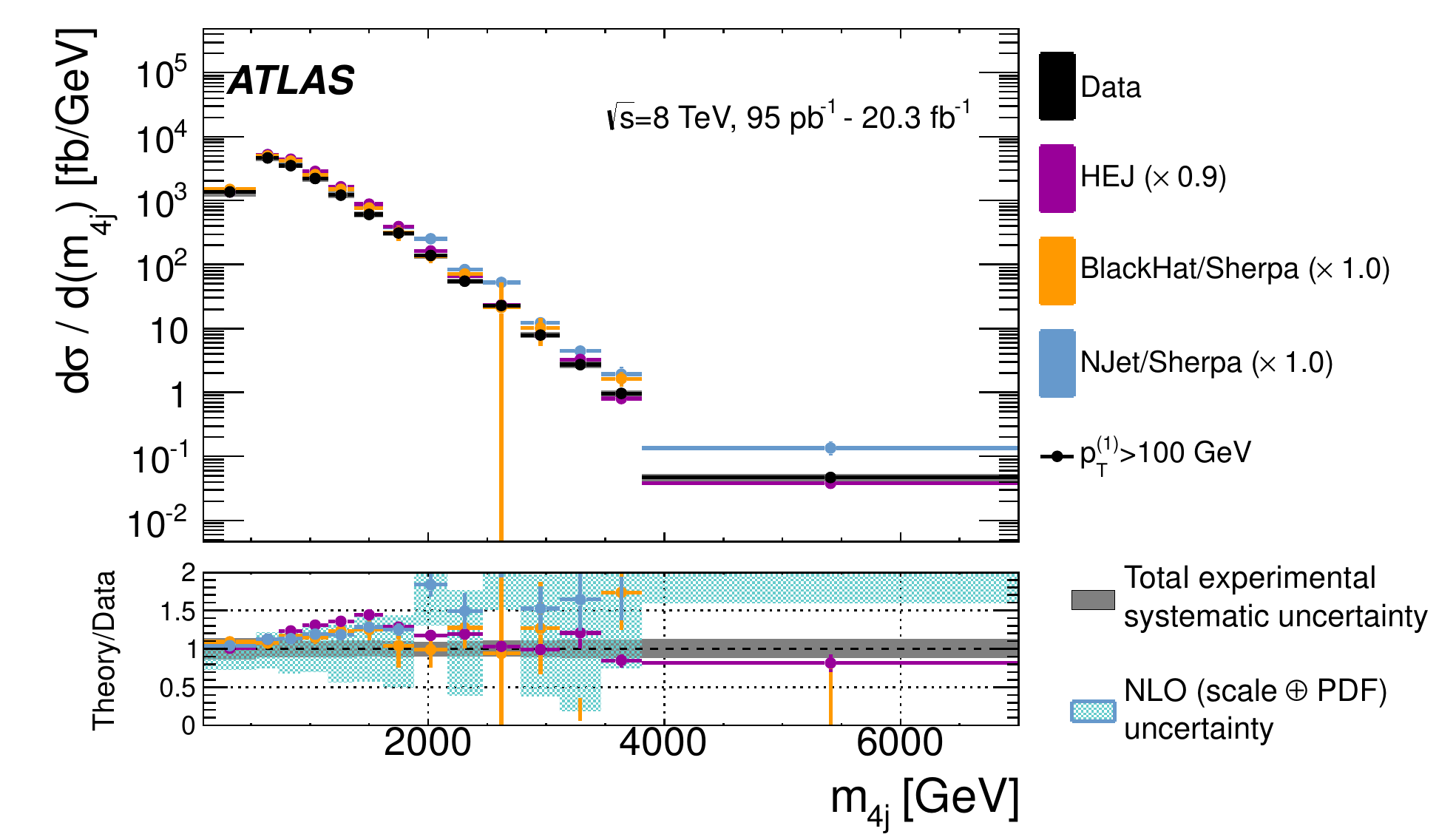}\\
\caption{Unfolded four-jet differential cross section as a function of \mjjjj,
compared to different theoretical predictions. \otherdetails \missNLO}
\label{fig:unfolded_mjjjj}
\end{center}
\end{figure}

\begin{figure}
\begin{center}
\includegraphics[width=1.0\textwidth]{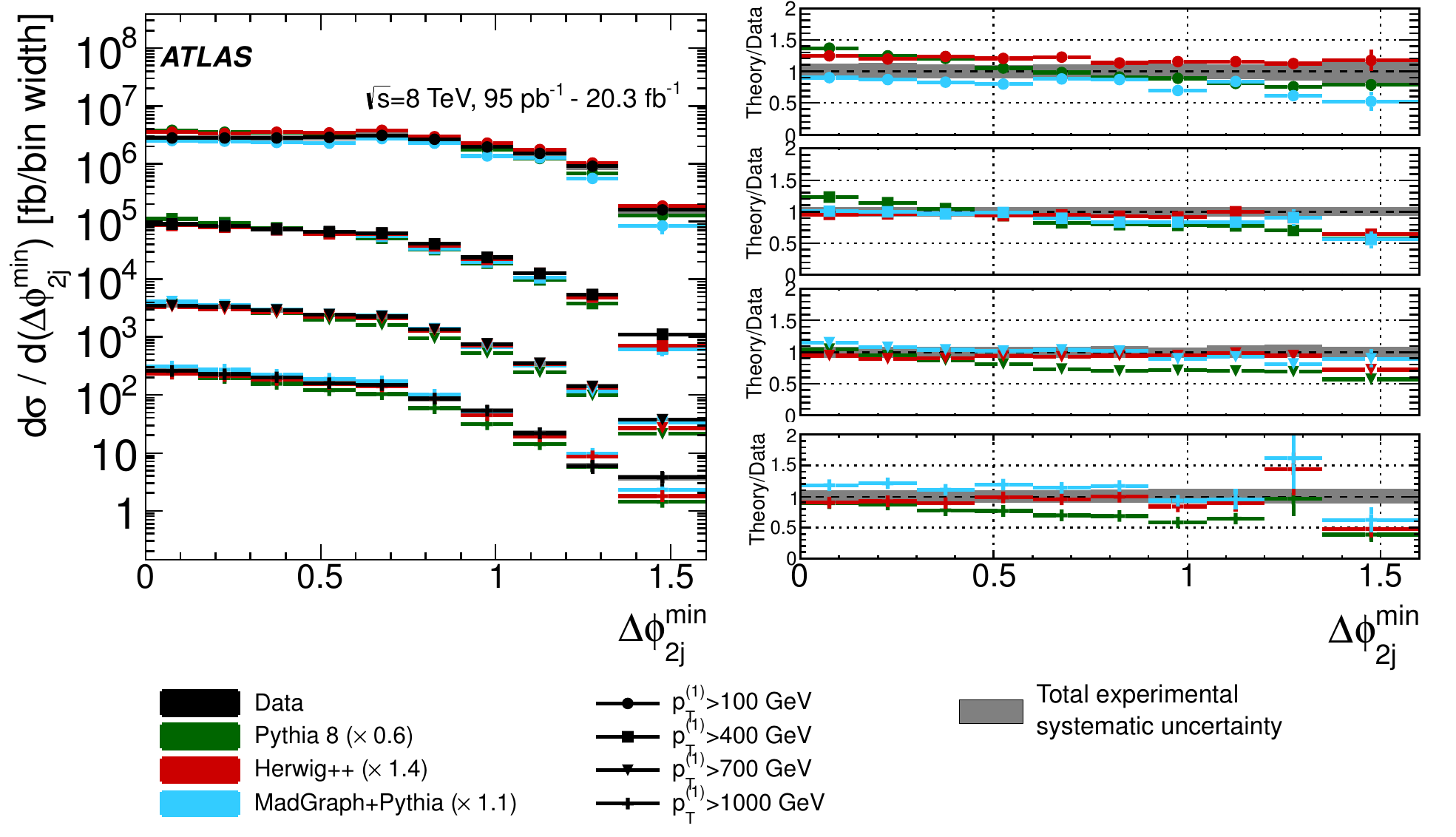}\\
\vspace{0.5cm}
\includegraphics[width=1.0\textwidth]{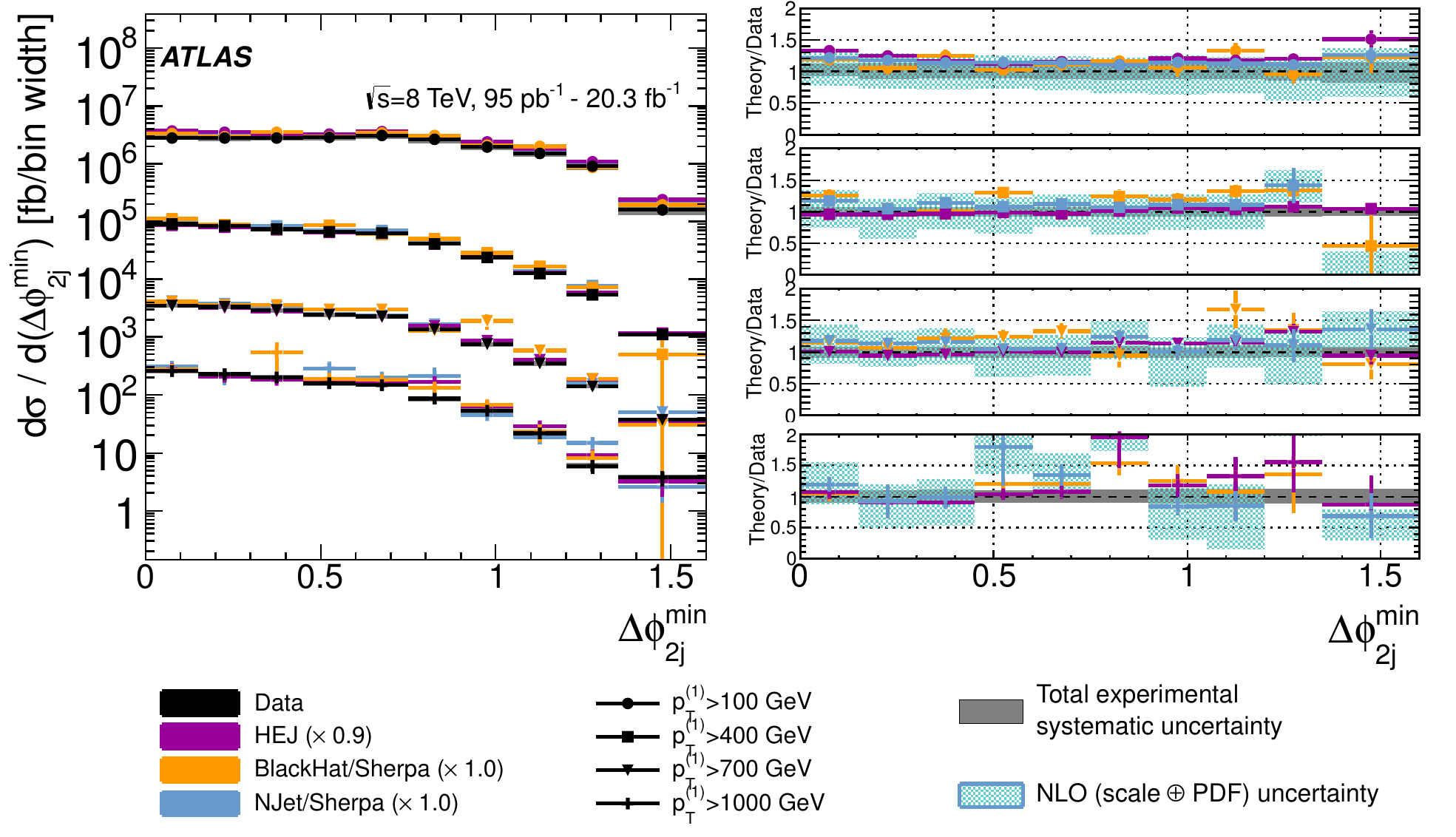}\\
\caption{Unfolded four-jet differential cross section as a function of
  \minDphiij, compared to different theoretical predictions. The other details
  are as for \figref{fig:unfolded_mjjminratio}, but here the multiple ratio
plots correspond to different selection criteria applied to \ptOne. \missNLO}
\label{fig:unfolded_minDphiij}
\end{center}
\end{figure}

\newcommand\otherdetailstwo{The other details are as for \figref{fig:unfolded_minDphiij}. }

\begin{figure}
\begin{center}
\includegraphics[width=1.0\textwidth]{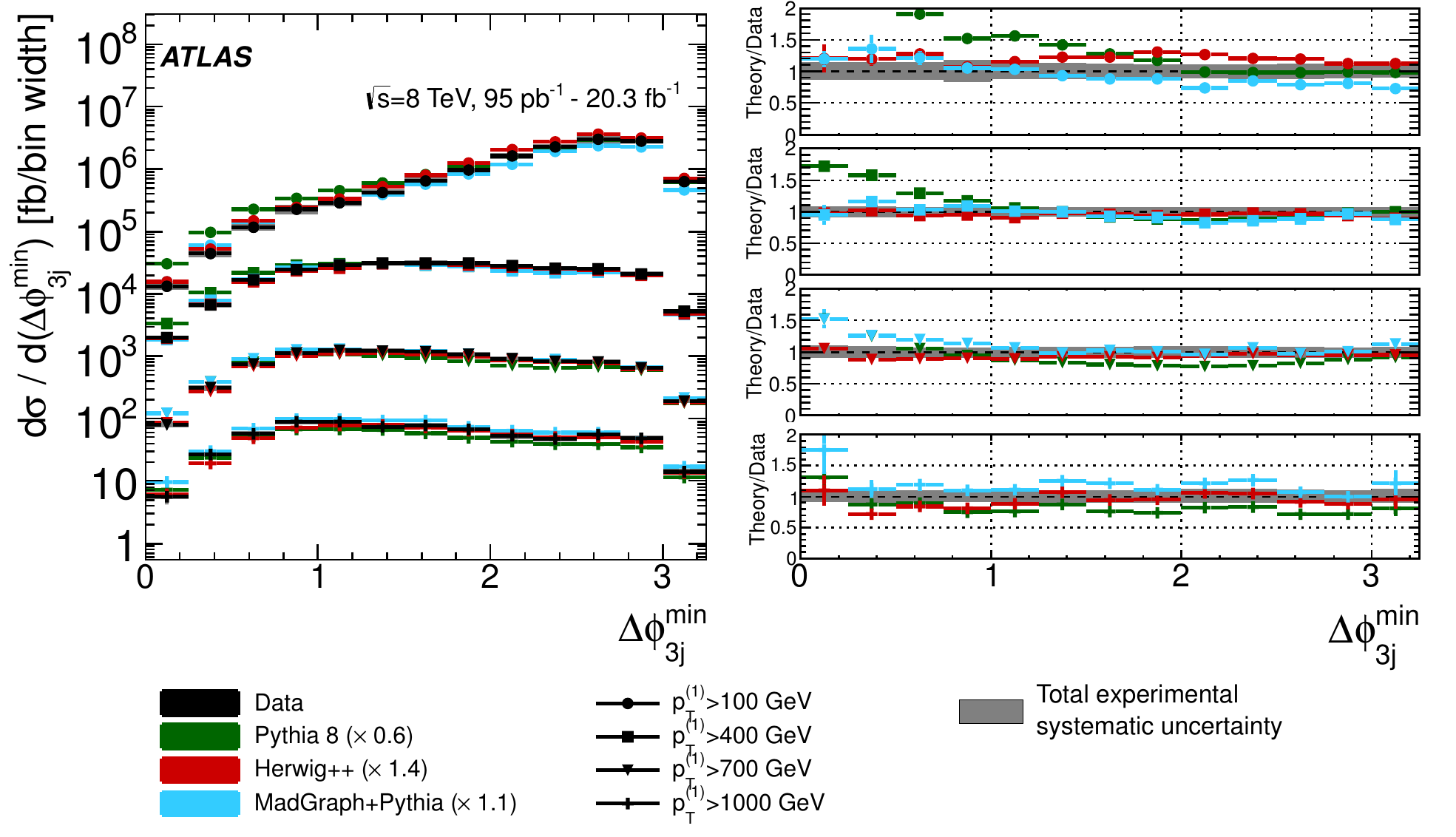}\\
\vspace{0.5cm}
\includegraphics[width=1.0\textwidth]{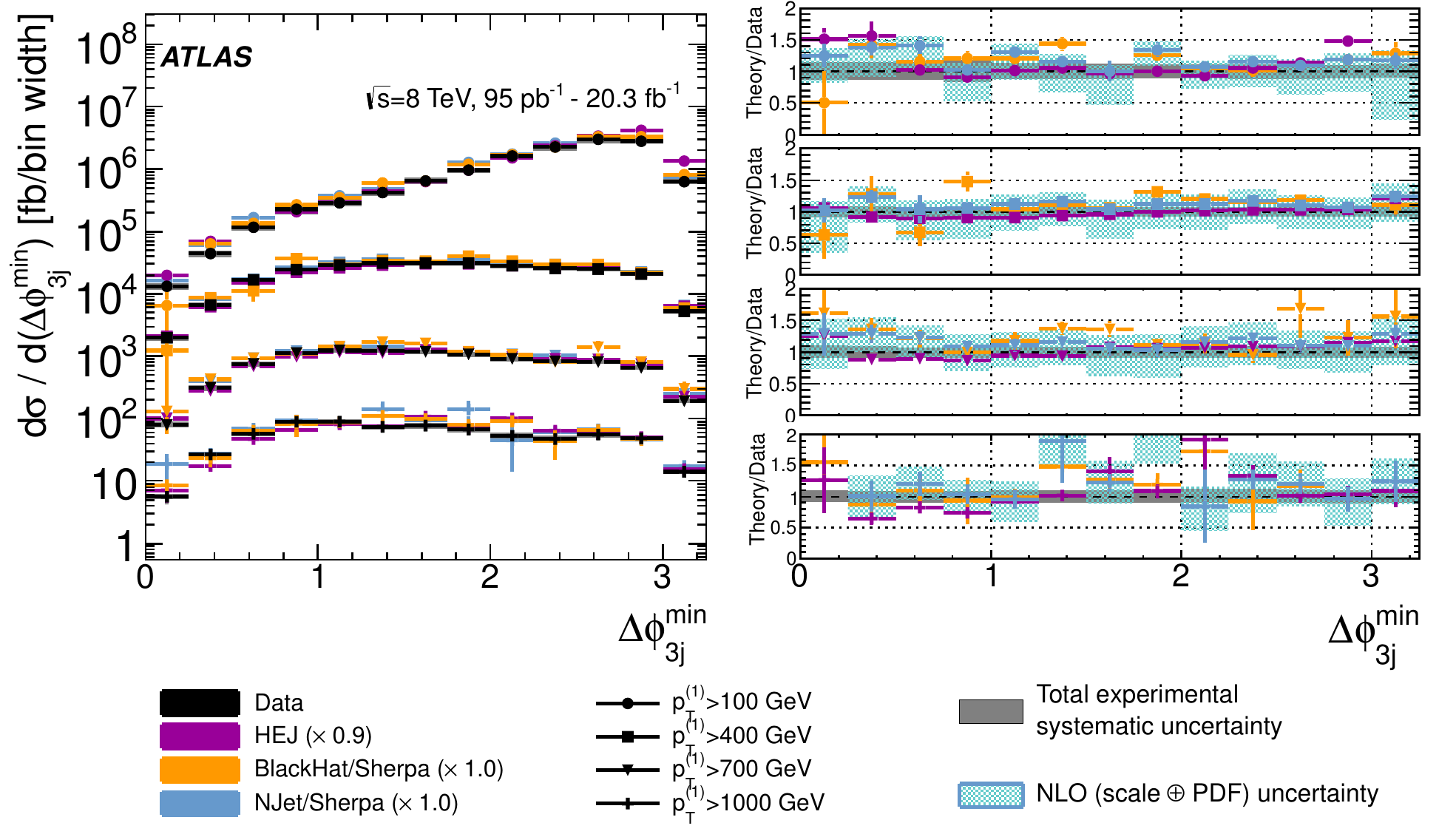}\\
\caption{Unfolded four-jet differential cross section as a function of
\minDphiijk, compared to different theoretical predictions. \otherdetailstwo 
 \missNLO}
\label{fig:unfolded_minDphiijk}
\end{center}
\end{figure}

\begin{figure}
\begin{center}
\includegraphics[width=1.0\textwidth]{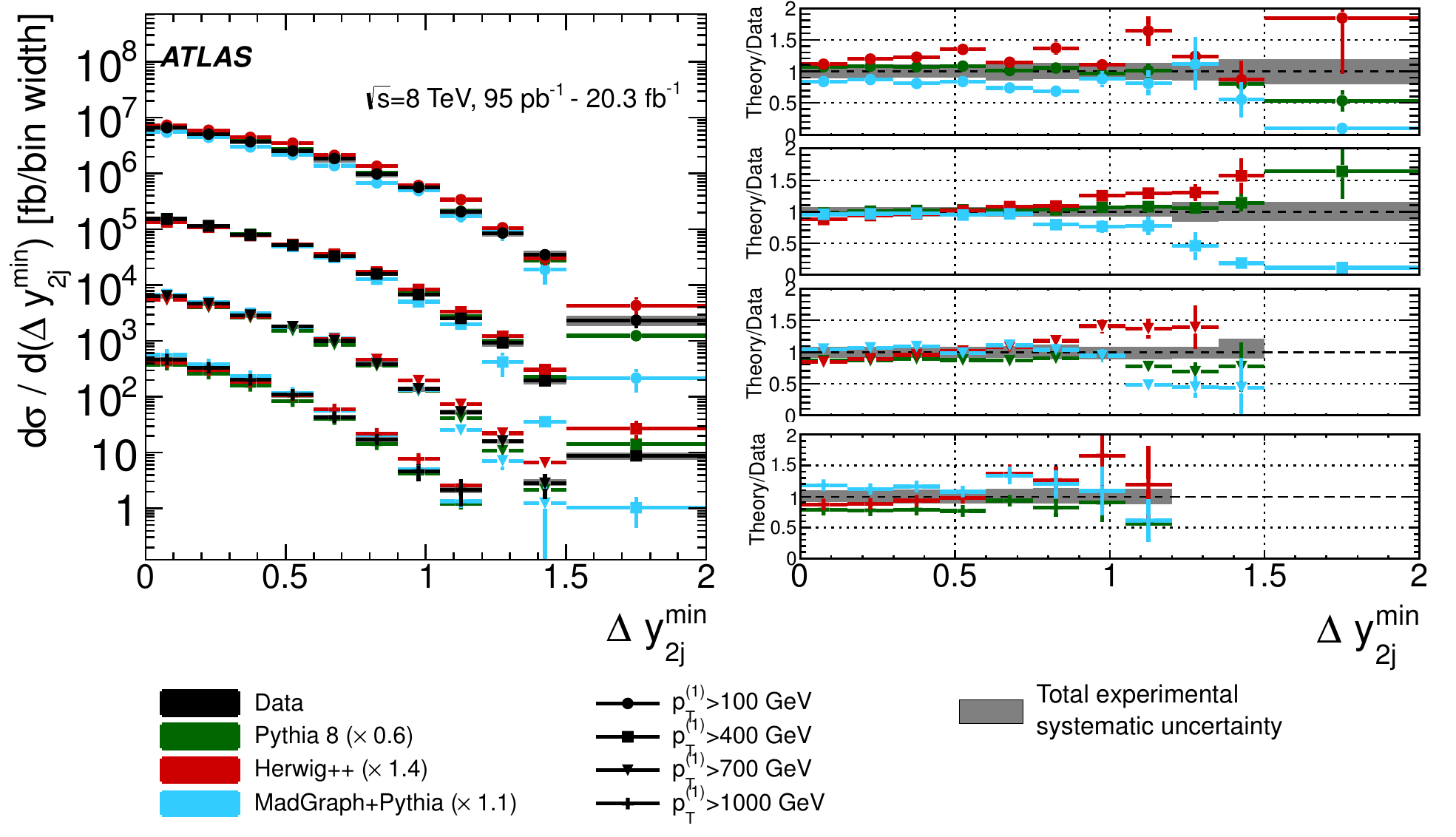}\\
\vspace{0.5cm}
\includegraphics[width=1.0\textwidth]{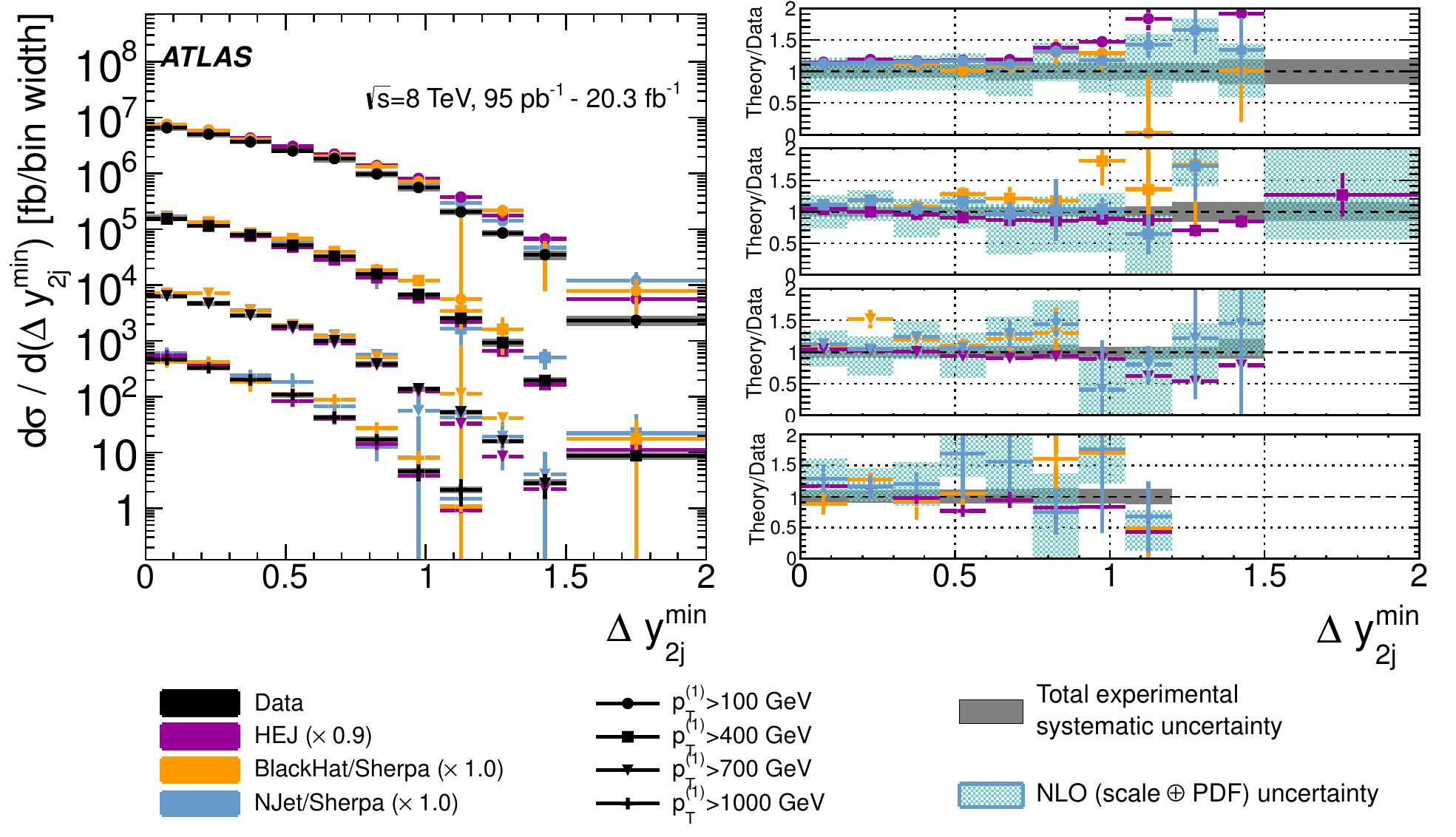}\\
\caption{Unfolded four-jet differential cross section as a function of
\minDrapij, compared to different theoretical predictions. \otherdetailstwo 
 \missNLO}
\label{fig:unfolded_minDrapij}
\end{center}
\end{figure}

\begin{figure}
\begin{center}
\includegraphics[width=1.0\textwidth]{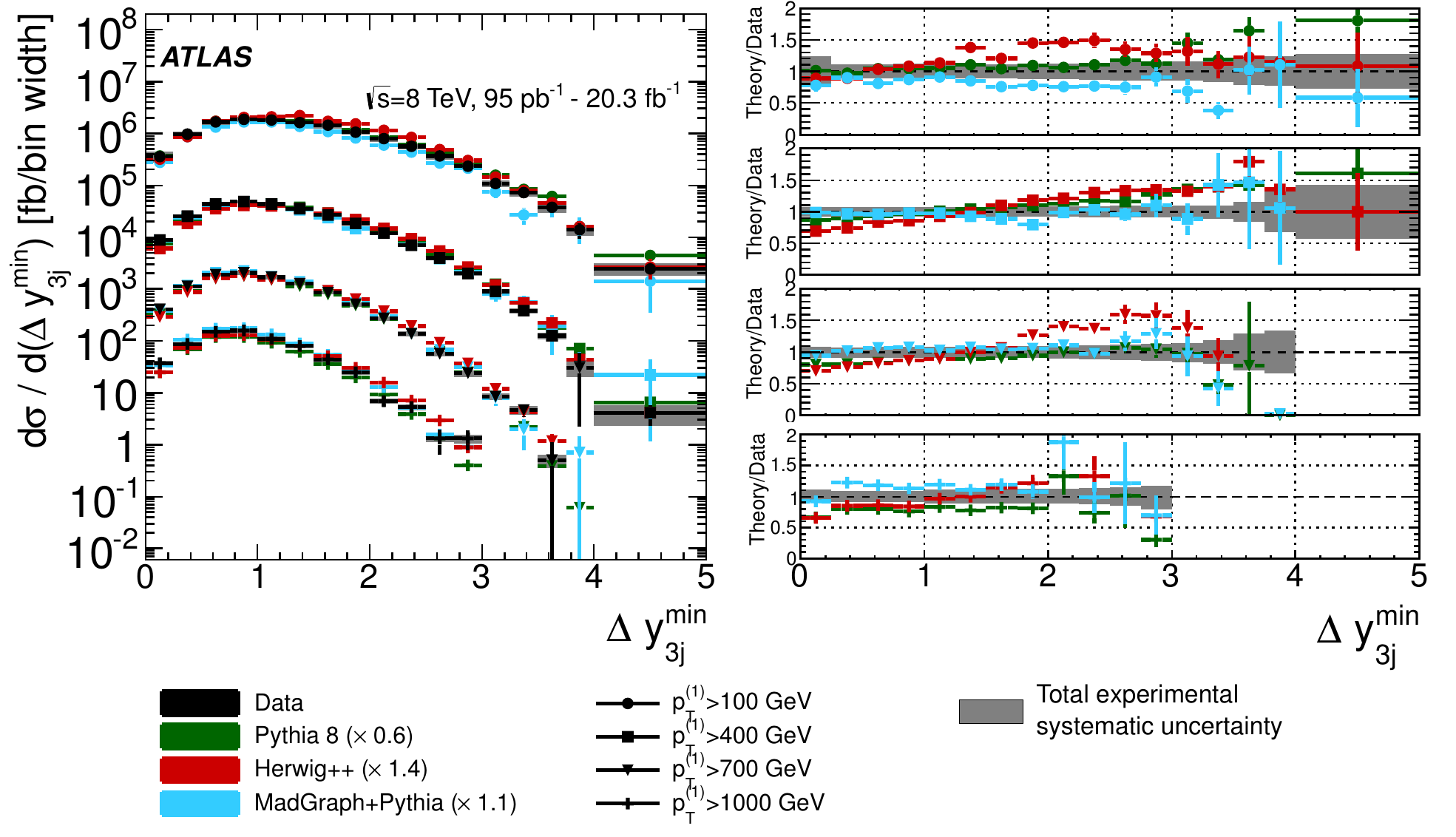}\\
\vspace{0.5cm}
\includegraphics[width=1.0\textwidth]{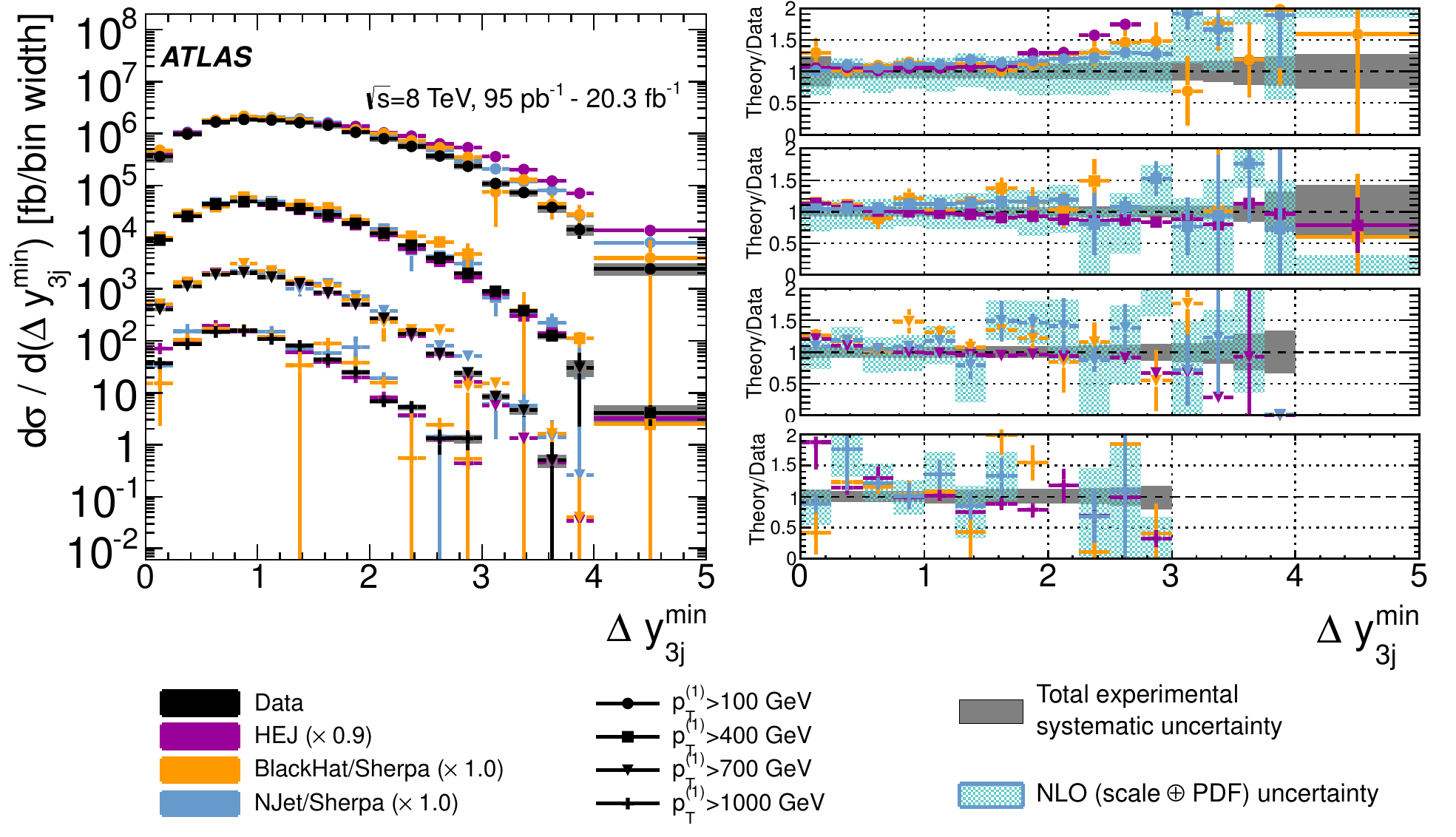}\\
\caption{Unfolded four-jet differential cross section as a function of
\minDrapijk, compared to different theoretical predictions. \otherdetailstwo 
 \missNLO}
\label{fig:unfolded_minDrapijk}
\end{center}
\end{figure}

\begin{figure}
\begin{center}
\includegraphics[width=1.0\textwidth]{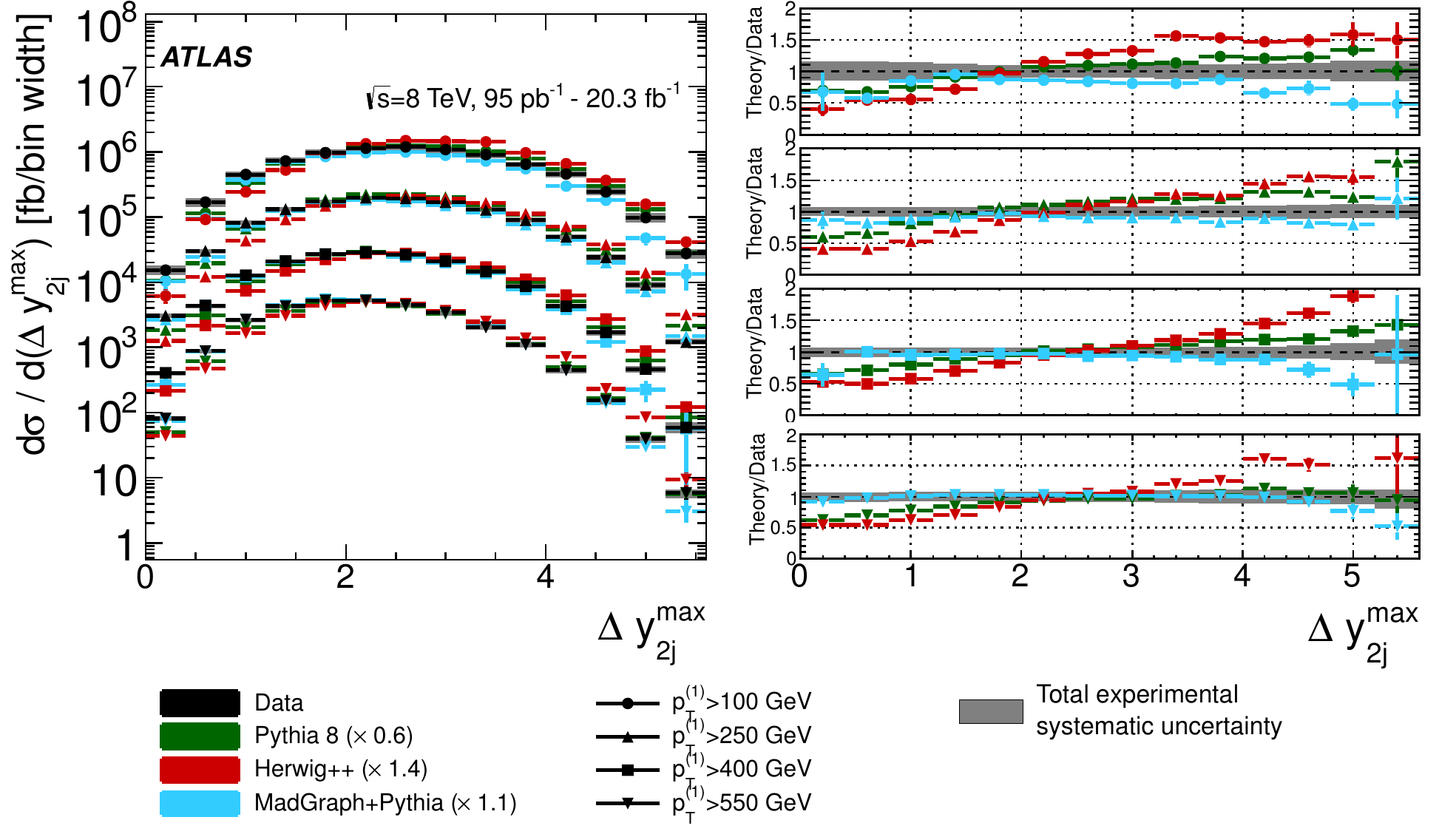}\\
\vspace{0.5cm}
\includegraphics[width=1.0\textwidth]{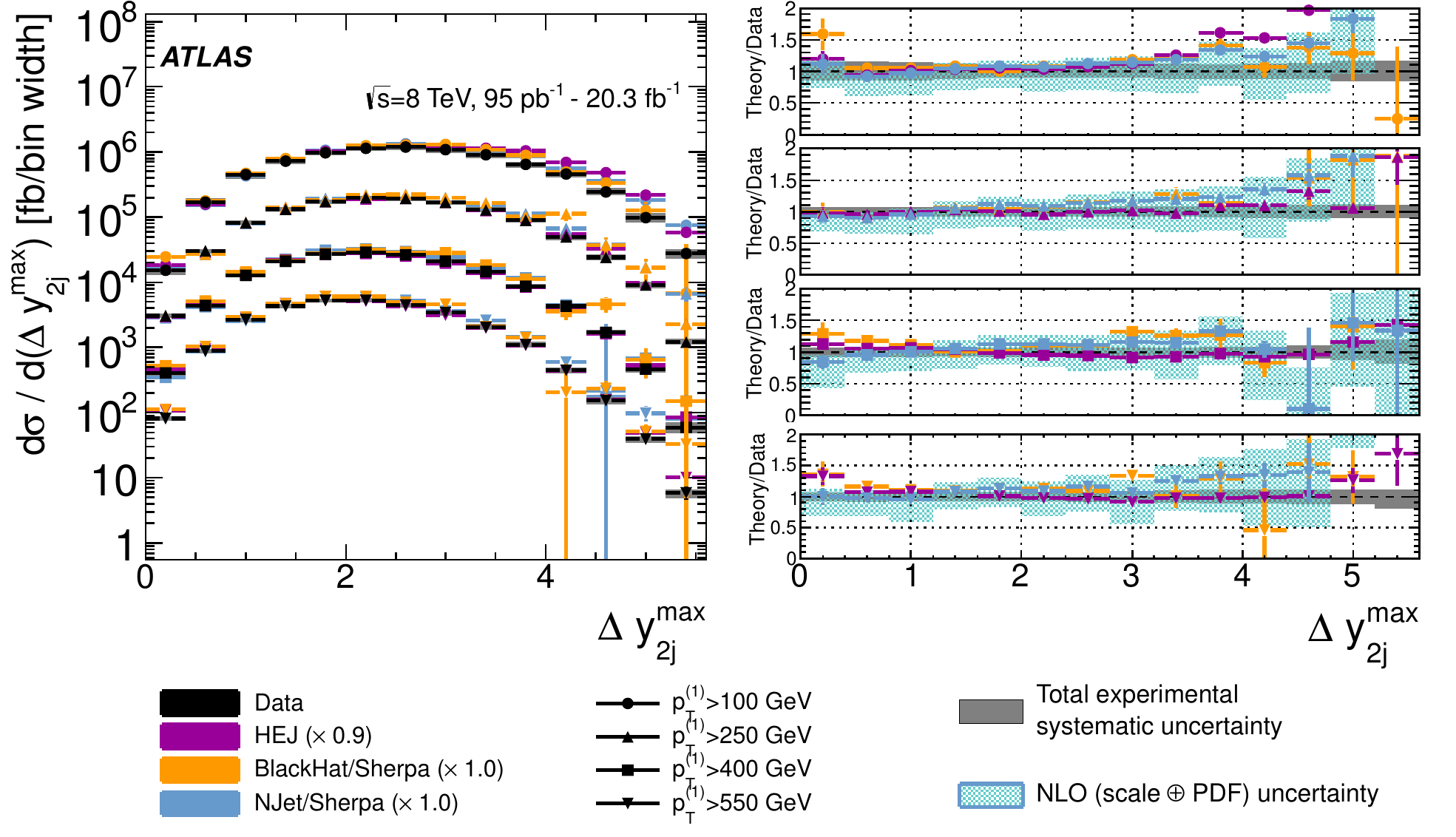}\\
\caption{Unfolded four-jet differential cross section as a function of
\maxDrapij, compared to different theoretical predictions. \otherdetailstwo
\missNLO}
\label{fig:unfolded_maxDrapij}
\end{center}
\end{figure}

\begin{figure}
\begin{center}
\includegraphics[width=1.0\textwidth]{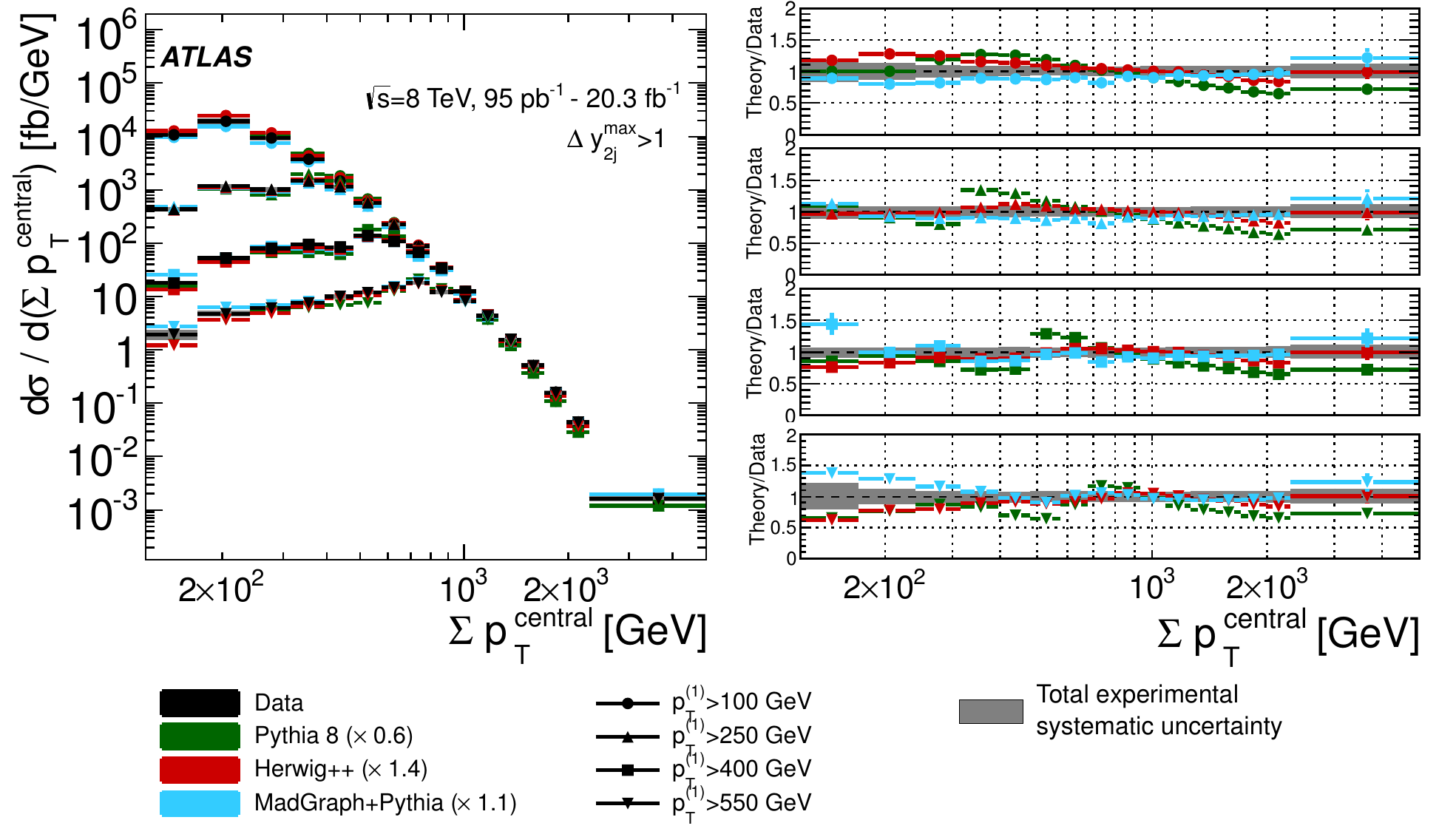}\\
\vspace{0.5cm}
\includegraphics[width=1.0\textwidth]{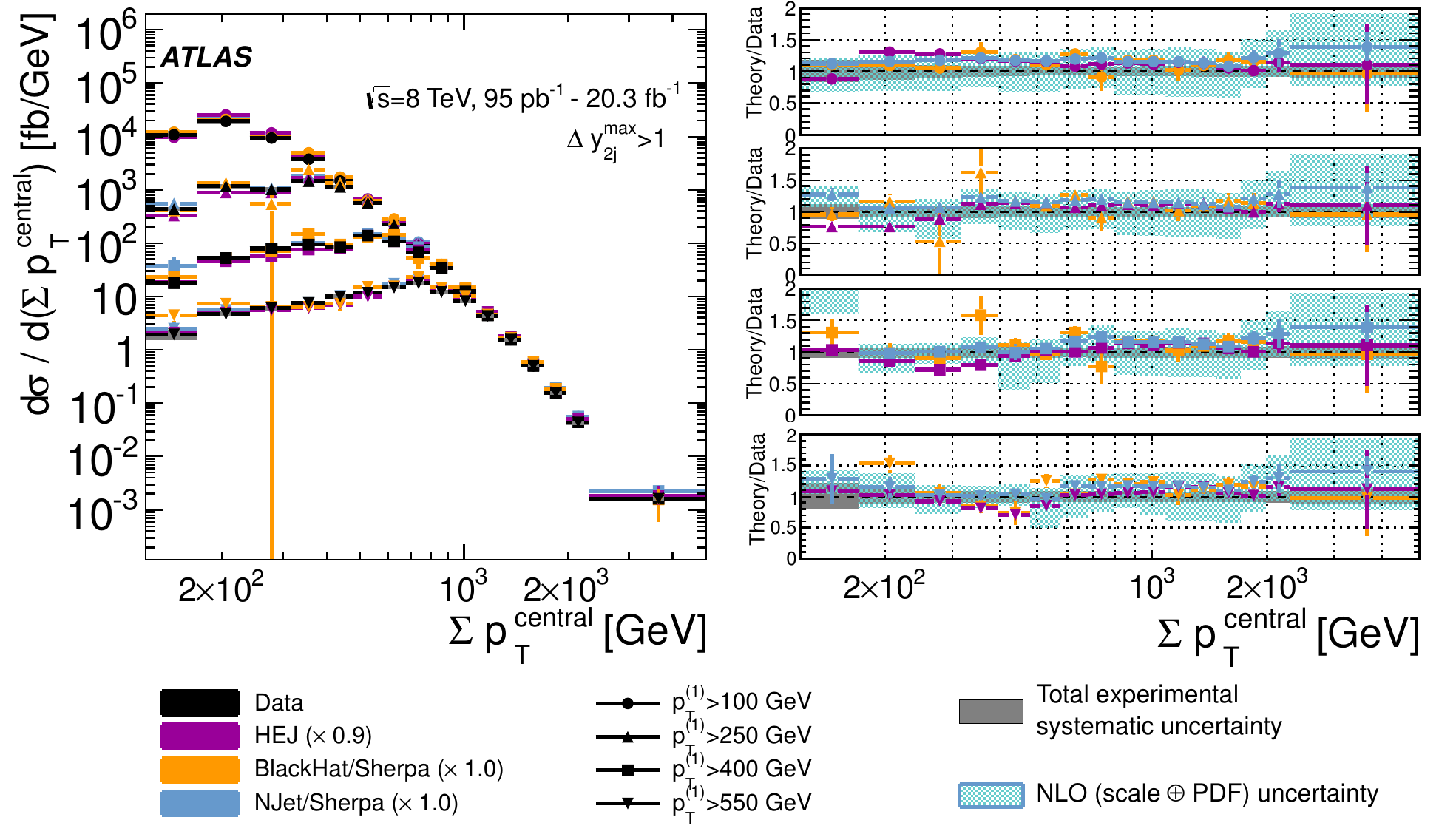}\\
\caption{Unfolded four-jet differential cross section as a function of \sumptcent\ with $\maxDrapij>1$, compared to different theoretical predictions. \otherdetailstwo}
\label{fig:unfolded_sumptcent1}
\end{center}
\end{figure}

\begin{figure}
\begin{center}
\includegraphics[width=1.0\textwidth]{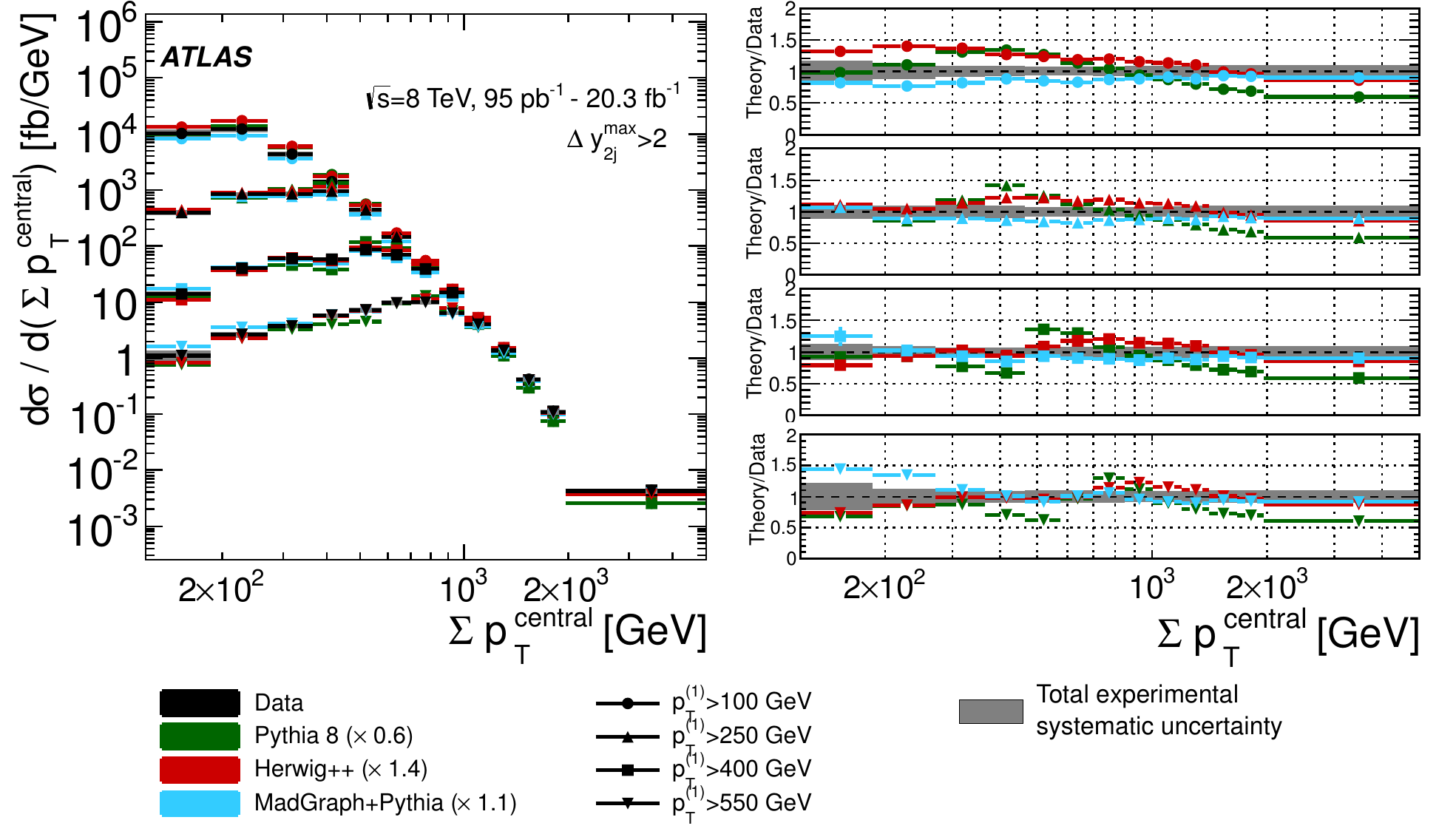}\\
\vspace{0.5cm}
\includegraphics[width=1.0\textwidth]{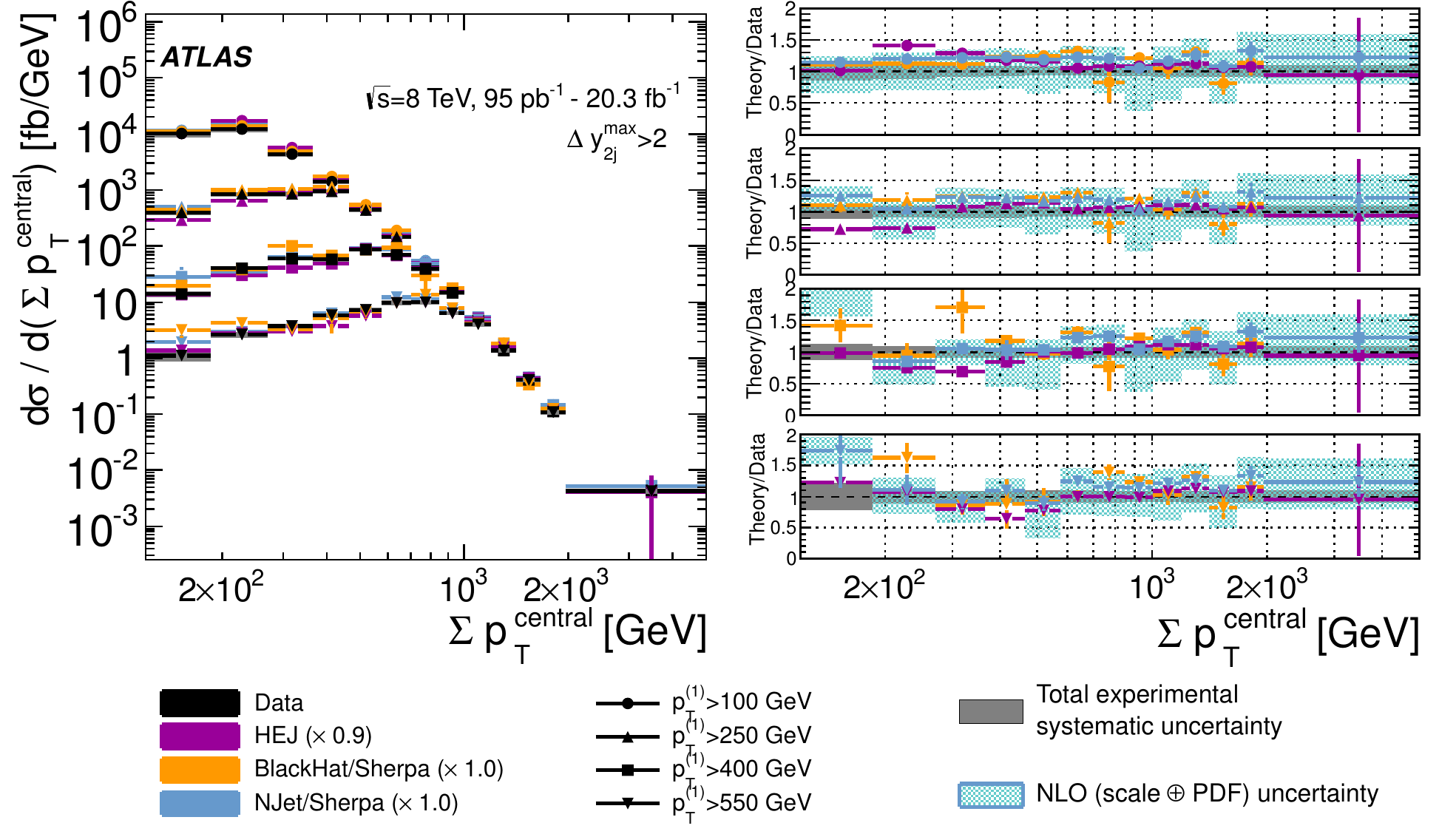}\\
\caption{Unfolded four-jet differential cross section as a function of \sumptcent\ with $\maxDrapij>2$, compared to different theoretical predictions. \otherdetailstwo}
\label{fig:unfolded_sumptcent2}
\end{center}
\end{figure}

\begin{figure}
\begin{center}
\includegraphics[width=1.0\textwidth]{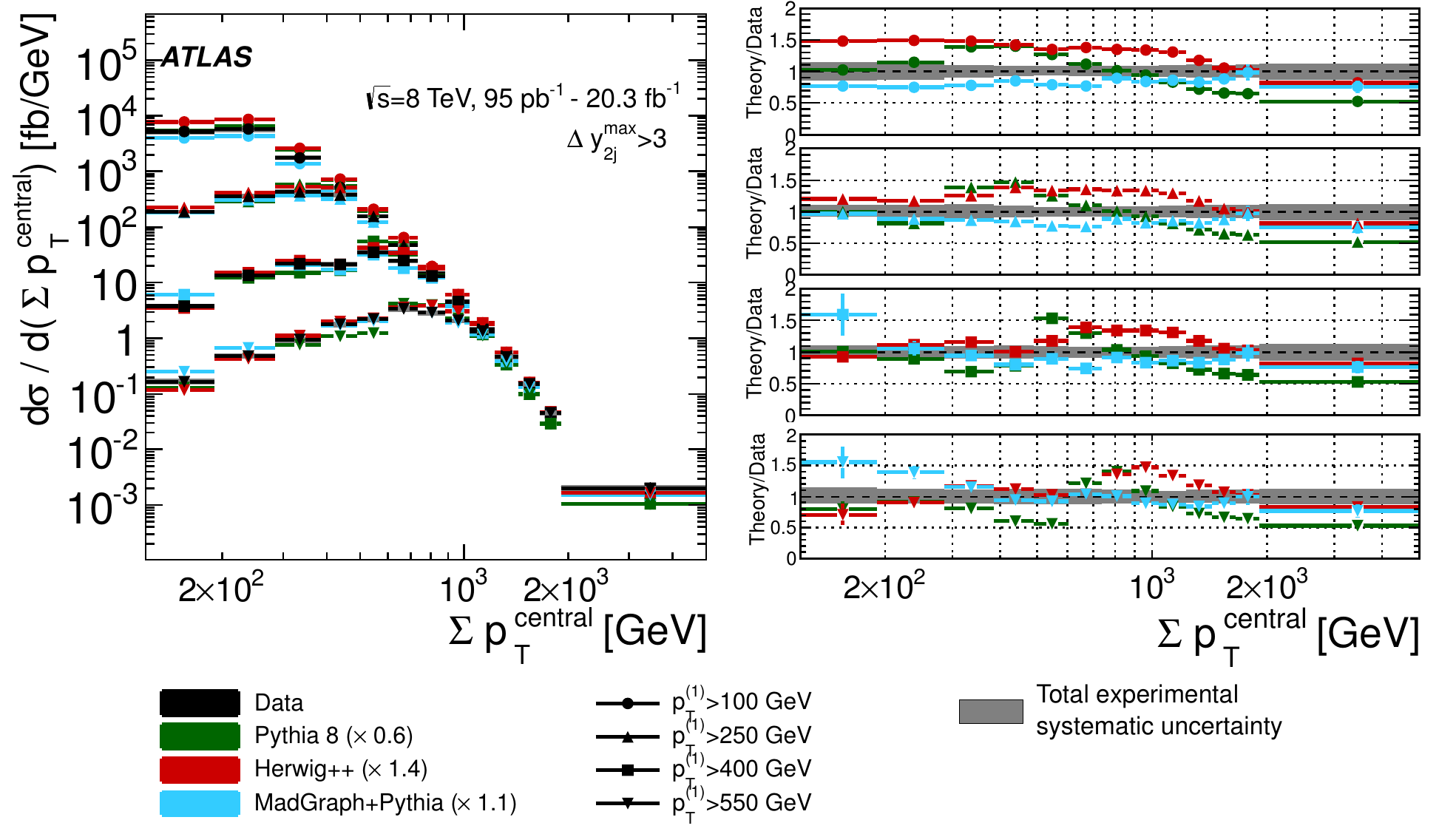}\\
\vspace{0.5cm}
\includegraphics[width=1.0\textwidth]{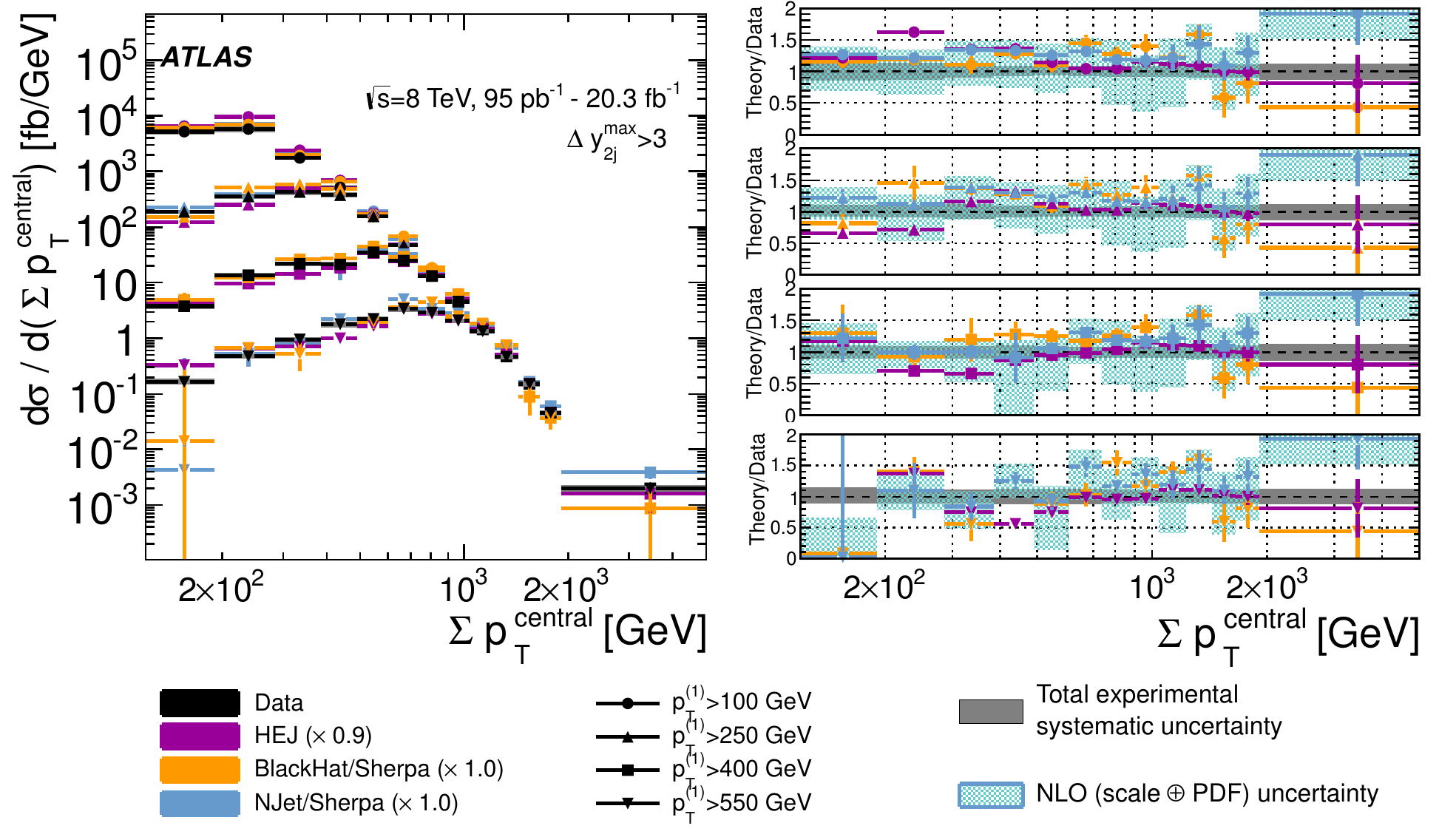}\\
\caption{Unfolded four-jet differential cross section as a function of \sumptcent\ with $\maxDrapij>3$, compared to different theoretical predictions. \otherdetailstwo}
\label{fig:unfolded_sumptcent3}
\end{center}
\end{figure}

\begin{figure}
\begin{center}
\includegraphics[width=1.0\textwidth]{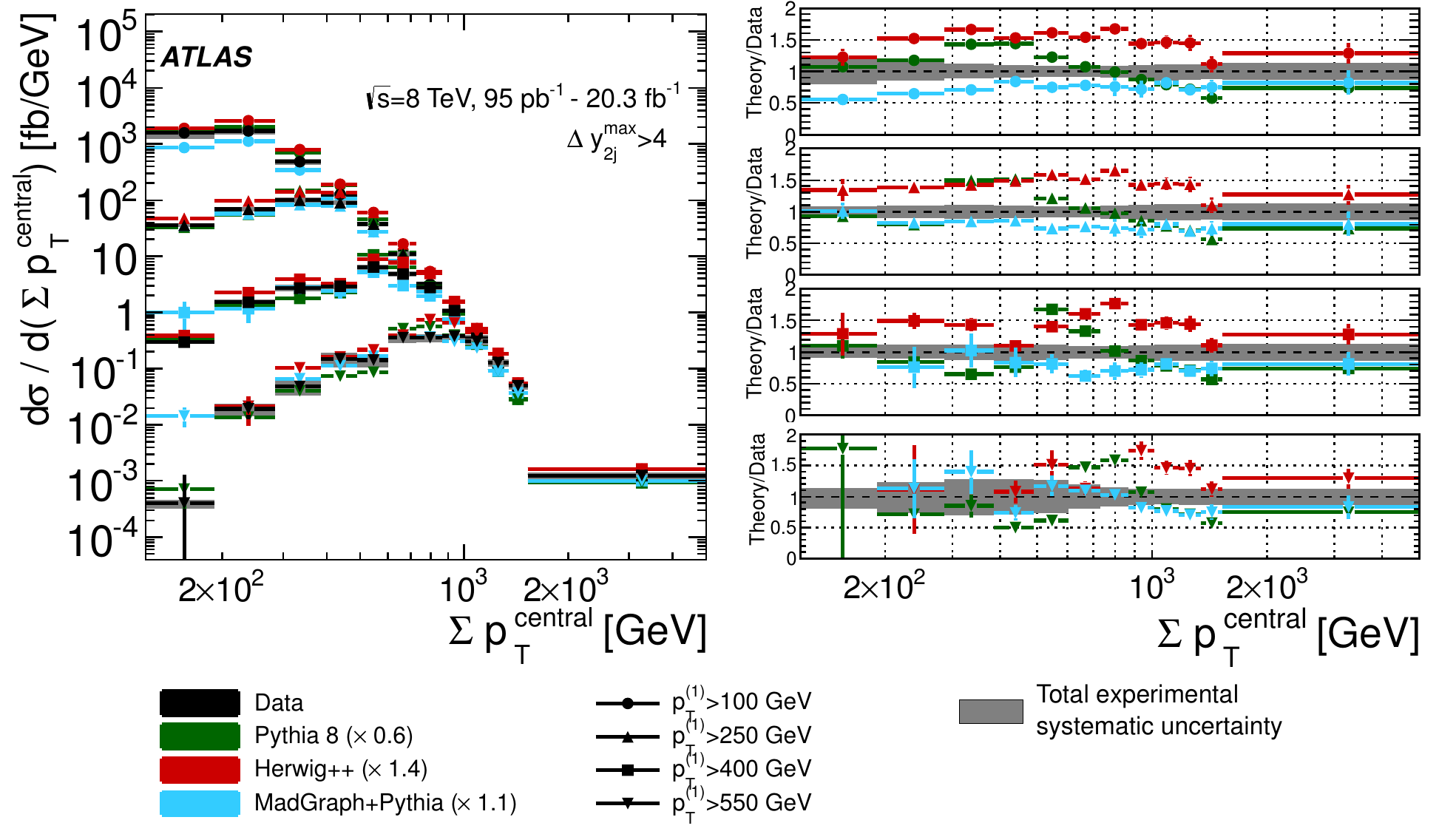}\\
\vspace{0.5cm}
\includegraphics[width=1.0\textwidth]{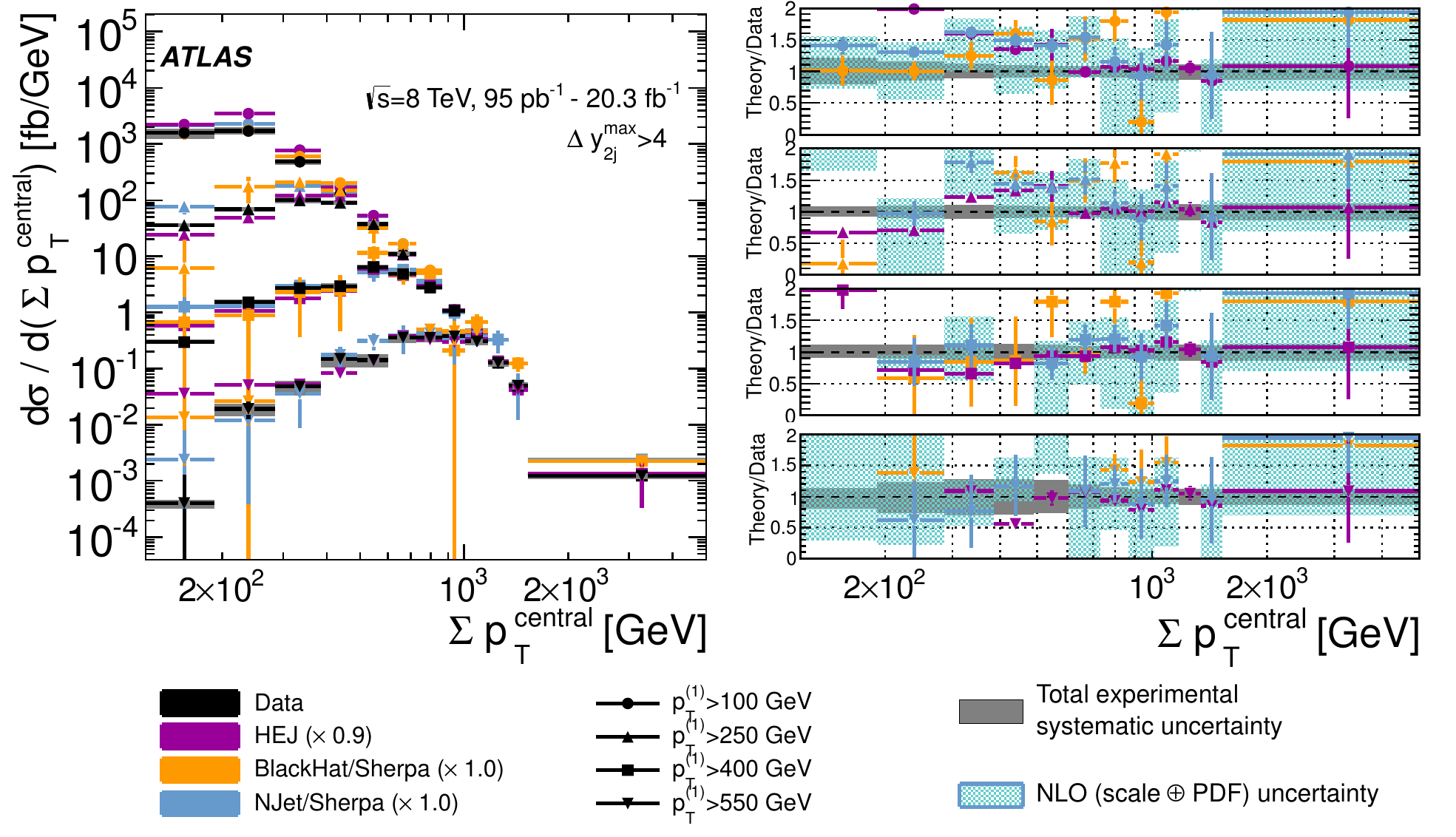}\\
\caption{Unfolded four-jet differential cross section as a function of
\sumptcent\ with $\maxDrapij>4$, compared to different theoretical predictions.
\otherdetailstwo  \missNLO}
\label{fig:unfolded_sumptcent4}
\end{center}
\end{figure}

%% file: tex/conclusion.tex
This paper presents unfolded differential cross sections of events with at least
four jets in $pp$ collisions at 8~TeV centre-of-mass energy. The cross
sections are studied as a function of a variety of kinematic and topological
variables which include momenta, masses and angles. 
Events are selected if the four \antikt\ $R=0.4$ jets with the largest
transverse momentum within the rapidity range $|y|<2.8$ are well separated
($\dRmin>0.65$), all have $\pT>64$~GeV, and include at least one jet
with $\pt>100$~GeV.
The results are obtained from the analysis of the full dataset collected by the ATLAS detector at the
LHC in 2012, which corresponds to a total integrated luminosity of 20.3~\ifb.
The total experimental systematic uncertainty is typically of the order of 10\%, and it is
dominated by the jet energy scale calibration uncertainty.

The measurements are compared to NLO \pQCD\ predictions provided by
\BlackHatSherpa\ and \NJETSherpa, as well as the all-orders calculation provided
by \HEJ. Three leading-order calculations are also considered, including two
$2\rightarrow 2$ PS samples (\Pythia{} and \Herwigpp) and a
multi-leg calculation with up to four partons in the ME matched to a
PS generated by \Pythia\ (\MadgraphPythia).

The LO cross sections and \HEJ\ are normalised by fixed factors to facilitate
the comparison of the spectra in the kinematic regions of interest; these
factors vary between $0.6$ and $1.4$ for the
different samples, where the \MadgraphPythia\ and \HEJ\ samples are the ones that need the
smallest corrections. The NLO predictions, \BlackHatSherpa\ and \NJETSherpa, are
almost always compatible with the data within their theoretical uncertainties, which
are found to be large ($\mathcal{O}(30\%)$ at low momenta) and asymmetric.
Within the normalisation scheme used, \MadgraphPythia\ also provides a good
description of the data, as does \HEJ, especially at high leading jet \pt. The
$2\to 2$ PS calculations generally describe the data relatively
poorly, although they are found to provide good predictions in some particular
cases: \Pythia\ gives a very good prediction of \ptFour\ and
\minDrapij, while \Herwigpp\ performs well in the azimuthal angle variables.

Looking at the individual distributions of the differential cross section, the
description of the jet momenta is compatible with previous measurements of the
multi-jet cross sections. It should be noted that 
\HEJ, \NJETSherpa\ and \BlackHatSherpa\ give a very good description of the
distributions of the leading jets but show some discrepancy with the data for \ptFour.  For
variables that are particularly sensitive to wide-angle configurations and
high-\pt radiation, such as masses or angles, \BlackHatSherpa, \NJETSherpa\ and
\MadgraphPythia\ do a remarkable job overall. \HEJ\ also provides a good description
of the data, the main exception being that it disagrees with the rapidity
measurements in events with low \ptOne. At high \ptOne\ the prediction is very good. These
measurements expose the shortcomings of $2 \to 2$ parton ME+PS predictions
in a variety of scenarios and highlight the importance of the more sophisticated
calculations.

%% file: acknowledgements/Acknowledgements.tex

We thank CERN for the very successful operation of the LHC, as well as the
support staff from our institutions without whom ATLAS could not be
operated efficiently.

We acknowledge the support of ANPCyT, Argentina; YerPhI, Armenia; ARC, Australia; BMWFW and FWF, Austria; ANAS, Azerbaijan; SSTC, Belarus; CNPq and FAPESP, Brazil; NSERC, NRC and CFI, Canada; CERN; CONICYT, Chile; CAS, MOST and NSFC, China; COLCIENCIAS, Colombia; MSMT CR, MPO CR and VSC CR, Czech Republic; DNRF, DNSRC and Lundbeck Foundation, Denmark; IN2P3-CNRS, CEA-DSM/IRFU, France; GNSF, Georgia; BMBF, HGF, and MPG, Germany; GSRT, Greece; RGC, Hong Kong SAR, China; ISF, I-CORE and Benoziyo Center, Israel; INFN, Italy; MEXT and JSPS, Japan; CNRST, Morocco; FOM and NWO, Netherlands; RCN, Norway; MNiSW and NCN, Poland; FCT, Portugal; MNE/IFA, Romania; MES of Russia and NRC KI, Russian Federation; JINR; MESTD, Serbia; MSSR, Slovakia; ARRS and MIZ\v{S}, Slovenia; DST/NRF, South Africa; MINECO, Spain; SRC and Wallenberg Foundation, Sweden; SERI, SNSF and Cantons of Bern and Geneva, Switzerland; MOST, Taiwan; TAEK, Turkey; STFC, United Kingdom; DOE and NSF, United States of America. In addition, individual groups and members have received support from BCKDF, the Canada Council, CANARIE, CRC, Compute Canada, FQRNT, and the Ontario Innovation Trust, Canada; EPLANET, ERC, FP7, Horizon 2020 and Marie Skłodowska-Curie Actions, European Union; Investissements d'Avenir Labex and Idex, ANR, Region Auvergne and Fondation Partager le Savoir, France; DFG and AvH Foundation, Germany; Herakleitos, Thales and Aristeia programmes co-financed by EU-ESF and the Greek NSRF; BSF, GIF and Minerva, Israel; BRF, Norway; the Royal Society and Leverhulme Trust, United Kingdom.

The crucial computing support from all WLCG partners is acknowledged
gratefully, in particular from CERN and the ATLAS Tier-1 facilities at
TRIUMF (Canada), NDGF (Denmark, Norway, Sweden), CC-IN2P3 (France),
KIT/GridKA (Germany), INFN-CNAF (Italy), NL-T1 (Netherlands), PIC (Spain),
ASGC (Taiwan), RAL (UK) and BNL (USA) and in the Tier-2 facilities
worldwide.

%% file: tex/xs_tables.tex
\begin{table*}[h!]
\centering
\small
\begin{tabular}{c c c c c c c c c c c c}
\hline
\hline
Bin  & Bin edges [GeV] & ${\rm d}\sigma/{\rm d}(\ptOne)$ [fb/GeV] & $\delta^{\rm data}_{\rm stat}$ [\%] & $\delta^{\rm MC}_{\rm stat}$ [\%] & $u_{\rm JES}$  [\%] & $u_{\rm JER}$  [\%] & $u_{\rm unfold}$  [\%] & $u_{\rm lumi}$ [\%]  \\
\hline
   1 & 100--155 & $2.62 \times 10^{4}$ & $0.3  $ & $1.4  $ & $^{+9.7}_{-8.7}$ & $5.2$ & $8.4$ & $2.8$ \\
   2 & 155--235 & $1.47 \times 10^{4}$ & $0.3  $ & $0.6  $ & $^{+8.2}_{-7.8}$ & $2.8$ & $6.4$ & $2.8$ \\
   3 & 235--325 & $4.89 \times 10^{3}$ & $0.4  $ & $0.4  $ & $^{+6.7}_{-6.7}$ & $1.5$ & $4.3$ & $2.8$ \\
   4 & 325--420 & $1.35 \times 10^{3}$ & $<0.1 $ & $0.3  $ & $^{+6.0}_{-6.0}$ & $1.2$ & $2.5$ & $2.8$ \\
   5 & 420--530 & $3.56 \times 10^{2}$ & $0.1  $ & $0.2  $ & $^{+6.4}_{-6.2}$ & $1.1$ & $1.6$ & $2.8$ \\
   6 & 530--650 & $9.2 \times 10^{1}$ & $0.2  $ & $0.3  $ & $^{+7.0}_{-6.9}$ & $1.3$ & $1.4$ & $2.8$ \\
   7 & 650--790 & $2.26 \times 10^{1}$ & $0.4  $ & $0.3  $ & $^{+7.5}_{-7.5}$ & $1.5$ & $1.3$ & $2.8$ \\
   8 & 790--950 & 5.34       & $0.8  $ & $0.2  $ & $^{+8.1}_{-7.8}$ & $1.7$ & $1.4$ & $2.8$ \\
   9 & 950--1130 & 1.19       & $2.1  $ & $0.2  $ & $^{+8.6}_{-8.3}$ & $1.8$ & $1.5$ & $2.8$ \\
  10 & 1130--1350 & $2.27 \times 10^{-1}$ & $3.8  $ & $0.3  $ & $^{+9.7}_{-9.0}$ & $2.0$ & $1.7$ & $2.8$ \\
  11 & 1350--1630 & $3.40 \times 10^{-2}$ & $9.1  $ & $0.3  $ & $^{+11.7}_{-10.7}$ & $2.1$ & $1.8$ & $2.8$ \\
  12 & 1630--4000 & $6.31 \times 10^{-4}$ & $21.9 $ & $0.4  $ & $^{+14.5}_{-13.2}$ & $2.5$ & $1.9$ & $2.8$ \\
\hline
\hline
\end{tabular}
\caption{\label{tab:xs_0JetPt}
Measured differential four-jet cross section for $R=0.4$ jets, in bins of \ptOne, along with the uncertainties in the measurement. The events are selected using the inclusive analysis cuts. All uncertainties are given in \%. $\delta^{\rm data}_{\rm stat}$ ($\delta^{\rm MC}_{\rm stat}$) are the statistical uncertainties due to the number of data (MC simulation) events. The other columns correspond to the experimental systematic uncertainties arising from JES, JER, unfolding and luminosity, respectively.
}
\end{table*}

\begin{table*}[h!]
\centering
\small
\begin{tabular}{c c c c c c c c c c c c}
\hline
\hline
Bin  & Bin edges [GeV] & ${\rm d}\sigma/{\rm d}(\ptTwo)$ [fb/GeV] & $\delta^{\rm data}_{\rm stat}$ [\%] & $\delta^{\rm MC}_{\rm stat}$ [\%] & $u_{\rm JES}$  [\%] & $u_{\rm JER}$  [\%] & $u_{\rm unfold}$  [\%] & $u_{\rm lumi}$ [\%]  \\
\hline
   1 & 64--145 & $2.85 \times 10^{4}$ & $0.2  $ & $0.8  $ & $^{+8.6}_{-8.1}$ & $3.7$ & $6.4$ & $2.8$ \\
   2 & 145--255 & $7.43 \times 10^{3}$ & $0.3  $ & $0.4  $ & $^{+7.5}_{-7.1}$ & $1.7$ & $4.2$ & $2.8$ \\
   3 & 255--385 & $7.61 \times 10^{2}$ & $0.4  $ & $0.3  $ & $^{+6.6}_{-6.8}$ & $0.8$ & $2.5$ & $2.8$ \\
   4 & 385--535 & $9.71 \times 10^{1}$ & $0.2  $ & $0.3  $ & $^{+6.7}_{-6.9}$ & $0.7$ & $2.0$ & $2.8$ \\
   5 & 535--715 & $1.36 \times 10^{1}$ & $0.6  $ & $0.3  $ & $^{+7.5}_{-7.5}$ & $0.8$ & $2.1$ & $2.8$ \\
   6 & 715--930 & 1.88       & $1.2  $ & $0.3  $ & $^{+8.4}_{-8.1}$ & $0.9$ & $2.3$ & $2.8$ \\
   7 & 930--1175 & $2.44 \times 10^{-1}$ & $3.1  $ & $0.4  $ & $^{+9.3}_{-8.5}$ & $1.0$ & $2.4$ & $2.8$ \\
   8 & 1175--3000 & $4.91 \times 10^{-3}$ & $9.2  $ & $0.4  $ & $^{+11.4}_{-10.5}$ & $1.0$ & $2.4$ & $2.8$ \\
\hline
\hline
\end{tabular}
\caption{\label{tab:xs_1JetPt}
Measured differential four-jet cross section for $R=0.4$ jets, in bins of \ptTwo, along with the uncertainties in the measurement. The events are selected using the inclusive analysis cuts. All other details are as for \tabref{tab:xs_0JetPt}.
}
\end{table*}

\begin{table*}[h!]
\centering
\small
\begin{tabular}{c c c c c c c c c c c c}
\hline
\hline
Bin  & Bin edges [GeV] & ${\rm d}\sigma/{\rm d}(\ptThree)$ [fb/GeV] & $\delta^{\rm data}_{\rm stat}$ [\%] & $\delta^{\rm MC}_{\rm stat}$ [\%] & $u_{\rm JES}$  [\%] & $u_{\rm JER}$  [\%] & $u_{\rm unfold}$  [\%] & $u_{\rm lumi}$ [\%]  \\
\hline
   1 & 64--120 & $4.85 \times 10^{4}$ & $0.2  $ & $0.7  $ & $^{+8.6}_{-8.0}$ & $3.4$ & $5.8$ & $2.8$ \\
   2 & 120--205 & $5.87 \times 10^{3}$ & $0.3  $ & $0.6  $ & $^{+7.3}_{-7.5}$ & $1.3$ & $2.9$ & $2.8$ \\
   3 & 205--305 & $3.0 \times 10^{2}$ & $0.6  $ & $0.7  $ & $^{+7.0}_{-6.8}$ & $0.4$ & $1.7$ & $2.8$ \\
   4 & 305--425 & $1.94 \times 10^{1}$ & $0.5  $ & $0.8  $ & $^{+7.4}_{-6.7}$ & $0.3$ & $1.9$ & $2.8$ \\
   5 & 425--570 & 1.39       & $1.9  $ & $1.1  $ & $^{+7.8}_{-7.3}$ & $0.2$ & $2.1$ & $2.8$ \\
   6 & 570--2000 & $1.16 \times 10^{-2}$ & $6.4  $ & $1.3  $ & $^{+10.8}_{-8.7}$ & $0.2$ & $2.7$ & $2.8$ \\
\hline
\hline
\end{tabular}
\caption{\label{tab:xs_2JetPt}
Measured differential four-jet cross section for $R=0.4$ jets, in bins of \ptThree, along with the uncertainties in the measurement. The events are selected using the inclusive analysis cuts. All other details are as for \tabref{tab:xs_0JetPt}.
}
\end{table*}

\begin{table*}[h!]
\centering
\small
\begin{tabular}{c c c c c c c c c c c c}
\hline
\hline
Bin  & Bin edges [GeV] & ${\rm d}\sigma/{\rm d}(\ptFour)$ [fb/GeV] & $\delta^{\rm data}_{\rm stat}$ [\%] & $\delta^{\rm MC}_{\rm stat}$ [\%] & $u_{\rm JES}$  [\%] & $u_{\rm JER}$  [\%] & $u_{\rm unfold}$  [\%] & $u_{\rm lumi}$ [\%]  \\
\hline
   1 & 64--85 & $1.18 \times 10^{5}$ & $0.2  $ & $0.7  $ & $^{+8.7}_{-8.0}$ & $3.6$ & $7.0$ & $2.8$ \\
   2 & 85--135 & $1.45 \times 10^{4}$ & $<0.1 $ & $0.8  $ & $^{+8.7}_{-8.1}$ & $2.9$ & $3.1$ & $2.8$ \\
   3 & 135--190 & $7.65 \times 10^{2}$ & $0.1  $ & $1.1  $ & $^{+7.0}_{-7.2}$ & $1.4$ & $1.5$ & $2.8$ \\
   4 & 190--255 & $6.24 \times 10^{1}$ & $0.4  $ & $2.2  $ & $^{+5.8}_{-5.9}$ & $0.8$ & $1.9$ & $2.8$ \\
   5 & 255--330 & 5.37       & $1.2  $ & $3.4  $ & $^{+5.6}_{-5.8}$ & $0.7$ & $2.1$ & $2.8$ \\
   6 & 330--415 & $4.58 \times 10^{-1}$ & $3.7  $ & $4.3  $ & $^{+6.1}_{-6.8}$ & $0.7$ & $2.1$ & $2.8$ \\
   7 & 415--1500 & $4.65 \times 10^{-3}$ & $11.9 $ & $6.4  $ & $^{+6.5}_{-7.5}$ & $0.7$ & $2.1$ & $2.8$ \\
\hline
\hline
\end{tabular}
\caption{\label{tab:xs_3JetPt}
Measured differential four-jet cross section for $R=0.4$ jets, in bins of \ptFour, along with the uncertainties in the measurement. The events are selected using the inclusive analysis cuts. All other details are as for \tabref{tab:xs_0JetPt}.
}
\end{table*}

\begin{table*}[h!]
\centering
\small
\begin{tabular}{c c c c c c c c c c c c}
\hline
\hline
Bin  & Bin edges [GeV] & ${\rm d}\sigma/{\rm d}(\HT)$ [fb/GeV] & $\delta^{\rm data}_{\rm stat}$ [\%] & $\delta^{\rm MC}_{\rm stat}$ [\%] & $u_{\rm JES}$  [\%] & $u_{\rm JER}$  [\%] & $u_{\rm unfold}$  [\%] & $u_{\rm lumi}$ [\%]  \\
\hline
   1 & 290--485 & $9.85 \times 10^{3}$ & $0.2  $ & $1.0  $ & $^{+8.8}_{-8.3}$ & $4.0$ & $7.8$ & $2.8$ \\
   2 & 485--705 & $4.73 \times 10^{3}$ & $0.3  $ & $0.4  $ & $^{+7.7}_{-7.3}$ & $1.7$ & $4.6$ & $2.8$ \\
   3 & 705--950 & $9.0 \times 10^{2}$ & $0.3  $ & $0.3  $ & $^{+6.7}_{-6.7}$ & $0.7$ & $2.2$ & $2.8$ \\
   4 & 950--1225 & $1.7 \times 10^{2}$ & $0.1  $ & $0.2  $ & $^{+6.6}_{-6.6}$ & $0.6$ & $1.5$ & $2.8$ \\
   5 & 1225--1530 & $3.35 \times 10^{1}$ & $0.2  $ & $0.3  $ & $^{+6.8}_{-7.0}$ & $0.6$ & $1.5$ & $2.8$ \\
   6 & 1530--1875 & 6.62       & $0.6  $ & $0.2  $ & $^{+7.3}_{-7.7}$ & $0.7$ & $1.5$ & $2.8$ \\
   7 & 1875--2265 & 1.29       & $1.1  $ & $0.3  $ & $^{+8.3}_{-8.2}$ & $0.8$ & $1.7$ & $2.8$ \\
   8 & 2265--2705 & $2.23 \times 10^{-1}$ & $2.8  $ & $0.3  $ & $^{+9.3}_{-8.4}$ & $0.8$ & $1.6$ & $2.8$ \\
   9 & 2705--7000 & $4.88 \times 10^{-3}$ & $6.4  $ & $0.3  $ & $^{+10.9}_{-10.6}$ & $1.1$ & $1.6$ & $2.8$ \\
\hline
\hline
\end{tabular}
\caption{\label{tab:xs_HT}
Measured differential four-jet cross section for $R=0.4$ jets, in bins of \HT, along with the uncertainties in the measurement. The events are selected using the inclusive analysis cuts. All other details are as for \tabref{tab:xs_0JetPt}.
}
\end{table*}

\begin{table*}[h!]
\centering
\small
\begin{tabular}{c c c c c c c c c c c c}
\hline
\hline
Bin  & Bin edges [GeV] & ${\rm d}\sigma/{\rm d}(\mjjjj)$ [fb/GeV] & $\delta^{\rm data}_{\rm stat}$ [\%] & $\delta^{\rm MC}_{\rm stat}$ [\%] & $u_{\rm JES}$  [\%] & $u_{\rm JER}$  [\%] & $u_{\rm unfold}$  [\%] & $u_{\rm lumi}$ [\%]  \\
\hline
   1 & 100--545 & $1.36 \times 10^{3}$ & $0.5  $ & $2.7  $ & $^{+8.8}_{-7.9}$ & $5.6$ & $9.5$ & $2.8$ \\
   2 & 545--735 & $4.66 \times 10^{3}$ & $0.4  $ & $1.4  $ & $^{+8.3}_{-7.9}$ & $3.5$ & $6.4$ & $2.8$ \\
   3 & 735--935 & $3.55 \times 10^{3}$ & $0.4  $ & $1.2  $ & $^{+8.7}_{-8.3}$ & $3.0$ & $5.0$ & $2.8$ \\
   4 & 935--1150 & $2.2 \times 10^{3}$ & $0.4  $ & $1.6  $ & $^{+8.7}_{-9.1}$ & $2.4$ & $4.4$ & $2.8$ \\
   5 & 1150--1375 & $1.22 \times 10^{3}$ & $0.6  $ & $1.6  $ & $^{+9.8}_{-9.2}$ & $2.1$ & $3.9$ & $2.8$ \\
   6 & 1375--1620 & $6.09 \times 10^{2}$ & $0.6  $ & $1.7  $ & $^{+10.0}_{-9.1}$ & $2.1$ & $2.9$ & $2.8$ \\
   7 & 1620--1880 & $3.09 \times 10^{2}$ & $0.7  $ & $2.2  $ & $^{+9.0}_{-9.6}$ & $2.1$ & $2.6$ & $2.8$ \\
   8 & 1880--2160 & $1.37 \times 10^{2}$ & $1.0  $ & $2.2  $ & $^{+8.3}_{-9.3}$ & $2.1$ & $2.6$ & $2.8$ \\
   9 & 2160--2460 & $5.53 \times 10^{1}$ & $1.4  $ & $2.7  $ & $^{+8.5}_{-9.0}$ & $2.1$ & $3.1$ & $2.8$ \\
  10 & 2460--2780 & $2.28 \times 10^{1}$ & $2.0  $ & $3.0  $ & $^{+8.8}_{-8.7}$ & $2.1$ & $4.5$ & $2.8$ \\
  11 & 2780--3115 & 7.97       & $3.3  $ & $5.0  $ & $^{+9.1}_{-8.8}$ & $2.1$ & $5.3$ & $2.8$ \\
  12 & 3115--3460 & 2.71       & $3.0  $ & $4.9  $ & $^{+10.2}_{-9.2}$ & $2.1$ & $5.8$ & $2.8$ \\
  13 & 3460--3810 & $9.55 \times 10^{-1}$ & $8.4  $ & $5.5  $ & $^{+10.6}_{-9.4}$ & $2.1$ & $6.0$ & $2.8$ \\
  14 & 3810--7000 & $4.66 \times 10^{-2}$ & $12.4 $ & $6.9  $ & $^{+11.6}_{-9.4}$ & $2.1$ & $6.0$ & $2.8$ \\
\hline
\hline
\end{tabular}
\caption{\label{tab:xs_Mjjjj}
Measured differential four-jet cross section for $R=0.4$ jets, in bins of \mjjjj, along with the uncertainties in the measurement. The events are selected using the inclusive analysis cuts. All other details are as for \tabref{tab:xs_0JetPt}.
}
\end{table*}

\begin{table*}[h!]
\centering
\small
\begin{tabular}{c c c c c c c c c c c c}
\hline
\hline
Bin  & Bin edges & ${\rm d}\sigma/{\rm d}(\mjjminratio)$ [fb] & $\delta^{\rm data}_{\rm stat}$ [\%]  & $\delta^{\rm MC}_{\rm stat}$ [\%] & $u_{\rm JES}$ [\%] & $u_{\rm JER}$ [\%] & $u_{\rm unfold}$ [\%] & $u_{\rm lumi}$ [\%] \\
\hline
   1 & 0--0.04 & $1.31 \times 10^{5}$ & $4.7  $ & $9.8  $ & $^{+9.5}_{-7.8}$ & $3.3$ & $7.0$ & $2.8$ \\
   2 & 0.04--0.08 & $4.74 \times 10^{6}$ & $0.7  $ & $2.5  $ & $^{+9.5}_{-8.1}$ & $3.3$ & $7.0$ & $2.8$ \\
   3 & 0.08--0.13 & $1.58 \times 10^{7}$ & $0.3  $ & $1.4  $ & $^{+8.7}_{-8.5}$ & $3.3$ & $7.0$ & $2.8$ \\
   4 & 0.13--0.17 & $1.85 \times 10^{7}$ & $0.3  $ & $1.3  $ & $^{+9.4}_{-9.6}$ & $3.3$ & $6.0$ & $2.8$ \\
   5 & 0.17--0.23 & $1.29 \times 10^{7}$ & $0.4  $ & $1.4  $ & $^{+8.5}_{-8.1}$ & $3.3$ & $4.0$ & $2.8$ \\
   6 & 0.23--0.3 & $4.69 \times 10^{6}$ & $0.5  $ & $2.5  $ & $^{+7.7}_{-8.9}$ & $3.3$ & $4.0$ & $2.8$ \\
   7 & 0.3--0.4 & $4.21 \times 10^{5}$ & $1.4  $ & $6.2  $ & $^{+11.0}_{-4.8}$ & $2.0$ & $10.2$ & $2.8$ \\
\hline
\hline
\end{tabular}
\caption{\label{tab:xs_Mjjmin_Mjjjj_pt500}
Measured differential four-jet cross section for $R=0.4$ jets, in bins of \mjjminratio, along with the uncertainties in the measurement. The events are selected using the inclusive analysis cuts, as well as $\mjjjj>500$ GeV. All other details are as for \tabref{tab:xs_0JetPt}.
}
\end{table*}

\begin{table*}[h!]
\centering
\small
\begin{tabular}{c c c c c c c c c c c c}
\hline
\hline
Bin  & Bin edges & ${\rm d}\sigma/{\rm d}(\mjjminratio)$ [fb] & $\delta^{\rm data}_{\rm stat}$ [\%]  & $\delta^{\rm MC}_{\rm stat}$ [\%] & $u_{\rm JES}$ [\%] & $u_{\rm JER}$ [\%] & $u_{\rm unfold}$ [\%] & $u_{\rm lumi}$ [\%] \\
\hline
   1 & 0--0.04 & $1.34 \times 10^{5}$ & $4.6  $ & $9.7  $ & $^{+9.7}_{-8.6}$ & $1.9$ & $10.5$ & $2.8$ \\
   2 & 0.04--0.08 & $4.01 \times 10^{6}$ & $0.7  $ & $2.2  $ & $^{+9.8}_{-8.9}$ & $1.9$ & $10.5$ & $2.8$ \\
   3 & 0.08--0.13 & $6.79 \times 10^{6}$ & $0.4  $ & $1.4  $ & $^{+8.9}_{-10.4}$ & $1.9$ & $5.8$ & $2.8$ \\
   4 & 0.13--0.17 & $5.35 \times 10^{6}$ & $0.5  $ & $1.8  $ & $^{+9.4}_{-7.9}$ & $1.9$ & $5.5$ & $2.8$ \\
   5 & 0.17--0.23 & $2.21 \times 10^{6}$ & $0.6  $ & $1.8  $ & $^{+8.2}_{-9.9}$ & $1.9$ & $3.1$ & $2.8$ \\
   6 & 0.23--0.3 & $3.97 \times 10^{5}$ & $1.0  $ & $2.2  $ & $^{+7.1}_{-7.5}$ & $1.9$ & $3.1$ & $2.8$ \\
   7 & 0.3--0.4 & $1.78 \times 10^{4}$ & $2.7  $ & $5.9  $ & $^{+6.7}_{-7.5}$ & $1.9$ & $3.1$ & $2.8$ \\
\hline
\hline
\end{tabular}
\caption{\label{tab:xs_Mjjmin_Mjjjj_pt1000}
Measured differential four-jet cross section for $R=0.4$ jets, in bins of \mjjminratio, along with the uncertainties in the measurement. The events are selected using the inclusive analysis cuts, as well as $\mjjjj>1000$ GeV. All other details are as for \tabref{tab:xs_0JetPt}.
}
\end{table*}

\begin{table*}[h!]
\centering
\small
\begin{tabular}{c c c c c c c c c c c c}
\hline
\hline
Bin  & Bin edges & ${\rm d}\sigma/{\rm d}(\mjjminratio)$ [fb] & $\delta^{\rm data}_{\rm stat}$ [\%]  & $\delta^{\rm MC}_{\rm stat}$ [\%] & $u_{\rm JES}$ [\%] & $u_{\rm JER}$ [\%] & $u_{\rm unfold}$ [\%] & $u_{\rm lumi}$ [\%] \\
\hline
   1 & 0--0.04 & $1.28 \times 10^{5}$ & $4.6  $ & $9.6  $ & $^{+8.2}_{-10.3}$ & $2.1$ & $4.8$ & $2.8$ \\
   2 & 0.04--0.08 & $1.67 \times 10^{6}$ & $1.1  $ & $2.2  $ & $^{+8.3}_{-11.1}$ & $2.1$ & $4.8$ & $2.8$ \\
   3 & 0.08--0.13 & $1.68 \times 10^{6}$ & $0.8  $ & $2.4  $ & $^{+8.3}_{-9.5}$ & $2.1$ & $4.8$ & $2.8$ \\
   4 & 0.13--0.17 & $8.81 \times 10^{5}$ & $0.9  $ & $2.5  $ & $^{+8.7}_{-8.9}$ & $2.1$ & $4.8$ & $2.8$ \\
   5 & 0.17--0.23 & $2.63 \times 10^{5}$ & $0.9  $ & $1.9  $ & $^{+8.7}_{-7.7}$ & $2.1$ & $5.2$ & $2.8$ \\
   6 & 0.23--0.3 & $4.04 \times 10^{4}$ & $1.7  $ & $2.6  $ & $^{+9.3}_{-7.1}$ & $2.1$ & $5.2$ & $2.8$ \\
   7 & 0.3--0.4 & $2.03 \times 10^{3}$ & $2.2  $ & $5.9  $ & $^{+8.9}_{-7.1}$ & $2.1$ & $3.1$ & $2.8$ \\
\hline
\hline
\end{tabular}
\caption{\label{tab:xs_Mjjmin_Mjjjj_pt1500}
Measured differential four-jet cross section for $R=0.4$ jets, in bins of \mjjminratio, along with the uncertainties in the measurement. The events are selected using the inclusive analysis cuts, as well as $\mjjjj>1500$ GeV. All other details are as for \tabref{tab:xs_0JetPt}.
}
\end{table*}

\begin{table*}[h!]
\centering
\small
\begin{tabular}{c c c c c c c c c c c c}
\hline
\hline
Bin  & Bin edges & ${\rm d}\sigma/{\rm d}(\mjjminratio)$ [fb] & $\delta^{\rm data}_{\rm stat}$ [\%]  & $\delta^{\rm MC}_{\rm stat}$ [\%] & $u_{\rm JES}$ [\%] & $u_{\rm JER}$ [\%] & $u_{\rm unfold}$ [\%] & $u_{\rm lumi}$ [\%] \\
\hline
   1 & 0--0.04 & $6.15 \times 10^{4}$ & $5.3  $ & $13.0 $ & $^{+9.9}_{-7.1}$ & $1.5$ & $13.2$ & $2.8$ \\
   2 & 0.04--0.08 & $4.66 \times 10^{5}$ & $1.6  $ & $2.8  $ & $^{+10.6}_{-10.4}$ & $1.5$ & $13.2$ & $2.8$ \\
   3 & 0.08--0.13 & $3.55 \times 10^{5}$ & $1.2  $ & $3.3  $ & $^{+10.6}_{-10.4}$ & $1.5$ & $9.7$ & $2.8$ \\
   4 & 0.13--0.17 & $1.29 \times 10^{5}$ & $0.8  $ & $3.0  $ & $^{+8.9}_{-9.0}$ & $1.5$ & $4.2$ & $2.8$ \\
   5 & 0.17--0.23 & $3.64 \times 10^{4}$ & $1.1  $ & $2.5  $ & $^{+8.6}_{-9.0}$ & $1.5$ & $4.2$ & $2.8$ \\
   6 & 0.23--0.3 & $5.61 \times 10^{3}$ & $1.3  $ & $4.0  $ & $^{+8.6}_{-9.0}$ & $1.5$ & $4.2$ & $2.8$ \\
   7 & 0.3--0.4 & $1.85 \times 10^{2}$ & $5.9  $ & $11.3 $ & $^{+8.6}_{-9.0}$ & $1.5$ & $16.4$ & $2.8$ \\
\hline
\hline
\end{tabular}
\caption{\label{tab:xs_Mjjmin_Mjjjj_pt2000}
Measured differential four-jet cross section for $R=0.4$ jets, in bins of \mjjminratio, along with the uncertainties in the measurement. The events are selected using the inclusive analysis cuts, as well as $\mjjjj>2000$ GeV. All other details are as for \tabref{tab:xs_0JetPt}.
}
\end{table*}

 \clearpage 
\begin{table*}[h!]
\centering
\small
\begin{tabular}{c c c c c c c c c c c c}
\hline
\hline
Bin  & Bin edges & ${\rm d}\sigma/{\rm d}(\minDphiij)$ [fb] & $\delta^{\rm data}_{\rm stat}$ [\%]  & $\delta^{\rm MC}_{\rm stat}$ [\%] & $u_{\rm JES}$ [\%] & $u_{\rm JER}$ [\%] & $u_{\rm unfold}$ [\%] & $u_{\rm lumi}$ [\%] \\
\hline
   1 & 0--0.15 & $2.79 \times 10^{6}$ & $0.5  $ & $1.8  $ & $^{+10.6}_{-6.5}$ & $3.8$ & $2.1$ & $2.8$ \\
   2 & 0.15--0.3 & $2.81 \times 10^{6}$ & $0.5  $ & $1.7  $ & $^{+9.6}_{-7.7}$ & $3.8$ & $7.7$ & $2.8$ \\
   3 & 0.3--0.45 & $2.81 \times 10^{6}$ & $0.5  $ & $1.6  $ & $^{+7.7}_{-8.6}$ & $3.5$ & $6.1$ & $2.8$ \\
   4 & 0.45--0.6 & $2.87 \times 10^{6}$ & $0.5  $ & $2.0  $ & $^{+7.2}_{-8.5}$ & $3.5$ & $6.1$ & $2.8$ \\
   5 & 0.6--0.75 & $3.12 \times 10^{6}$ & $0.4  $ & $1.7  $ & $^{+7.2}_{-8.4}$ & $3.5$ & $7.6$ & $2.8$ \\
   6 & 0.75--0.9 & $2.62 \times 10^{6}$ & $0.5  $ & $2.3  $ & $^{+8.9}_{-9.4}$ & $3.7$ & $7.5$ & $2.8$ \\
   7 & 0.9--1.05 & $1.95 \times 10^{6}$ & $0.6  $ & $2.5  $ & $^{+9.5}_{-8.8}$ & $3.7$ & $6.6$ & $2.8$ \\
   8 & 1.05--1.2 & $1.52 \times 10^{6}$ & $0.8  $ & $3.1  $ & $^{+7.8}_{-8.6}$ & $3.7$ & $6.6$ & $2.8$ \\
   9 & 1.2--1.35 & $9.07 \times 10^{5}$ & $1.0  $ & $4.8  $ & $^{+7.8}_{-8.6}$ & $3.7$ & $11.0$ & $2.8$ \\
  10 & 1.35--1.6 & $1.6 \times 10^{5}$ & $1.9  $ & $9.1  $ & $^{+7.5}_{-14.4}$ & $3.7$ & $11.0$ & $2.8$ \\
\hline
\hline
\end{tabular}
\caption{\label{tab:xs_minDphiij}
Measured differential four-jet cross section for $R=0.4$ jets, in bins of \minDphiij, along with the uncertainties in the measurement. The events are selected using the inclusive analysis cuts. All other details are as for \tabref{tab:xs_0JetPt}.
}
\end{table*}

\begin{table*}[h!]
\centering
\small
\begin{tabular}{c c c c c c c c c c c c}
\hline
\hline
Bin  & Bin edges & ${\rm d}\sigma/{\rm d}(\minDphiij)$ [fb] & $\delta^{\rm data}_{\rm stat}$ [\%]  & $\delta^{\rm MC}_{\rm stat}$ [\%] & $u_{\rm JES}$ [\%] & $u_{\rm JER}$ [\%] & $u_{\rm unfold}$ [\%] & $u_{\rm lumi}$ [\%] \\
\hline
   1 & 0--0.15 & $8.98 \times 10^{4}$ & $0.2  $ & $0.4  $ & $^{+6.4}_{-6.5}$ & $0.9$ & $3.3$ & $2.8$ \\
   2 & 0.15--0.3 & $8.31 \times 10^{4}$ & $0.2  $ & $0.5  $ & $^{+6.4}_{-6.2}$ & $0.9$ & $3.0$ & $2.8$ \\
   3 & 0.3--0.45 & $7.37 \times 10^{4}$ & $0.2  $ & $0.5  $ & $^{+6.6}_{-6.2}$ & $1.5$ & $2.4$ & $2.8$ \\
   4 & 0.45--0.6 & $6.61 \times 10^{4}$ & $0.3  $ & $0.5  $ & $^{+5.9}_{-6.3}$ & $1.5$ & $3.7$ & $2.8$ \\
   5 & 0.6--0.75 & $6.28 \times 10^{4}$ & $0.3  $ & $0.6  $ & $^{+6.1}_{-6.7}$ & $1.3$ & $2.5$ & $2.8$ \\
   6 & 0.75--0.9 & $4.07 \times 10^{4}$ & $0.3  $ & $0.6  $ & $^{+6.4}_{-6.4}$ & $1.3$ & $2.2$ & $2.8$ \\
   7 & 0.9--1.05 & $2.41 \times 10^{4}$ & $0.4  $ & $0.9  $ & $^{+5.9}_{-6.8}$ & $1.8$ & $3.6$ & $2.8$ \\
   8 & 1.05--1.2 & $1.26 \times 10^{4}$ & $0.7  $ & $1.3  $ & $^{+5.9}_{-6.3}$ & $1.8$ & $2.8$ & $2.8$ \\
   9 & 1.2--1.35 & $5.47 \times 10^{3}$ & $0.9  $ & $2.1  $ & $^{+5.9}_{-5.6}$ & $1.8$ & $2.8$ & $2.8$ \\
  10 & 1.35--1.6 & $1.11 \times 10^{3}$ & $1.5  $ & $3.7  $ & $^{+6.4}_{-5.1}$ & $1.8$ & $2.8$ & $2.8$ \\
\hline
\hline
\end{tabular}
\caption{\label{tab:xs_minDphiij_pt400}
Measured differential four-jet cross section for $R=0.4$ jets, in bins of \minDphiij, along with the uncertainties in the measurement. The events are selected using the inclusive analysis cuts, as well as $\ptOne>400$ GeV. All other details are as for \tabref{tab:xs_0JetPt}.
}
\end{table*}

\begin{table*}[h!]
\centering
\small
\begin{tabular}{c c c c c c c c c c c c}
\hline
\hline
Bin  & Bin edges & ${\rm d}\sigma/{\rm d}(\minDphiij)$ [fb] & $\delta^{\rm data}_{\rm stat}$ [\%]  & $\delta^{\rm MC}_{\rm stat}$ [\%] & $u_{\rm JES}$ [\%] & $u_{\rm JER}$ [\%] & $u_{\rm unfold}$ [\%] & $u_{\rm lumi}$ [\%] \\
\hline
   1 & 0--0.15 & $3.57 \times 10^{3}$ & $1.1  $ & $0.4  $ & $^{+7.4}_{-7.8}$ & $1.8$ & $2.2$ & $2.8$ \\
   2 & 0.15--0.3 & $3.37 \times 10^{3}$ & $1.2  $ & $0.4  $ & $^{+7.8}_{-7.6}$ & $1.3$ & $3.7$ & $2.8$ \\
   3 & 0.3--0.45 & $2.94 \times 10^{3}$ & $1.3  $ & $0.5  $ & $^{+7.1}_{-7.4}$ & $1.7$ & $3.3$ & $2.8$ \\
   4 & 0.45--0.6 & $2.45 \times 10^{3}$ & $1.5  $ & $0.5  $ & $^{+7.6}_{-7.5}$ & $1.9$ & $2.8$ & $2.8$ \\
   5 & 0.6--0.75 & $2.26 \times 10^{3}$ & $1.3  $ & $0.6  $ & $^{+7.6}_{-7.3}$ & $2.1$ & $2.5$ & $2.8$ \\
   6 & 0.75--0.9 & $1.37 \times 10^{3}$ & $2.0  $ & $0.8  $ & $^{+8.6}_{-8.1}$ & $1.6$ & $2.7$ & $2.8$ \\
   7 & 0.9--1.05 & $7.54 \times 10^{2}$ & $2.5  $ & $0.9  $ & $^{+6.8}_{-9.0}$ & $1.6$ & $2.7$ & $2.8$ \\
   8 & 1.05--1.2 & $3.5 \times 10^{2}$ & $3.4  $ & $1.6  $ & $^{+9.0}_{-7.5}$ & $1.6$ & $5.3$ & $2.8$ \\
   9 & 1.2--1.35 & $1.4 \times 10^{2}$ & $5.7  $ & $2.3  $ & $^{+10.6}_{-6.4}$ & $1.6$ & $3.2$ & $2.8$ \\
  10 & 1.35--1.6 & $3.72 \times 10^{1}$ & $9.3  $ & $4.7  $ & $^{+8.0}_{-6.4}$ & $1.6$ & $3.2$ & $2.8$ \\
\hline
\hline
\end{tabular}
\caption{\label{tab:xs_minDphiij_pt700}
Measured differential four-jet cross section for $R=0.4$ jets, in bins of \minDphiij, along with the uncertainties in the measurement. The events are selected using the inclusive analysis cuts, as well as $\ptOne>700$ GeV. All other details are as for \tabref{tab:xs_0JetPt}.
}
\end{table*}

\begin{table*}[h!]
\centering
\small
\begin{tabular}{c c c c c c c c c c c c}
\hline
\hline
Bin  & Bin edges & ${\rm d}\sigma/{\rm d}(\minDphiij)$ [fb] & $\delta^{\rm data}_{\rm stat}$ [\%]  & $\delta^{\rm MC}_{\rm stat}$ [\%] & $u_{\rm JES}$ [\%] & $u_{\rm JER}$ [\%] & $u_{\rm unfold}$ [\%] & $u_{\rm lumi}$ [\%] \\
\hline
   1 & 0--0.15 & $2.6 \times 10^{2}$ & $4.4  $ & $0.4  $ & $^{+8.9}_{-8.4}$ & $1.9$ & $2.5$ & $2.8$ \\
   2 & 0.15--0.3 & $2.27 \times 10^{2}$ & $4.4  $ & $0.6  $ & $^{+8.6}_{-8.5}$ & $2.1$ & $3.1$ & $2.8$ \\
   3 & 0.3--0.45 & $2.02 \times 10^{2}$ & $4.8  $ & $0.7  $ & $^{+9.5}_{-7.9}$ & $2.1$ & $3.6$ & $2.8$ \\
   4 & 0.45--0.6 & $1.59 \times 10^{2}$ & $5.8  $ & $0.7  $ & $^{+9.0}_{-8.0}$ & $2.1$ & $3.5$ & $2.8$ \\
   5 & 0.6--0.75 & $1.5 \times 10^{2}$ & $6.2  $ & $0.7  $ & $^{+8.6}_{-8.0}$ & $2.1$ & $3.2$ & $2.8$ \\
   6 & 0.75--0.9 & $8.7 \times 10^{1}$ & $6.7  $ & $0.9  $ & $^{+9.8}_{-8.3}$ & $2.8$ & $2.5$ & $2.8$ \\
   7 & 0.9--1.05 & $5.33 \times 10^{1}$ & $8.8  $ & $1.2  $ & $^{+10.1}_{-8.5}$ & $2.8$ & $5.4$ & $2.8$ \\
   8 & 1.05--1.2 & $2.18 \times 10^{1}$ & $14.4 $ & $2.2  $ & $^{+10.0}_{-10.2}$ & $2.8$ & $4.8$ & $2.8$ \\
   9 & 1.2--1.35 & 5.96       & $27.4 $ & $5.6  $ & $^{+9.5}_{-7.9}$ & $4.6$ & $4.8$ & $2.8$ \\
  10 & 1.35--1.6 & 3.71       & $29.4 $ & $3.5  $ & $^{+9.5}_{-7.9}$ & $4.6$ & $6.7$ & $2.8$ \\
\hline
\hline
\end{tabular}
\caption{\label{tab:xs_minDphiij_pt1000}
Measured differential four-jet cross section for $R=0.4$ jets, in bins of \minDphiij, along with the uncertainties in the measurement. The events are selected using the inclusive analysis cuts, as well as $\ptOne>1000$ GeV. All other details are as for \tabref{tab:xs_0JetPt}.
}
\end{table*}

\begin{table*}[h!]
\centering
\small
\begin{tabular}{c c c c c c c c c c c c}
\hline
\hline
Bin  & Bin edges & ${\rm d}\sigma/{\rm d}(\minDphiijk)$ [fb] & $\delta^{\rm data}_{\rm stat}$ [\%]  & $\delta^{\rm MC}_{\rm stat}$ [\%] & $u_{\rm JES}$ [\%] & $u_{\rm JER}$ [\%] & $u_{\rm unfold}$ [\%] & $u_{\rm lumi}$ [\%] \\
\hline
   1 & 0--0.25 & $1.31 \times 10^{4}$ & $5.6  $ & $8.0  $ & $^{+9.9}_{-8.0}$ & $5.2$ & $7.9$ & $2.8$ \\
   2 & 0.25--0.5 & $4.41 \times 10^{4}$ & $3.1  $ & $3.3  $ & $^{+10.0}_{-8.3}$ & $5.2$ & $7.9$ & $2.8$ \\
   3 & 0.5--0.75 & $1.18 \times 10^{5}$ & $2.0  $ & $4.4  $ & $^{+10.0}_{-8.1}$ & $5.2$ & $7.9$ & $2.8$ \\
   4 & 0.75--1 & $2.25 \times 10^{5}$ & $1.4  $ & $2.3  $ & $^{+10.0}_{-7.2}$ & $5.2$ & $13.9$ & $2.8$ \\
   5 & 1--1.25 & $2.89 \times 10^{5}$ & $1.3  $ & $3.5  $ & $^{+9.9}_{-7.4}$ & $4.1$ & $8.0$ & $2.8$ \\
   6 & 1.25--1.5 & $4.21 \times 10^{5}$ & $1.2  $ & $2.2  $ & $^{+9.3}_{-8.0}$ & $4.1$ & $7.7$ & $2.8$ \\
   7 & 1.5--1.75 & $6.48 \times 10^{5}$ & $0.9  $ & $2.4  $ & $^{+8.9}_{-8.6}$ & $4.0$ & $6.8$ & $2.8$ \\
   8 & 1.75--2 & $9.53 \times 10^{5}$ & $0.7  $ & $2.1  $ & $^{+8.4}_{-7.9}$ & $3.8$ & $5.6$ & $2.8$ \\
   9 & 2--2.25 & $1.62 \times 10^{6}$ & $0.6  $ & $2.0  $ & $^{+8.3}_{-8.4}$ & $3.7$ & $5.6$ & $2.8$ \\
  10 & 2.25--2.5 & $2.24 \times 10^{6}$ & $0.5  $ & $1.7  $ & $^{+8.4}_{-8.5}$ & $3.7$ & $6.6$ & $2.8$ \\
  11 & 2.5--2.75 & $3.0 \times 10^{6}$ & $0.4  $ & $1.6  $ & $^{+8.1}_{-7.9}$ & $3.3$ & $8.7$ & $2.8$ \\
  12 & 2.75--3 & $2.81 \times 10^{6}$ & $0.4  $ & $1.8  $ & $^{+8.6}_{-8.9}$ & $3.3$ & $4.1$ & $2.8$ \\
  13 & 3--3.25 & $6.3 \times 10^{5}$ & $0.9  $ & $4.0  $ & $^{+10.9}_{-7.8}$ & $3.3$ & $4.1$ & $2.8$ \\
\hline
\hline
\end{tabular}
\caption{\label{tab:xs_minDphiijk}
Measured differential four-jet cross section for $R=0.4$ jets, in bins of \minDphiijk, along with the uncertainties in the measurement. The events are selected using the inclusive analysis cuts. All other details are as for \tabref{tab:xs_0JetPt}.
}
\end{table*}

\begin{table*}[h!]
\centering
\small
\begin{tabular}{c c c c c c c c c c c c}
\hline
\hline
Bin  & Bin edges & ${\rm d}\sigma/{\rm d}(\minDphiijk)$ [fb] & $\delta^{\rm data}_{\rm stat}$ [\%]  & $\delta^{\rm MC}_{\rm stat}$ [\%] & $u_{\rm JES}$ [\%] & $u_{\rm JER}$ [\%] & $u_{\rm unfold}$ [\%] & $u_{\rm lumi}$ [\%] \\
\hline
   1 & 0--0.25 & $1.97 \times 10^{3}$ & $1.2  $ & $2.3  $ & $^{+6.2}_{-6.5}$ & $0.8$ & $5.7$ & $2.8$ \\
   2 & 0.25--0.5 & $6.68 \times 10^{3}$ & $0.6  $ & $1.0  $ & $^{+6.2}_{-6.3}$ & $0.8$ & $5.7$ & $2.8$ \\
   3 & 0.5--0.75 & $1.68 \times 10^{4}$ & $0.4  $ & $0.6  $ & $^{+6.7}_{-6.2}$ & $0.8$ & $3.6$ & $2.8$ \\
   4 & 0.75--1 & $2.46 \times 10^{4}$ & $0.3  $ & $0.6  $ & $^{+6.2}_{-6.0}$ & $0.8$ & $3.2$ & $2.8$ \\
   5 & 1--1.25 & $2.87 \times 10^{4}$ & $0.3  $ & $0.6  $ & $^{+6.5}_{-6.2}$ & $0.8$ & $4.5$ & $2.8$ \\
   6 & 1.25--1.5 & $3.07 \times 10^{4}$ & $0.3  $ & $0.5  $ & $^{+6.4}_{-6.2}$ & $0.9$ & $4.0$ & $2.8$ \\
   7 & 1.5--1.75 & $3.13 \times 10^{4}$ & $0.3  $ & $0.6  $ & $^{+6.5}_{-6.2}$ & $1.1$ & $3.2$ & $2.8$ \\
   8 & 1.75--2 & $3.08 \times 10^{4}$ & $0.3  $ & $0.6  $ & $^{+6.3}_{-6.4}$ & $1.8$ & $2.5$ & $2.8$ \\
   9 & 2--2.25 & $2.79 \times 10^{4}$ & $0.3  $ & $0.6  $ & $^{+6.2}_{-6.5}$ & $1.6$ & $2.2$ & $2.8$ \\
  10 & 2.25--2.5 & $2.55 \times 10^{4}$ & $0.3  $ & $0.6  $ & $^{+6.3}_{-6.6}$ & $1.4$ & $2.2$ & $2.8$ \\
  11 & 2.5--2.75 & $2.48 \times 10^{4}$ & $0.4  $ & $0.7  $ & $^{+5.9}_{-7.3}$ & $1.6$ & $3.1$ & $2.8$ \\
  12 & 2.75--3 & $2.1 \times 10^{4}$ & $0.4  $ & $0.7  $ & $^{+6.1}_{-6.2}$ & $1.6$ & $3.3$ & $2.8$ \\
  13 & 3--3.25 & $5.29 \times 10^{3}$ & $0.7  $ & $1.4  $ & $^{+5.5}_{-7.2}$ & $1.1$ & $2.0$ & $2.8$ \\
\hline
\hline
\end{tabular}
\caption{\label{tab:xs_minDphiijk_pt400}
Measured differential four-jet cross section for $R=0.4$ jets, in bins of \minDphiijk, along with the uncertainties in the measurement. The events are selected using the inclusive analysis cuts, as well as $\ptOne>400$ GeV. All other details are as for \tabref{tab:xs_0JetPt}.
}
\end{table*}

\begin{table*}[h!]
\centering
\small
\begin{tabular}{c c c c c c c c c c c c}
\hline
\hline
Bin  & Bin edges & ${\rm d}\sigma/{\rm d}(\minDphiijk)$ [fb] & $\delta^{\rm data}_{\rm stat}$ [\%]  & $\delta^{\rm MC}_{\rm stat}$ [\%] & $u_{\rm JES}$ [\%] & $u_{\rm JER}$ [\%] & $u_{\rm unfold}$ [\%] & $u_{\rm lumi}$ [\%] \\
\hline
   1 & 0--0.25 & $8.01 \times 10^{1}$ & $5.8  $ & $2.1  $ & $^{+8.3}_{-7.1}$ & $2.4$ & $5.5$ & $2.8$ \\
   2 & 0.25--0.5 & $3.1 \times 10^{2}$ & $2.8  $ & $0.9  $ & $^{+7.7}_{-6.7}$ & $2.4$ & $5.5$ & $2.8$ \\
   3 & 0.5--0.75 & $7.61 \times 10^{2}$ & $2.2  $ & $0.7  $ & $^{+7.0}_{-8.7}$ & $1.1$ & $2.5$ & $2.8$ \\
   4 & 0.75--1 & $1.11 \times 10^{3}$ & $1.5  $ & $0.5  $ & $^{+7.0}_{-7.3}$ & $1.1$ & $2.3$ & $2.8$ \\
   5 & 1--1.25 & $1.22 \times 10^{3}$ & $1.5  $ & $0.6  $ & $^{+7.4}_{-7.5}$ & $1.2$ & $2.8$ & $2.8$ \\
   6 & 1.25--1.5 & $1.22 \times 10^{3}$ & $1.6  $ & $0.5  $ & $^{+7.6}_{-7.4}$ & $1.5$ & $3.2$ & $2.8$ \\
   7 & 1.5--1.75 & $1.17 \times 10^{3}$ & $1.6  $ & $0.5  $ & $^{+7.7}_{-7.4}$ & $1.7$ & $3.6$ & $2.8$ \\
   8 & 1.75--2 & $1.07 \times 10^{3}$ & $1.6  $ & $0.6  $ & $^{+8.1}_{-7.6}$ & $1.8$ & $4.4$ & $2.8$ \\
   9 & 2--2.25 & $9.19 \times 10^{2}$ & $1.8  $ & $0.6  $ & $^{+8.3}_{-8.0}$ & $2.0$ & $3.3$ & $2.8$ \\
  10 & 2.25--2.5 & $8.33 \times 10^{2}$ & $1.9  $ & $0.7  $ & $^{+8.5}_{-8.1}$ & $2.6$ & $2.4$ & $2.8$ \\
  11 & 2.5--2.75 & $8.17 \times 10^{2}$ & $1.9  $ & $0.7  $ & $^{+7.6}_{-7.7}$ & $1.9$ & $2.2$ & $2.8$ \\
  12 & 2.75--3 & $6.6 \times 10^{2}$ & $2.4  $ & $0.7  $ & $^{+7.0}_{-8.0}$ & $1.9$ & $1.9$ & $2.8$ \\
  13 & 3--3.25 & $1.92 \times 10^{2}$ & $4.0  $ & $1.4  $ & $^{+7.1}_{-6.8}$ & $3.2$ & $1.9$ & $2.8$ \\
\hline
\hline
\end{tabular}
\caption{\label{tab:xs_minDphiijk_pt700}
Measured differential four-jet cross section for $R=0.4$ jets, in bins of \minDphiijk, along with the uncertainties in the measurement. The events are selected using the inclusive analysis cuts, as well as $\ptOne>700$ GeV. All other details are as for \tabref{tab:xs_0JetPt}.
}
\end{table*}

\begin{table*}[h!]
\centering
\small
\begin{tabular}{c c c c c c c c c c c c}
\hline
\hline
Bin  & Bin edges & ${\rm d}\sigma/{\rm d}(\minDphiijk)$ [fb] & $\delta^{\rm data}_{\rm stat}$ [\%]  & $\delta^{\rm MC}_{\rm stat}$ [\%] & $u_{\rm JES}$ [\%] & $u_{\rm JER}$ [\%] & $u_{\rm unfold}$ [\%] & $u_{\rm lumi}$ [\%] \\
\hline
   1 & 0--0.25 & 5.54       & $21.2 $ & $2.5  $ & $^{+9.0}_{-7.8}$ & $1.6$ & $2.4$ & $2.8$ \\
   2 & 0.25--0.5 & $2.68 \times 10^{1}$ & $11.6 $ & $1.3  $ & $^{+7.6}_{-8.6}$ & $1.6$ & $2.4$ & $2.8$ \\
   3 & 0.5--0.75 & $5.75 \times 10^{1}$ & $7.4  $ & $0.9  $ & $^{+9.0}_{-8.6}$ & $1.6$ & $2.3$ & $2.8$ \\
   4 & 0.75--1 & $8.98 \times 10^{1}$ & $5.7  $ & $0.7  $ & $^{+7.9}_{-8.2}$ & $1.6$ & $2.3$ & $2.8$ \\
   5 & 1--1.25 & $8.83 \times 10^{1}$ & $6.0  $ & $0.8  $ & $^{+8.7}_{-8.2}$ & $1.6$ & $3.8$ & $2.8$ \\
   6 & 1.25--1.5 & $7.41 \times 10^{1}$ & $6.3  $ & $0.6  $ & $^{+8.9}_{-8.1}$ & $1.7$ & $3.4$ & $2.8$ \\
   7 & 1.5--1.75 & $7.68 \times 10^{1}$ & $6.3  $ & $0.7  $ & $^{+8.9}_{-8.4}$ & $1.7$ & $3.5$ & $2.8$ \\
   8 & 1.75--2 & $6.68 \times 10^{1}$ & $7.0  $ & $0.7  $ & $^{+9.4}_{-8.3}$ & $2.1$ & $3.8$ & $2.8$ \\
   9 & 2--2.25 & $5.29 \times 10^{1}$ & $7.3  $ & $1.0  $ & $^{+10.1}_{-7.9}$ & $3.1$ & $4.5$ & $2.8$ \\
  10 & 2.25--2.5 & $4.76 \times 10^{1}$ & $7.9  $ & $1.0  $ & $^{+10.3}_{-8.1}$ & $3.6$ & $3.7$ & $2.8$ \\
  11 & 2.5--2.75 & $5.53 \times 10^{1}$ & $6.7  $ & $1.0  $ & $^{+10.1}_{-8.1}$ & $2.8$ & $3.3$ & $2.8$ \\
  12 & 2.75--3 & $4.83 \times 10^{1}$ & $8.4  $ & $1.0  $ & $^{+9.1}_{-9.2}$ & $2.8$ & $3.3$ & $2.8$ \\
  13 & 3--3.25 & $1.41 \times 10^{1}$ & $15.1 $ & $1.7  $ & $^{+10.7}_{-6.7}$ & $4.3$ & $5.6$ & $2.8$ \\
\hline
\hline
\end{tabular}
\caption{\label{tab:xs_minDphiijk_pt1000}
Measured differential four-jet cross section for $R=0.4$ jets, in bins of \minDphiijk, along with the uncertainties in the measurement. The events are selected using the inclusive analysis cuts, as well as $\ptOne>1000$ GeV. All other details are as for \tabref{tab:xs_0JetPt}.
}
\end{table*}

\begin{table*}[h!]
\centering
\small
\begin{tabular}{c c c c c c c c c c c c}
\hline
\hline
Bin  & Bin edges & ${\rm d}\sigma/{\rm d}(\minDrapij)$ [fb] & $\delta^{\rm data}_{\rm stat}$ [\%]  & $\delta^{\rm MC}_{\rm stat}$ [\%] & $u_{\rm JES}$ [\%] & $u_{\rm JER}$ [\%] & $u_{\rm unfold}$ [\%] & $u_{\rm lumi}$ [\%] \\
\hline
   1 & 0--0.15 & $6.61 \times 10^{6}$ & $0.4  $ & $1.2  $ & $^{+8.4}_{-7.2}$ & $3.5$ & $7.4$ & $2.8$ \\
   2 & 0.15--0.3 & $5.03 \times 10^{6}$ & $0.4  $ & $1.5  $ & $^{+8.2}_{-8.6}$ & $3.5$ & $4.6$ & $2.8$ \\
   3 & 0.3--0.45 & $3.68 \times 10^{6}$ & $0.4  $ & $1.5  $ & $^{+8.9}_{-8.3}$ & $3.5$ & $4.6$ & $2.8$ \\
   4 & 0.45--0.6 & $2.58 \times 10^{6}$ & $0.6  $ & $2.0  $ & $^{+9.3}_{-8.3}$ & $3.7$ & $4.6$ & $2.8$ \\
   5 & 0.6--0.75 & $1.86 \times 10^{6}$ & $0.6  $ & $2.6  $ & $^{+10.8}_{-8.8}$ & $3.7$ & $10.2$ & $2.8$ \\
   6 & 0.75--0.9 & $9.92 \times 10^{5}$ & $0.9  $ & $3.7  $ & $^{+9.6}_{-8.8}$ & $3.7$ & $7.3$ & $2.8$ \\
   7 & 0.9--1.05 & $5.56 \times 10^{5}$ & $1.2  $ & $4.4  $ & $^{+9.6}_{-10.2}$ & $3.7$ & $7.3$ & $2.8$ \\
   8 & 1.05--1.2 & $2.08 \times 10^{5}$ & $1.9  $ & $9.4  $ & $^{+9.6}_{-10.2}$ & $3.7$ & $7.3$ & $2.8$ \\
   9 & 1.2--1.35 & $8.57 \times 10^{4}$ & $3.0  $ & $12.6 $ & $^{+7.9}_{-10.2}$ & $6.9$ & $7.3$ & $2.8$ \\
  10 & 1.35--1.5 & $3.49 \times 10^{4}$ & $6.3  $ & $19.3 $ & $^{+7.9}_{-10.2}$ & $6.9$ & $15.9$ & $2.8$ \\
  11 & 1.5--2 & $2.33 \times 10^{3}$ & $10.9 $ & $22.6 $ & $^{+7.9}_{-10.2}$ & $6.9$ & $15.9$ & $2.8$ \\
\hline
\hline
\end{tabular}
\caption{\label{tab:xs_minDrapij}
Measured differential four-jet cross section for $R=0.4$ jets, in bins of \minDrapij, along with the uncertainties in the measurement. The events are selected using the inclusive analysis cuts. All other details are as for \tabref{tab:xs_0JetPt}.
}
\end{table*}

\begin{table*}[h!]
\centering
\small
\begin{tabular}{c c c c c c c c c c c c}
\hline
\hline
Bin  & Bin edges & ${\rm d}\sigma/{\rm d}(\minDrapij)$ [fb] & $\delta^{\rm data}_{\rm stat}$ [\%]  & $\delta^{\rm MC}_{\rm stat}$ [\%] & $u_{\rm JES}$ [\%] & $u_{\rm JER}$ [\%] & $u_{\rm unfold}$ [\%] & $u_{\rm lumi}$ [\%] \\
\hline
   1 & 0--0.15 & $1.54 \times 10^{5}$ & $0.2  $ & $0.3  $ & $^{+6.4}_{-6.3}$ & $1.4$ & $2.7$ & $2.8$ \\
   2 & 0.15--0.3 & $1.15 \times 10^{5}$ & $0.2  $ & $0.3  $ & $^{+6.2}_{-6.3}$ & $0.9$ & $2.5$ & $2.8$ \\
   3 & 0.3--0.45 & $7.96 \times 10^{4}$ & $0.2  $ & $0.4  $ & $^{+6.3}_{-6.2}$ & $0.9$ & $3.2$ & $2.8$ \\
   4 & 0.45--0.6 & $5.23 \times 10^{4}$ & $0.3  $ & $0.6  $ & $^{+6.3}_{-6.3}$ & $1.7$ & $3.1$ & $2.8$ \\
   5 & 0.6--0.75 & $3.27 \times 10^{4}$ & $0.4  $ & $0.7  $ & $^{+6.0}_{-7.0}$ & $1.0$ & $2.4$ & $2.8$ \\
   6 & 0.75--0.9 & $1.59 \times 10^{4}$ & $0.5  $ & $0.9  $ & $^{+6.2}_{-6.5}$ & $1.0$ & $2.8$ & $2.8$ \\
   7 & 0.9--1.05 & $6.68 \times 10^{3}$ & $0.8  $ & $1.5  $ & $^{+7.3}_{-6.5}$ & $1.0$ & $2.8$ & $2.8$ \\
   8 & 1.05--1.2 & $2.57 \times 10^{3}$ & $1.4  $ & $2.6  $ & $^{+7.6}_{-6.5}$ & $1.0$ & $2.8$ & $2.8$ \\
   9 & 1.2--1.35 & $9.3 \times 10^{2}$ & $2.5  $ & $5.5  $ & $^{+8.3}_{-7.0}$ & $1.0$ & $14.0$ & $2.8$ \\
  10 & 1.35--1.5 & $1.97 \times 10^{2}$ & $4.6  $ & $8.3  $ & $^{+8.3}_{-7.0}$ & $1.0$ & $12.9$ & $2.8$ \\
  11 & 1.5--2 & 8.66       & $9.9  $ & $20.2 $ & $^{+8.3}_{-7.0}$ & $1.0$ & $12.9$ & $2.8$ \\
\hline
\hline
\end{tabular}
\caption{\label{tab:xs_minDrapij_pt400}
Measured differential four-jet cross section for $R=0.4$ jets, in bins of \minDrapij, along with the uncertainties in the measurement. The events are selected using the inclusive analysis cuts, as well as $\ptOne>400$ GeV. All other details are as for \tabref{tab:xs_0JetPt}.
}
\end{table*}

\begin{table*}[h!]
\centering
\small
\begin{tabular}{c c c c c c c c c c c c}
\hline
\hline
Bin  & Bin edges & ${\rm d}\sigma/{\rm d}(\minDrapij)$ [fb] & $\delta^{\rm data}_{\rm stat}$ [\%]  & $\delta^{\rm MC}_{\rm stat}$ [\%] & $u_{\rm JES}$ [\%] & $u_{\rm JER}$ [\%] & $u_{\rm unfold}$ [\%] & $u_{\rm lumi}$ [\%] \\
\hline
   1 & 0--0.15 & $6.37 \times 10^{3}$ & $0.8  $ & $0.3  $ & $^{+7.3}_{-7.4}$ & $1.4$ & $2.7$ & $2.8$ \\
   2 & 0.15--0.3 & $4.64 \times 10^{3}$ & $1.1  $ & $0.3  $ & $^{+7.6}_{-7.4}$ & $1.6$ & $2.9$ & $2.8$ \\
   3 & 0.3--0.45 & $2.89 \times 10^{3}$ & $1.4  $ & $0.6  $ & $^{+7.7}_{-7.8}$ & $1.9$ & $3.3$ & $2.8$ \\
   4 & 0.45--0.6 & $1.77 \times 10^{3}$ & $1.8  $ & $0.6  $ & $^{+7.7}_{-7.7}$ & $1.9$ & $2.8$ & $2.8$ \\
   5 & 0.6--0.75 & $9.97 \times 10^{2}$ & $2.4  $ & $0.8  $ & $^{+7.9}_{-8.3}$ & $2.6$ & $3.3$ & $2.8$ \\
   6 & 0.75--0.9 & $3.91 \times 10^{2}$ & $3.5  $ & $1.4  $ & $^{+10.1}_{-9.3}$ & $2.6$ & $5.8$ & $2.8$ \\
   7 & 0.9--1.05 & $1.38 \times 10^{2}$ & $6.0  $ & $2.3  $ & $^{+7.5}_{-9.7}$ & $0.8$ & $4.6$ & $2.8$ \\
   8 & 1.05--1.2 & $5.34 \times 10^{1}$ & $9.5  $ & $4.7  $ & $^{+7.5}_{-9.7}$ & $0.8$ & $4.6$ & $2.8$ \\
   9 & 1.2--1.35 & $1.6 \times 10^{1}$ & $20.9 $ & $6.6  $ & $^{+7.5}_{-8.8}$ & $0.8$ & $4.6$ & $2.8$ \\
  10 & 1.35--1.5 & 2.82       & $45.2 $ & $17.5 $ & $^{+20.3}_{-8.8}$ & $0.8$ & $4.6$ & $2.8$ \\
\hline
\hline
\end{tabular}
\caption{\label{tab:xs_minDrapij_pt700}
Measured differential four-jet cross section for $R=0.4$ jets, in bins of \minDrapij, along with the uncertainties in the measurement. The events are selected using the inclusive analysis cuts, as well as $\ptOne>700$ GeV. All other details are as for \tabref{tab:xs_0JetPt}.
}
\end{table*}

\begin{table*}[h!]
\centering
\small
\begin{tabular}{c c c c c c c c c c c c}
\hline
\hline
Bin  & Bin edges & ${\rm d}\sigma/{\rm d}(\minDrapij)$ [fb] & $\delta^{\rm data}_{\rm stat}$ [\%]  & $\delta^{\rm MC}_{\rm stat}$ [\%] & $u_{\rm JES}$ [\%] & $u_{\rm JER}$ [\%] & $u_{\rm unfold}$ [\%] & $u_{\rm lumi}$ [\%] \\
\hline
   1 & 0--0.15 & $4.69 \times 10^{2}$ & $3.2  $ & $0.4  $ & $^{+8.6}_{-7.7}$ & $1.7$ & $2.5$ & $2.8$ \\
   2 & 0.15--0.3 & $3.33 \times 10^{2}$ & $4.1  $ & $0.5  $ & $^{+8.7}_{-8.5}$ & $2.5$ & $2.7$ & $2.8$ \\
   3 & 0.3--0.45 & $1.98 \times 10^{2}$ & $5.0  $ & $0.7  $ & $^{+9.0}_{-8.1}$ & $2.1$ & $5.3$ & $2.8$ \\
   4 & 0.45--0.6 & $1.08 \times 10^{2}$ & $6.8  $ & $0.8  $ & $^{+9.3}_{-9.2}$ & $2.1$ & $3.9$ & $2.8$ \\
   5 & 0.6--0.75 & $4.24 \times 10^{1}$ & $10.3 $ & $1.3  $ & $^{+11.7}_{-8.7}$ & $4.1$ & $4.3$ & $2.8$ \\
   6 & 0.75--0.9 & $1.72 \times 10^{1}$ & $16.8 $ & $2.5  $ & $^{+13.0}_{-9.5}$ & $4.1$ & $3.8$ & $2.8$ \\
   7 & 0.9--1.05 & 4.62       & $34.1 $ & $4.0  $ & $^{+12.3}_{-9.6}$ & $1.0$ & $3.8$ & $2.8$ \\
   8 & 1.05--1.2 & 2.17       & $50.6 $ & $7.0  $ & $^{+12.3}_{-11.3}$ & $1.0$ & $3.8$ & $2.8$ \\
\hline
\hline
\end{tabular}
\caption{\label{tab:xs_minDrapij_pt1000}
Measured differential four-jet cross section for $R=0.4$ jets, in bins of \minDrapij, along with the uncertainties in the measurement. The events are selected using the inclusive analysis cuts, as well as $\ptOne>1000$ GeV. All other details are as for \tabref{tab:xs_0JetPt}.
}
\end{table*}

 \clearpage 
\begin{table*}[h!]
\centering
\small
\begin{tabular}{c c c c c c c c c c c c}
\hline
\hline
Bin  & Bin edges & ${\rm d}\sigma/{\rm d}(\minDrapijk)$ [fb] & $\delta^{\rm data}_{\rm stat}$ [\%]  & $\delta^{\rm MC}_{\rm stat}$ [\%] & $u_{\rm JES}$ [\%] & $u_{\rm JER}$ [\%] & $u_{\rm unfold}$ [\%] & $u_{\rm lumi}$ [\%] \\
\hline
   1 & 0--0.25 & $3.61 \times 10^{5}$ & $1.3  $ & $4.6  $ & $^{+7.5}_{-7.1}$ & $3.4$ & $23.0$ & $2.8$ \\
   2 & 0.25--0.5 & $9.65 \times 10^{5}$ & $0.7  $ & $2.6  $ & $^{+7.6}_{-7.3}$ & $3.4$ & $6.9$ & $2.8$ \\
   3 & 0.5--0.75 & $1.64 \times 10^{6}$ & $0.5  $ & $1.9  $ & $^{+7.6}_{-7.4}$ & $3.4$ & $6.9$ & $2.8$ \\
   4 & 0.75--1 & $1.89 \times 10^{6}$ & $0.5  $ & $1.9  $ & $^{+8.8}_{-7.9}$ & $3.4$ & $6.9$ & $2.8$ \\
   5 & 1--1.25 & $1.83 \times 10^{6}$ & $0.5  $ & $1.6  $ & $^{+9.0}_{-8.7}$ & $2.8$ & $6.6$ & $2.8$ \\
   6 & 1.25--1.5 & $1.62 \times 10^{6}$ & $0.5  $ & $2.3  $ & $^{+9.3}_{-8.4}$ & $2.8$ & $6.6$ & $2.8$ \\
   7 & 1.5--1.75 & $1.42 \times 10^{6}$ & $0.5  $ & $1.9  $ & $^{+8.8}_{-8.1}$ & $2.8$ & $6.6$ & $2.8$ \\
\hline
\hline
\end{tabular}
\caption{\label{tab:xs_minDrapijk}
Measured differential four-jet cross section for $R=0.4$ jets, in bins of \minDrapijk, along with the uncertainties in the measurement. The events are selected using the inclusive analysis cuts. All other details are as for \tabref{tab:xs_0JetPt}.
}
\end{table*}

\begin{table*}[h!]
\centering
\small
\begin{tabular}{c c c c c c c c c c c c}
\hline
\hline
Bin  & Bin edges & ${\rm d}\sigma/{\rm d}(\minDrapijk)$ [fb] & $\delta^{\rm data}_{\rm stat}$ [\%]  & $\delta^{\rm MC}_{\rm stat}$ [\%] & $u_{\rm JES}$ [\%] & $u_{\rm JER}$ [\%] & $u_{\rm unfold}$ [\%] & $u_{\rm lumi}$ [\%] \\
\hline
   1 & 0--0.25 & $8.7 \times 10^{3}$ & $0.6  $ & $0.8  $ & $^{+6.1}_{-6.2}$ & $1.7$ & $2.8$ & $2.8$ \\
   2 & 0.25--0.5 & $2.49 \times 10^{4}$ & $0.3  $ & $0.7  $ & $^{+6.3}_{-6.3}$ & $1.7$ & $3.0$ & $2.8$ \\
   3 & 0.5--0.75 & $4.28 \times 10^{4}$ & $0.3  $ & $0.5  $ & $^{+5.9}_{-6.5}$ & $1.0$ & $2.0$ & $2.8$ \\
   4 & 0.75--1 & $4.83 \times 10^{4}$ & $0.2  $ & $0.4  $ & $^{+6.3}_{-5.7}$ & $1.0$ & $2.4$ & $2.8$ \\
   5 & 1--1.25 & $4.34 \times 10^{4}$ & $0.3  $ & $0.4  $ & $^{+6.2}_{-6.3}$ & $1.2$ & $3.1$ & $2.8$ \\
   6 & 1.25--1.5 & $3.58 \times 10^{4}$ & $0.3  $ & $0.5  $ & $^{+6.1}_{-6.3}$ & $1.2$ & $3.3$ & $2.8$ \\
   7 & 1.5--1.75 & $2.69 \times 10^{4}$ & $0.4  $ & $0.6  $ & $^{+6.0}_{-6.4}$ & $1.4$ & $3.1$ & $2.8$ \\
\hline
\hline
\end{tabular}
\caption{\label{tab:xs_minDrapijk_pt400}
Measured differential four-jet cross section for $R=0.4$ jets, in bins of \minDrapijk, along with the uncertainties in the measurement. The events are selected using the inclusive analysis cuts, as well as $\ptOne>400$ GeV. All other details are as for \tabref{tab:xs_0JetPt}.
}
\end{table*}

\begin{table*}[h!]
\centering
\small
\begin{tabular}{c c c c c c c c c c c c}
\hline
\hline
Bin  & Bin edges & ${\rm d}\sigma/{\rm d}(\minDrapijk)$ [fb] & $\delta^{\rm data}_{\rm stat}$ [\%]  & $\delta^{\rm MC}_{\rm stat}$ [\%] & $u_{\rm JES}$ [\%] & $u_{\rm JER}$ [\%] & $u_{\rm unfold}$ [\%] & $u_{\rm lumi}$ [\%] \\
\hline
   1 & 0--0.25 & $4.06 \times 10^{2}$ & $2.3  $ & $1.1  $ & $^{+6.7}_{-7.6}$ & $1.5$ & $3.9$ & $2.8$ \\
   2 & 0.25--0.5 & $1.14 \times 10^{3}$ & $1.6  $ & $0.5  $ & $^{+6.6}_{-7.1}$ & $1.5$ & $3.2$ & $2.8$ \\
   3 & 0.5--0.75 & $1.91 \times 10^{3}$ & $1.4  $ & $0.5  $ & $^{+7.3}_{-6.9}$ & $1.5$ & $3.7$ & $2.8$ \\
   4 & 0.75--1 & $2.06 \times 10^{3}$ & $1.0  $ & $0.4  $ & $^{+7.5}_{-7.3}$ & $1.3$ & $2.3$ & $2.8$ \\
   5 & 1--1.25 & $1.71 \times 10^{3}$ & $1.3  $ & $0.5  $ & $^{+7.4}_{-8.0}$ & $1.5$ & $2.9$ & $2.8$ \\
   6 & 1.25--1.5 & $1.28 \times 10^{3}$ & $1.4  $ & $0.6  $ & $^{+7.7}_{-7.8}$ & $2.0$ & $2.9$ & $2.8$ \\
   7 & 1.5--1.75 & $8.57 \times 10^{2}$ & $1.7  $ & $0.7  $ & $^{+8.4}_{-8.0}$ & $2.2$ & $2.6$ & $2.8$ \\
   8 & 1.75--2 & $5.06 \times 10^{2}$ & $2.7  $ & $0.8  $ & $^{+9.0}_{-8.1}$ & $2.4$ & $2.1$ & $2.8$ \\
   9 & 2--2.25 & $2.69 \times 10^{2}$ & $3.4  $ & $1.3  $ & $^{+8.9}_{-9.3}$ & $2.5$ & $1.9$ & $2.8$ \\
  10 & 2.25--2.5 & $1.39 \times 10^{2}$ & $5.0  $ & $1.9  $ & $^{+9.2}_{-10.6}$ & $2.6$ & $2.6$ & $2.8$ \\
  11 & 2.5--2.75 & $5.72 \times 10^{1}$ & $8.0  $ & $2.7  $ & $^{+9.6}_{-11.0}$ & $2.6$ & $4.7$ & $2.8$ \\
  12 & 2.75--3 & $2.39 \times 10^{1}$ & $12.0 $ & $4.6  $ & $^{+10.9}_{-12.4}$ & $2.6$ & $5.5$ & $2.8$ \\
  13 & 3--3.25 & 8.54       & $16.3 $ & $8.3  $ & $^{+13.7}_{-14.0}$ & $2.6$ & $5.8$ & $2.8$ \\
  14 & 3.25--3.5 & 4.60       & $24.3 $ & $9.2  $ & $^{+14.7}_{-14.2}$ & $2.6$ & $10.9$ & $2.8$ \\
  15 & 3.5--3.75 & $4.97 \times 10^{-1}$ & $80.2 $ & $88.5 $ & $^{+14.7}_{-14.2}$ & $2.6$ & $24.9$ & $2.8$ \\
  16 & 3.75--4 & $3.03 \times 10^{1}$ & $82.6 $ & $81.5 $ & $^{+14.7}_{-14.1}$ & $2.6$ & $30.0$ & $2.8$ \\
\hline
\hline
\end{tabular}
\caption{\label{tab:xs_minDrapijk_pt700}
Measured differential four-jet cross section for $R=0.4$ jets, in bins of \minDrapijk, along with the uncertainties in the measurement. The events are selected using the inclusive analysis cuts, as well as $\ptOne>700$ GeV. All other details are as for \tabref{tab:xs_0JetPt}.
}
\end{table*}

\begin{table*}[h!]
\centering
\small
\begin{tabular}{c c c c c c c c c c c c}
\hline
\hline
Bin  & Bin edges & ${\rm d}\sigma/{\rm d}(\minDrapijk)$ [fb] & $\delta^{\rm data}_{\rm stat}$ [\%]  & $\delta^{\rm MC}_{\rm stat}$ [\%] & $u_{\rm JES}$ [\%] & $u_{\rm JER}$ [\%] & $u_{\rm unfold}$ [\%] & $u_{\rm lumi}$ [\%] \\
\hline
   1 & 0--0.25 & $3.74 \times 10^{1}$ & $8.2  $ & $1.4  $ & $^{+9.7}_{-6.5}$ & $1.5$ & $4.0$ & $2.8$ \\
   2 & 0.25--0.5 & $8.69 \times 10^{1}$ & $5.1  $ & $0.7  $ & $^{+8.3}_{-7.4}$ & $1.5$ & $2.2$ & $2.8$ \\
   3 & 0.5--0.75 & $1.48 \times 10^{2}$ & $4.8  $ & $0.5  $ & $^{+8.1}_{-8.0}$ & $1.5$ & $2.2$ & $2.8$ \\
   4 & 0.75--1 & $1.56 \times 10^{2}$ & $4.4  $ & $0.5  $ & $^{+8.1}_{-8.0}$ & $1.9$ & $2.9$ & $2.8$ \\
   5 & 1--1.25 & $1.11 \times 10^{2}$ & $6.2  $ & $0.6  $ & $^{+9.7}_{-8.8}$ & $2.2$ & $3.8$ & $2.8$ \\
   6 & 1.25--1.5 & $8.07 \times 10^{1}$ & $6.5  $ & $0.7  $ & $^{+10.0}_{-8.8}$ & $2.7$ & $4.0$ & $2.8$ \\
   7 & 1.5--1.75 & $4.44 \times 10^{1}$ & $8.2  $ & $1.0  $ & $^{+10.8}_{-8.9}$ & $3.4$ & $4.2$ & $2.8$ \\
   8 & 1.75--2 & $2.48 \times 10^{1}$ & $10.3 $ & $1.3  $ & $^{+10.8}_{-9.5}$ & $3.8$ & $3.6$ & $2.8$ \\
   9 & 2--2.25 & 6.87       & $22.3 $ & $2.3  $ & $^{+11.6}_{-10.0}$ & $3.8$ & $3.3$ & $2.8$ \\
  10 & 2.25--2.5 & 5.32       & $23.6 $ & $3.0  $ & $^{+12.8}_{-11.1}$ & $3.6$ & $3.3$ & $2.8$ \\
  11 & 2.5--2.75 & 1.31       & $50.7 $ & $6.8  $ & $^{+14.3}_{-13.8}$ & $3.5$ & $3.3$ & $2.8$ \\
  12 & 2.75--3 & 1.32       & $39.5 $ & $7.9  $ & $^{+16.5}_{-20.0}$ & $3.4$ & $3.3$ & $2.8$ \\
\hline
\hline
\end{tabular}
\caption{\label{tab:xs_minDrapijk_pt1000}
Measured differential four-jet cross section for $R=0.4$ jets, in bins of \minDrapijk, along with the uncertainties in the measurement. The events are selected using the inclusive analysis cuts, as well as $\ptOne>1000$ GeV. All other details are as for \tabref{tab:xs_0JetPt}.
}
\end{table*}

\begin{table*}[h!]
\centering
\small
\begin{tabular}{c c c c c c c c c c c c}
\hline
\hline
Bin  & Bin edges & ${\rm d}\sigma/{\rm d}(\maxDrapij)$ [fb] & $\delta^{\rm data}_{\rm stat}$ [\%]  & $\delta^{\rm MC}_{\rm stat}$ [\%] & $u_{\rm JES}$ [\%] & $u_{\rm JER}$ [\%] & $u_{\rm unfold}$ [\%] & $u_{\rm lumi}$ [\%] \\
\hline
   1 & 0--0.4 & $1.55 \times 10^{4}$ & $4.4  $ & $10.2 $ & $^{+8.3}_{-8.4}$ & $3.3$ & $11.8$ & $2.8$ \\
   2 & 0.4--0.8 & $1.69 \times 10^{5}$ & $1.6  $ & $5.5  $ & $^{+8.4}_{-9.1}$ & $3.3$ & $11.8$ & $2.8$ \\
   3 & 0.8--1.2 & $4.44 \times 10^{5}$ & $0.9  $ & $3.3  $ & $^{+8.4}_{-8.5}$ & $3.3$ & $10.8$ & $2.8$ \\
   4 & 1.2--1.6 & $7.29 \times 10^{5}$ & $0.6  $ & $2.6  $ & $^{+8.1}_{-7.5}$ & $3.3$ & $8.4$ & $2.8$ \\
   5 & 1.6--2 & $9.77 \times 10^{5}$ & $0.5  $ & $2.2  $ & $^{+7.1}_{-7.1}$ & $3.3$ & $5.7$ & $2.8$ \\
   6 & 2--2.4 & $1.15 \times 10^{6}$ & $0.6  $ & $1.7  $ & $^{+7.1}_{-7.3}$ & $3.4$ & $4.9$ & $2.8$ \\
   7 & 2.4--2.8 & $1.18 \times 10^{6}$ & $0.5  $ & $1.7  $ & $^{+8.5}_{-7.4}$ & $3.7$ & $4.7$ & $2.8$ \\
   8 & 2.8--3.2 & $1.09 \times 10^{6}$ & $0.6  $ & $2.0  $ & $^{+9.9}_{-8.0}$ & $3.8$ & $5.1$ & $2.8$ \\
   9 & 3.2--3.6 & $9.05 \times 10^{5}$ & $0.7  $ & $2.0  $ & $^{+9.5}_{-9.4}$ & $3.7$ & $5.9$ & $2.8$ \\
  10 & 3.6--4 & $6.39 \times 10^{5}$ & $0.6  $ & $2.5  $ & $^{+9.1}_{-9.8}$ & $3.6$ & $5.4$ & $2.8$ \\
  11 & 4--4.4 & $4.55 \times 10^{5}$ & $0.7  $ & $3.3  $ & $^{+10.4}_{-9.5}$ & $3.5$ & $5.1$ & $2.8$ \\
  12 & 4.4--4.8 & $2.46 \times 10^{5}$ & $1.0  $ & $4.2  $ & $^{+11.7}_{-10.3}$ & $3.5$ & $5.1$ & $2.8$ \\
  13 & 4.8--5.2 & $9.91 \times 10^{4}$ & $1.8  $ & $5.2  $ & $^{+13.3}_{-13.3}$ & $3.5$ & $8.9$ & $2.8$ \\
  14 & 5.2--5.6 & $2.78 \times 10^{4}$ & $3.7  $ & $10.3 $ & $^{+13.3}_{-13.3}$ & $3.5$ & $8.9$ & $2.8$ \\
\hline
\hline
\end{tabular}
\caption{\label{tab:xs_maxDrapij}
Measured differential four-jet cross section for $R=0.4$ jets, in bins of \maxDrapij, along with the uncertainties in the measurement. The events are selected using the inclusive analysis cuts. All other details are as for \tabref{tab:xs_0JetPt}.
}
\end{table*}

\begin{table*}[h!]
\centering
\small
\begin{tabular}{c c c c c c c c c c c c}
\hline
\hline
Bin  & Bin edges & ${\rm d}\sigma/{\rm d}(\maxDrapij)$ [fb] & $\delta^{\rm data}_{\rm stat}$ [\%]  & $\delta^{\rm MC}_{\rm stat}$ [\%] & $u_{\rm JES}$ [\%] & $u_{\rm JER}$ [\%] & $u_{\rm unfold}$ [\%] & $u_{\rm lumi}$ [\%] \\
\hline
   1 & 0--0.4 & $3.08 \times 10^{3}$ & $6.4  $ & $7.6  $ & $^{+6.2}_{-3.8}$ & $1.5$ & $4.0$ & $2.8$ \\
   2 & 0.4--0.8 & $3.0 \times 10^{4}$ & $2.2  $ & $2.4  $ & $^{+6.2}_{-5.5}$ & $1.5$ & $4.0$ & $2.8$ \\
   3 & 0.8--1.2 & $8.29 \times 10^{4}$ & $1.4  $ & $1.3  $ & $^{+6.5}_{-6.0}$ & $1.5$ & $4.0$ & $2.8$ \\
   4 & 1.2--1.6 & $1.33 \times 10^{5}$ & $0.9  $ & $0.8  $ & $^{+6.4}_{-5.9}$ & $1.5$ & $4.0$ & $2.8$ \\
   5 & 1.6--2 & $1.73 \times 10^{5}$ & $0.8  $ & $0.7  $ & $^{+6.0}_{-6.0}$ & $1.5$ & $4.0$ & $2.8$ \\
   6 & 2--2.4 & $1.99 \times 10^{5}$ & $0.9  $ & $0.8  $ & $^{+5.9}_{-6.3}$ & $1.5$ & $4.0$ & $2.8$ \\
   7 & 2.4--2.8 & $1.92 \times 10^{5}$ & $0.8  $ & $0.7  $ & $^{+6.0}_{-6.4}$ & $1.5$ & $4.1$ & $2.8$ \\
   8 & 2.8--3.2 & $1.68 \times 10^{5}$ & $0.9  $ & $0.8  $ & $^{+6.8}_{-6.7}$ & $1.5$ & $4.2$ & $2.8$ \\
   9 & 3.2--3.6 & $1.29 \times 10^{5}$ & $1.0  $ & $0.9  $ & $^{+7.7}_{-6.7}$ & $1.6$ & $4.1$ & $2.8$ \\
  10 & 3.6--4 & $9.03 \times 10^{4}$ & $1.4  $ & $1.1  $ & $^{+8.3}_{-7.0}$ & $1.9$ & $4.1$ & $2.8$ \\
  11 & 4--4.4 & $4.93 \times 10^{4}$ & $1.8  $ & $1.4  $ & $^{+8.1}_{-7.7}$ & $2.0$ & $5.2$ & $2.8$ \\
  12 & 4.4--4.8 & $2.43 \times 10^{4}$ & $3.0  $ & $2.3  $ & $^{+7.9}_{-8.3}$ & $2.0$ & $6.1$ & $2.8$ \\
  13 & 4.8--5.2 & $9.11 \times 10^{3}$ & $5.1  $ & $3.0  $ & $^{+9.2}_{-7.9}$ & $2.0$ & $6.4$ & $2.8$ \\
  14 & 5.2--5.6 & $1.22 \times 10^{3}$ & $13.8 $ & $11.9 $ & $^{+8.9}_{-8.2}$ & $2.0$ & $5.0$ & $2.8$ \\
\hline
\hline
\end{tabular}
\caption{\label{tab:xs_maxDrapij_pt250}
Measured differential four-jet cross section for $R=0.4$ jets, in bins of \maxDrapij, along with the uncertainties in the measurement. The events are selected using the inclusive analysis cuts, as well as $\ptOne>250$ GeV. All other details are as for \tabref{tab:xs_0JetPt}.
}
\end{table*}

\begin{table*}[h!]
\centering
\small
\begin{tabular}{c c c c c c c c c c c c}
\hline
\hline
Bin  & Bin edges & ${\rm d}\sigma/{\rm d}(\maxDrapij)$ [fb] & $\delta^{\rm data}_{\rm stat}$ [\%]  & $\delta^{\rm MC}_{\rm stat}$ [\%] & $u_{\rm JES}$ [\%] & $u_{\rm JER}$ [\%] & $u_{\rm unfold}$ [\%] & $u_{\rm lumi}$ [\%] \\
\hline
   1 & 0--0.4 & $4.07 \times 10^{2}$ & $2.2  $ & $4.2  $ & $^{+6.5}_{-6.5}$ & $0.8$ & $3.6$ & $2.8$ \\
   2 & 0.4--0.8 & $4.4 \times 10^{3}$ & $0.7  $ & $1.0  $ & $^{+6.5}_{-6.5}$ & $0.8$ & $3.6$ & $2.8$ \\
   3 & 0.8--1.2 & $1.28 \times 10^{4}$ & $0.4  $ & $0.6  $ & $^{+6.0}_{-5.5}$ & $0.8$ & $3.2$ & $2.8$ \\
   4 & 1.2--1.6 & $2.12 \times 10^{4}$ & $0.3  $ & $0.5  $ & $^{+5.9}_{-5.7}$ & $0.8$ & $2.6$ & $2.8$ \\
   5 & 1.6--2 & $2.73 \times 10^{4}$ & $0.3  $ & $0.4  $ & $^{+5.7}_{-6.1}$ & $0.8$ & $2.6$ & $2.8$ \\
   6 & 2--2.4 & $2.88 \times 10^{4}$ & $0.2  $ & $0.4  $ & $^{+5.9}_{-6.3}$ & $1.1$ & $3.2$ & $2.8$ \\
   7 & 2.4--2.8 & $2.66 \times 10^{4}$ & $0.3  $ & $0.4  $ & $^{+6.3}_{-6.3}$ & $1.8$ & $3.1$ & $2.8$ \\
   8 & 2.8--3.2 & $2.13 \times 10^{4}$ & $0.3  $ & $0.5  $ & $^{+6.4}_{-6.3}$ & $1.7$ & $2.9$ & $2.8$ \\
   9 & 3.2--3.6 & $1.46 \times 10^{4}$ & $0.4  $ & $0.6  $ & $^{+7.0}_{-6.9}$ & $1.2$ & $2.2$ & $2.8$ \\
  10 & 3.6--4 & $8.75 \times 10^{3}$ & $0.4  $ & $0.8  $ & $^{+7.6}_{-7.7}$ & $1.0$ & $2.1$ & $2.8$ \\
  11 & 4--4.4 & $4.32 \times 10^{3}$ & $0.7  $ & $1.2  $ & $^{+8.3}_{-8.5}$ & $1.0$ & $2.6$ & $2.8$ \\
  12 & 4.4--4.8 & $1.69 \times 10^{3}$ & $1.0  $ & $1.8  $ & $^{+9.1}_{-9.1}$ & $1.0$ & $4.1$ & $2.8$ \\
  13 & 4.8--5.2 & $4.64 \times 10^{2}$ & $2.0  $ & $3.3  $ & $^{+9.4}_{-8.3}$ & $1.0$ & $9.2$ & $2.8$ \\
  14 & 5.2--5.6 & $5.92 \times 10^{1}$ & $5.4  $ & $12.6 $ & $^{+9.4}_{-8.3}$ & $1.0$ & $16.9$ & $2.8$ \\
\hline
\hline
\end{tabular}
\caption{\label{tab:xs_maxDrapij_pt400}
Measured differential four-jet cross section for $R=0.4$ jets, in bins of \maxDrapij, along with the uncertainties in the measurement. The events are selected using the inclusive analysis cuts, as well as $\ptOne>400$ GeV. All other details are as for \tabref{tab:xs_0JetPt}.
}
\end{table*}

\begin{table*}[h!]
\centering
\small
\begin{tabular}{c c c c c c c c c c c c}
\hline
\hline
Bin  & Bin edges & ${\rm d}\sigma/{\rm d}(\maxDrapij)$ [fb] & $\delta^{\rm data}_{\rm stat}$ [\%]  & $\delta^{\rm MC}_{\rm stat}$ [\%] & $u_{\rm JES}$ [\%] & $u_{\rm JER}$ [\%] & $u_{\rm unfold}$ [\%] & $u_{\rm lumi}$ [\%] \\
\hline
   1 & 0--0.4 & $8.16 \times 10^{1}$ & $5.1  $ & $4.5  $ & $^{+5.5}_{-6.1}$ & $1.9$ & $3.0$ & $2.8$ \\
   2 & 0.4--0.8 & $8.85 \times 10^{2}$ & $1.6  $ & $1.3  $ & $^{+5.6}_{-6.8}$ & $1.9$ & $3.0$ & $2.8$ \\
   3 & 0.8--1.2 & $2.65 \times 10^{3}$ & $0.8  $ & $0.8  $ & $^{+6.1}_{-6.3}$ & $1.9$ & $2.9$ & $2.8$ \\
   4 & 1.2--1.6 & $4.34 \times 10^{3}$ & $0.7  $ & $0.6  $ & $^{+6.6}_{-6.6}$ & $1.7$ & $2.8$ & $2.8$ \\
   5 & 1.6--2 & $5.3 \times 10^{3}$ & $0.6  $ & $0.5  $ & $^{+6.6}_{-6.7}$ & $1.7$ & $2.4$ & $2.8$ \\
   6 & 2--2.4 & $5.3 \times 10^{3}$ & $0.6  $ & $0.5  $ & $^{+6.9}_{-6.7}$ & $1.6$ & $2.6$ & $2.8$ \\
   7 & 2.4--2.8 & $4.5 \times 10^{3}$ & $0.7  $ & $0.5  $ & $^{+7.2}_{-7.0}$ & $1.8$ & $2.8$ & $2.8$ \\
   8 & 2.8--3.2 & $3.4 \times 10^{3}$ & $0.7  $ & $0.7  $ & $^{+7.6}_{-7.8}$ & $2.0$ & $3.0$ & $2.8$ \\
   9 & 3.2--3.6 & $2.06 \times 10^{3}$ & $0.9  $ & $0.8  $ & $^{+8.2}_{-8.4}$ & $2.0$ & $3.8$ & $2.8$ \\
  10 & 3.6--4 & $1.1 \times 10^{3}$ & $1.2  $ & $1.2  $ & $^{+9.6}_{-9.1}$ & $1.9$ & $4.8$ & $2.8$ \\
  11 & 4--4.4 & $4.47 \times 10^{2}$ & $2.0  $ & $1.6  $ & $^{+10.1}_{-10.1}$ & $2.1$ & $4.2$ & $2.8$ \\
  12 & 4.4--4.8 & $1.54 \times 10^{2}$ & $3.5  $ & $3.1  $ & $^{+9.4}_{-11.4}$ & $2.4$ & $3.9$ & $2.8$ \\
  13 & 4.8--5.2 & $3.92 \times 10^{1}$ & $6.9  $ & $8.4  $ & $^{+9.1}_{-12.0}$ & $2.6$ & $3.9$ & $2.8$ \\
  14 & 5.2--5.6 & 5.91       & $17.1 $ & $13.9 $ & $^{+9.1}_{-17.7}$ & $2.6$ & $3.9$ & $2.8$ \\
\hline
\hline
\end{tabular}
\caption{\label{tab:xs_maxDrapij_pt550}
Measured differential four-jet cross section for $R=0.4$ jets, in bins of \maxDrapij, along with the uncertainties in the measurement. The events are selected using the inclusive analysis cuts, as well as $\ptOne>550$ GeV. All other details are as for \tabref{tab:xs_0JetPt}.
}
\end{table*}

\begin{table*}[h!]
\centering
\small
\begin{tabular}{c c c c c c c c c c c c}
\hline
\hline
Bin  & Bin edges [GeV] & ${\rm d}\sigma/{\rm d}(\sumptcent)$ [fb/GeV] & $\delta^{\rm data}_{\rm stat}$ [\%] & $\delta^{\rm MC}_{\rm stat}$ [\%] & $u_{\rm JES}$  [\%] & $u_{\rm JER}$  [\%] & $u_{\rm unfold}$  [\%] & $u_{\rm lumi}$ [\%]  \\
\hline
   1 & 120--170 & $1.07 \times 10^{4}$ & $0.6  $ & $2.4  $ & $^{+9.9}_{-7.7}$ & $5.6$ & $13.5$ & $2.8$ \\
   2 & 170--240 & $1.93 \times 10^{4}$ & $0.3  $ & $1.2  $ & $^{+9.2}_{-8.3}$ & $4.6$ & $8.4$ & $2.8$ \\
   3 & 240--315 & $9.21 \times 10^{3}$ & $0.4  $ & $1.0  $ & $^{+7.8}_{-8.0}$ & $1.9$ & $5.5$ & $2.8$ \\
   4 & 315--395 & $3.76 \times 10^{3}$ & $0.4  $ & $0.7  $ & $^{+7.2}_{-7.3}$ & $0.8$ & $4.6$ & $2.8$ \\
   5 & 395--480 & $1.46 \times 10^{3}$ & $0.5  $ & $0.7  $ & $^{+6.4}_{-6.5}$ & $0.7$ & $4.2$ & $2.8$ \\
   6 & 480--575 & $5.75 \times 10^{2}$ & $0.6  $ & $0.5  $ & $^{+6.3}_{-6.1}$ & $0.7$ & $2.6$ & $2.8$ \\
   7 & 575--680 & $2.23 \times 10^{2}$ & $0.5  $ & $0.5  $ & $^{+6.6}_{-6.2}$ & $0.7$ & $2.1$ & $2.8$ \\
\hline
\hline
\end{tabular}
\caption{\label{tab:xs_RapGap1SumPtCentral}
Measured differential four-jet cross section for $R=0.4$ jets, in bins of \sumptcent, along with the uncertainties in the measurement. The events are selected using the inclusive analysis cuts and $\maxDrapij>1$. All other details are as for \tabref{tab:xs_0JetPt}.
}
\end{table*}

\begin{table*}[h!]
\centering
\small
\begin{tabular}{c c c c c c c c c c c c}
\hline
\hline
Bin  & Bin edges [GeV] & ${\rm d}\sigma/{\rm d}(\sumptcent)$ [fb/GeV] & $\delta^{\rm data}_{\rm stat}$ [\%] & $\delta^{\rm MC}_{\rm stat}$ [\%] & $u_{\rm JES}$  [\%] & $u_{\rm JER}$  [\%] & $u_{\rm unfold}$  [\%] & $u_{\rm lumi}$ [\%]  \\
\hline
   1 & 120--170 & $4.3 \times 10^{2}$ & $2.5  $ & $2.3  $ & $^{+5.9}_{-4.9}$ & $2.4$ & $9.2$ & $2.8$ \\
   2 & 170--240 & $1.15 \times 10^{3}$ & $0.9  $ & $0.9  $ & $^{+6.4}_{-5.8}$ & $2.2$ & $5.8$ & $2.8$ \\
   3 & 240--315 & $1.01 \times 10^{3}$ & $0.8  $ & $0.8  $ & $^{+6.9}_{-6.6}$ & $1.6$ & $4.5$ & $2.8$ \\
   4 & 315--395 & $1.48 \times 10^{3}$ & $0.7  $ & $0.7  $ & $^{+6.8}_{-6.9}$ & $1.4$ & $6.3$ & $2.8$ \\
   5 & 395--480 & $1.13 \times 10^{3}$ & $0.6  $ & $0.5  $ & $^{+6.6}_{-6.6}$ & $1.3$ & $6.5$ & $2.8$ \\
   6 & 480--575 & $5.73 \times 10^{2}$ & $0.6  $ & $0.5  $ & $^{+6.6}_{-6.2}$ & $1.0$ & $3.7$ & $2.8$ \\
   7 & 575--680 & $2.25 \times 10^{2}$ & $0.5  $ & $0.5  $ & $^{+6.7}_{-6.3}$ & $0.9$ & $2.3$ & $2.8$ \\
\hline
\hline
\end{tabular}
\caption{\label{tab:xs_RapGap1SumPtCentral_pt250}
Measured differential four-jet cross section for $R=0.4$ jets, in bins of \sumptcent, along with the uncertainties in the measurement. The events are selected using the inclusive analysis cuts and $\maxDrapij>1$, as well as $\ptOne>250$ GeV. All other details are as for \tabref{tab:xs_0JetPt}.
}
\end{table*}

\begin{table*}[h!]
\centering
\small
\begin{tabular}{c c c c c c c c c c c c}
\hline
\hline
Bin  & Bin edges [GeV] & ${\rm d}\sigma/{\rm d}(\sumptcent)$ [fb/GeV] & $\delta^{\rm data}_{\rm stat}$ [\%] & $\delta^{\rm MC}_{\rm stat}$ [\%] & $u_{\rm JES}$  [\%] & $u_{\rm JER}$  [\%] & $u_{\rm unfold}$  [\%] & $u_{\rm lumi}$ [\%]  \\
\hline
   1 & 120--170 & $1.78 \times 10^{1}$ & $0.8  $ & $2.5  $ & $^{+6.2}_{-6.6}$ & $1.4$ & $3.3$ & $2.8$ \\
   2 & 170--240 & $5.29 \times 10^{1}$ & $0.5  $ & $1.0  $ & $^{+5.9}_{-6.6}$ & $1.4$ & $3.8$ & $2.8$ \\
   3 & 240--315 & $7.88 \times 10^{1}$ & $0.3  $ & $0.7  $ & $^{+5.8}_{-6.5}$ & $1.4$ & $4.6$ & $2.8$ \\
   4 & 315--395 & $9.39 \times 10^{1}$ & $0.3  $ & $0.7  $ & $^{+6.3}_{-6.3}$ & $1.4$ & $4.3$ & $2.8$ \\
   5 & 395--480 & $8.49 \times 10^{1}$ & $0.3  $ & $0.8  $ & $^{+6.4}_{-6.3}$ & $1.4$ & $4.9$ & $2.8$ \\
   6 & 480--575 & $1.37 \times 10^{2}$ & $0.2  $ & $0.4  $ & $^{+6.2}_{-6.3}$ & $1.2$ & $4.1$ & $2.8$ \\
   7 & 575--680 & $1.09 \times 10^{2}$ & $0.2  $ & $0.4  $ & $^{+6.7}_{-6.3}$ & $1.1$ & $3.6$ & $2.8$ \\
\hline
\hline
\end{tabular}
\caption{\label{tab:xs_RapGap1SumPtCentral_pt400}
Measured differential four-jet cross section for $R=0.4$ jets, in bins of \sumptcent, along with the uncertainties in the measurement. The events are selected using the inclusive analysis cuts and $\maxDrapij>1$, as well as $\ptOne>400$ GeV. All other details are as for \tabref{tab:xs_0JetPt}.
}
\end{table*}

\begin{table*}[h!]
\centering
\small
\begin{tabular}{c c c c c c c c c c c c}
\hline
\hline
Bin  & Bin edges [GeV] & ${\rm d}\sigma/{\rm d}(\sumptcent)$ [fb/GeV] & $\delta^{\rm data}_{\rm stat}$ [\%] & $\delta^{\rm MC}_{\rm stat}$ [\%] & $u_{\rm JES}$  [\%] & $u_{\rm JER}$  [\%] & $u_{\rm unfold}$  [\%] & $u_{\rm lumi}$ [\%]  \\
\hline
   1 & 120--170 & 1.94       & $2.6  $ & $3.5  $ & $^{+11.7}_{-5.8}$ & $10.1$ & $16.8$ & $2.8$ \\
   2 & 170--240 & 4.79       & $1.5  $ & $1.6  $ & $^{+8.0}_{-6.6}$ & $5.5$ & $8.6$ & $2.8$ \\
   3 & 240--315 & 6.03       & $1.4  $ & $1.4  $ & $^{+6.9}_{-7.5}$ & $3.2$ & $4.8$ & $2.8$ \\
   4 & 315--395 & 7.51       & $1.1  $ & $1.2  $ & $^{+7.3}_{-7.7}$ & $2.2$ & $3.9$ & $2.8$ \\
   5 & 395--480 & 9.99       & $0.9  $ & $1.0  $ & $^{+6.8}_{-7.7}$ & $1.9$ & $3.7$ & $2.8$ \\
   6 & 480--575 & $1.19 \times 10^{1}$ & $0.8  $ & $0.9  $ & $^{+7.1}_{-7.1}$ & $1.9$ & $3.7$ & $2.8$ \\
   7 & 575--680 & $1.49 \times 10^{1}$ & $0.6  $ & $0.8  $ & $^{+7.3}_{-6.8}$ & $1.8$ & $3.6$ & $2.8$ \\
   8 & 680--795 & $1.81 \times 10^{1}$ & $0.6  $ & $0.5  $ & $^{+7.4}_{-7.1}$ & $1.6$ & $3.3$ & $2.8$ \\
   9 & 795--930 & $1.21 \times 10^{1}$ & $0.6  $ & $0.5  $ & $^{+7.3}_{-7.1}$ & $1.4$ & $3.4$ & $2.8$ \\
  10 & 930--1085 & 8.24       & $0.8  $ & $0.6  $ & $^{+7.0}_{-7.3}$ & $1.3$ & $3.3$ & $2.8$ \\
  11 & 1085--1260 & 4.41       & $1.0  $ & $0.7  $ & $^{+7.2}_{-7.6}$ & $1.3$ & $2.9$ & $2.8$ \\
  12 & 1260--1465 & 1.54       & $1.4  $ & $0.5  $ & $^{+7.9}_{-7.7}$ & $1.4$ & $2.8$ & $2.8$ \\
  13 & 1465--1705 & $5.01 \times 10^{-1}$ & $2.1  $ & $0.5  $ & $^{+7.9}_{-7.8}$ & $1.5$ & $2.9$ & $2.8$ \\
  14 & 1705--1980 & $1.57 \times 10^{-1}$ & $3.7  $ & $0.6  $ & $^{+8.0}_{-7.8}$ & $1.8$ & $2.6$ & $2.8$ \\
  15 & 1980--2300 & $4.38 \times 10^{-2}$ & $7.1  $ & $0.7  $ & $^{+8.2}_{-8.3}$ & $1.5$ & $2.4$ & $2.8$ \\
  16 & 2300--5000 & $1.62 \times 10^{-3}$ & $11.8 $ & $0.6  $ & $^{+11.1}_{-9.5}$ & $1.2$ & $3.7$ & $2.8$ \\
\hline
\hline
\end{tabular}
\caption{\label{tab:xs_RapGap1SumPtCentral_pt550}
Measured differential four-jet cross section for $R=0.4$ jets, in bins of \sumptcent, along with the uncertainties in the measurement. The events are selected using the inclusive analysis cuts and $\maxDrapij>1$, as well as $\ptOne>550$ GeV. All other details are as for \tabref{tab:xs_0JetPt}.
}
\end{table*}

 \clearpage 
\begin{table*}[h!]
\centering
\small
\begin{tabular}{c c c c c c c c c c c c}
\hline
\hline
Bin  & Bin edges [GeV] & ${\rm d}\sigma/{\rm d}(\sumptcent)$ [fb/GeV] & $\delta^{\rm data}_{\rm stat}$ [\%] & $\delta^{\rm MC}_{\rm stat}$ [\%] & $u_{\rm JES}$  [\%] & $u_{\rm JER}$  [\%] & $u_{\rm unfold}$  [\%] & $u_{\rm lumi}$ [\%]  \\
\hline
   1 & 120--185 & $1.01 \times 10^{4}$ & $0.5  $ & $2.2  $ & $^{+10.9}_{-8.3}$ & $4.6$ & $11.1$ & $2.8$ \\
   2 & 185--270 & $1.23 \times 10^{4}$ & $0.3  $ & $1.1  $ & $^{+8.9}_{-8.9}$ & $3.9$ & $6.7$ & $2.8$ \\
   3 & 270--365 & $4.38 \times 10^{3}$ & $0.4  $ & $1.1  $ & $^{+8.0}_{-7.5}$ & $1.7$ & $4.4$ & $2.8$ \\
   4 & 365--465 & $1.39 \times 10^{3}$ & $0.6  $ & $0.7  $ & $^{+6.8}_{-6.6}$ & $0.8$ & $3.9$ & $2.8$ \\
   5 & 465--575 & $4.43 \times 10^{2}$ & $0.7  $ & $0.5  $ & $^{+6.2}_{-6.4}$ & $0.8$ & $2.5$ & $2.8$ \\
   6 & 575--700 & $1.44 \times 10^{2}$ & $0.6  $ & $0.5  $ & $^{+6.5}_{-6.5}$ & $0.8$ & $2.0$ & $2.8$ \\
   7 & 700--845 & $4.56 \times 10^{1}$ & $0.3  $ & $0.5  $ & $^{+6.5}_{-6.8}$ & $0.8$ & $1.9$ & $2.8$ \\
   8 & 845--1005 & $1.46 \times 10^{1}$ & $0.6  $ & $0.5  $ & $^{+7.3}_{-7.1}$ & $0.8$ & $2.2$ & $2.8$ \\
   9 & 1005--1195 & 4.58       & $0.9  $ & $0.8  $ & $^{+7.7}_{-7.6}$ & $0.8$ & $2.9$ & $2.8$ \\
  10 & 1195--1410 & 1.38       & $1.6  $ & $0.7  $ & $^{+8.1}_{-7.8}$ & $0.8$ & $3.6$ & $2.8$ \\
  11 & 1410--1665 & $4.12 \times 10^{-1}$ & $2.4  $ & $0.6  $ & $^{+8.7}_{-8.0}$ & $0.9$ & $3.6$ & $2.8$ \\
  12 & 1665--1960 & $1.10 \times 10^{-1}$ & $4.4  $ & $0.7  $ & $^{+9.0}_{-8.4}$ & $1.3$ & $3.4$ & $2.8$ \\
  13 & 1960--5000 & $4.31 \times 10^{-3}$ & $7.5  $ & $0.7  $ & $^{+8.9}_{-9.3}$ & $1.6$ & $3.0$ & $2.8$ \\
\hline
\hline
\end{tabular}
\caption{\label{tab:xs_RapGap2SumPtCentral}
Measured differential four-jet cross section for $R=0.4$ jets, in bins of \sumptcent, along with the uncertainties in the measurement. The events are selected using the inclusive analysis cuts and $\maxDrapij>2$. All other details are as for \tabref{tab:xs_0JetPt}.
}
\end{table*}

\begin{table*}[h!]
\centering
\small
\begin{tabular}{c c c c c c c c c c c c}
\hline
\hline
Bin  & Bin edges [GeV] & ${\rm d}\sigma/{\rm d}(\sumptcent)$ [fb/GeV] & $\delta^{\rm data}_{\rm stat}$ [\%] & $\delta^{\rm MC}_{\rm stat}$ [\%] & $u_{\rm JES}$  [\%] & $u_{\rm JER}$  [\%] & $u_{\rm unfold}$  [\%] & $u_{\rm lumi}$ [\%]  \\
\hline
   1 & 120--185 & $3.96 \times 10^{2}$ & $2.1  $ & $1.9  $ & $^{+5.7}_{-6.5}$ & $1.8$ & $8.4$ & $2.8$ \\
   2 & 185--270 & $8.47 \times 10^{2}$ & $0.9  $ & $0.8  $ & $^{+6.8}_{-6.6}$ & $1.7$ & $6.1$ & $2.8$ \\
   3 & 270--365 & $8.57 \times 10^{2}$ & $0.9  $ & $1.0  $ & $^{+7.1}_{-6.8}$ & $1.3$ & $7.3$ & $2.8$ \\
   4 & 365--465 & $9.37 \times 10^{2}$ & $0.7  $ & $0.5  $ & $^{+7.0}_{-6.8}$ & $1.1$ & $6.9$ & $2.8$ \\
   5 & 465--575 & $4.35 \times 10^{2}$ & $0.7  $ & $0.5  $ & $^{+6.6}_{-6.5}$ & $1.1$ & $3.8$ & $2.8$ \\
   6 & 575--700 & $1.45 \times 10^{2}$ & $0.6  $ & $0.5  $ & $^{+6.7}_{-6.6}$ & $1.1$ & $2.4$ & $2.8$ \\
   7 & 700--845 & $4.59 \times 10^{1}$ & $0.3  $ & $0.5  $ & $^{+6.7}_{-6.9}$ & $1.3$ & $2.0$ & $2.8$ \\
   8 & 845--1005 & $1.46 \times 10^{1}$ & $0.6  $ & $0.5  $ & $^{+7.4}_{-7.2}$ & $1.4$ & $2.2$ & $2.8$ \\
   9 & 1005--1195 & 4.61       & $0.9  $ & $0.7  $ & $^{+7.8}_{-7.6}$ & $1.4$ & $2.9$ & $2.8$ \\
  10 & 1195--1410 & 1.39       & $1.6  $ & $0.7  $ & $^{+8.2}_{-7.8}$ & $1.5$ & $3.6$ & $2.8$ \\
  11 & 1410--1665 & $4.14 \times 10^{-1}$ & $2.4  $ & $0.6  $ & $^{+8.8}_{-8.0}$ & $1.5$ & $3.6$ & $2.8$ \\
  12 & 1665--1960 & $1.11 \times 10^{-1}$ & $4.4  $ & $0.6  $ & $^{+9.1}_{-8.4}$ & $1.8$ & $3.4$ & $2.8$ \\
  13 & 1960--5000 & $4.33 \times 10^{-3}$ & $7.4  $ & $0.7  $ & $^{+8.9}_{-9.3}$ & $2.0$ & $3.0$ & $2.8$ \\
\hline
\hline
\end{tabular}
\caption{\label{tab:xs_RapGap2SumPtCentral_pt250}
Measured differential four-jet cross section for $R=0.4$ jets, in bins of \sumptcent, along with the uncertainties in the measurement. The events are selected using the inclusive analysis cuts and $\maxDrapij>2$, as well as $\ptOne>250$ GeV. All other details are as for \tabref{tab:xs_0JetPt}.
}
\end{table*}

\begin{table*}[h!]
\centering
\small
%
\caption{\label{tab:xs_RapGap3SumPtCentral}
Measured differential four-jet cross section for $R=0.4$ jets, in bins of \sumptcent, along with the uncertainties in the measurement. The events are selected using the inclusive analysis cuts and $\maxDrapij>3$. All other details are as for \tabref{tab:xs_0JetPt}.
}
\end{table*}

\begin{table*}[h!]
\centering
\small
%
\caption{\label{tab:xs_RapGap4SumPtCentral_pt550}
Measured differential four-jet cross section for $R=0.4$ jets, in bins of \sumptcent, along with the uncertainties in the measurement. The events are selected using the inclusive analysis cuts and $\maxDrapij>4$, as well as $\ptOne>550$ GeV. All other details are as for \tabref{tab:xs_0JetPt}.
}
\end{table*}

 \clearpage 

%% file: atlas_authlist.tex
\begin{flushleft}
{\Large The ATLAS Collaboration}

\bigskip

G.~Aad$^{\rm 85}$,
B.~Abbott$^{\rm 113}$,
J.~Abdallah$^{\rm 151}$,
O.~Abdinov$^{\rm 11}$,
R.~Aben$^{\rm 107}$,
M.~Abolins$^{\rm 90}$,
O.S.~AbouZeid$^{\rm 158}$,
H.~Abramowicz$^{\rm 153}$,
H.~Abreu$^{\rm 152}$,
R.~Abreu$^{\rm 116}$
% ...
\bigskip
\\
% ...
$^{11}$ Institute of Physics, Azerbaijan Academy of Sciences, Baku, Azerbaijan\\
% ...
$^{85}$ CPPM, Aix-Marseille Universit{\'e} and CNRS/IN2P3, Marseille, France\\
% ...
$^{90}$ Department of Physics and Astronomy, Michigan State University, East Lansing MI, United States of America\\
% ...
$^{107}$ Nikhef National Institute for Subatomic Physics and University of Amsterdam, Amsterdam, Netherlands\\
% ...
$^{113}$ Homer L. Dodge Department of Physics and Astronomy, University of Oklahoma, Norman OK, United States of America\\
% ...
$^{116}$ Center for High Energy Physics, University of Oregon, Eugene OR, United States of America\\
% ...
$^{151}$ Institute of Physics, Academia Sinica, Taipei, Taiwan\\
$^{152}$ Department of Physics, Technion: Israel Institute of Technology, Haifa, Israel\\
$^{153}$ Raymond and Beverly Sackler School of Physics and Astronomy, Tel Aviv University, Tel Aviv, Israel\\
% ...
$^{158}$ Department of Physics, University of Toronto, Toronto ON, Canada\\
% ...
\end{flushleft}
